\documentclass[reqno, 11pt]{gen-m-l}
\usepackage{latexsym, amssymb}
\usepackage{hyperref}
\usepackage{amsmath}
\usepackage{amsthm}
\usepackage{epsfig}
\usepackage[mathscr]{eucal}
\numberwithin{equation}{section}
\allowdisplaybreaks[1]

\allowdisplaybreaks[1]

\newtheorem{Def}{Def.}[section]
\newtheorem{Thm}[Def]{Theorem}
\newtheorem{Prp}[Def]{Proposition}
\newtheorem{Lemma}[Def]{Lemma}
\newtheorem{Remark}[Def]{Remark}

\newcommand{\Proof}{{\em{Proof. }}}
\newcommand{\QED}{\ \hfill $\FBox$ \\[1em]}
\newcommand{\spc}{\;\;\;\;\;\;\;\;\;\;}
\newcommand{\bra}{\mbox{$< \!\!$ \nolinebreak}}
\newcommand{\ket}{\mbox{\nolinebreak $>$}}
\newcommand{\Bra}{\langle}
\newcommand{\Ket}{\rangle}
\newfont{\Bb}{msbm10 scaled 1095}
\newcommand\C{{\mbox{\Bb C}}}

\newcommand\sC{{\mbox{\bf{\scriptsize{C}}}}}
\newcommand\R{{\mbox{\Bb R}}}

\newcommand\Q{{\mbox{\Bb Q}}}
\newcommand{\1}{\mbox{\rm 1 \hspace{-1.05 em} 1}}

\newcommand{\sR}{\mbox{\rm \scriptsize I \hspace{-.8 em} R}}

\newcommand{\Pdd}{\mbox{$\partial$ \hspace{-1.2 em} $/$}}
\newcommand{\slsh}{\mbox{ \hspace{-1.1 em} $/$}}
\newcommand{\Tr}{\mbox{\rm{Tr}\/}}
\newcommand{\tr}{\mbox{tr\/}}
\newcommand{\FBox}{\rule{2mm}{2.25mm}}

\newcommand{\Sl}{\mbox{$\prec \!\!$ \nolinebreak}}
\newcommand{\Sr}{\mbox{\nolinebreak $\succ$}}
\newcommand{\Aslsh}{\mbox{ $\!\!A$ \hspace{-1.2 em} $/$}}
\newcommand{\gslsh}{\mbox{ $\!\!g$ \hspace{-1.2 em} $/$}}
\newcommand{\intL}{{\mbox{ \footnotesize{L}}} \hspace{-1.4 em} \displaystyle \int}
\newcommand{\intR}{{\mbox{ \footnotesize{R}}} \hspace{-1.4 em} \displaystyle \int}
\newcommand{\intc}{{\mbox{ \footnotesize{c}}} \hspace{-1.2 em} \displaystyle \int}
\newcommand{\intcb}{{\mbox{ \footnotesize{$\overline{c}$}}} \hspace{-1.2 em}
\displaystyle \int}

\newcommand{\g}{{\mathfrak{g}}}
\newcommand{\Pexp}{\mbox{\rm{Pexp}}}
\newcommand{\eff}{{\mbox{\scriptsize{eff}}}}
\newcommand{\beq}{\begin{equation}}
\newcommand{\eeq}{\end{equation}}

\renewcommand{\H}{\mathscr{H}}

\makeindex

\begin{document}

\frontmatter
\title{{{\vspace*{2cm}The Principle of the Fermionic Projector}} \\[.5cm] \Large{Chapters 5-8}}
\author{\Large{Felix Finster} \\[2cm]
{\normalsize{\bf{Abstract}}} \\[.3cm]
\begin{quote}
\pagestyle{empty}
\footnotesize
The  ``principle of the fermionic projector'' provides a new mathematical 
framework for the formulation of physical theories and is a promising 
approach for physics beyond the standard model. The book begins with a 
brief review of relativity, relativistic quantum mechanics and classical 
gauge theories, with the emphasis on the basic physical concepts and the 
mathematical foundations. The external field problem and Klein's paradox 
are discussed and then resolved by introducing the so-called fermionic 
projector, a global object in space-time which generalizes the notion of 
the Dirac sea. The mathematical core of the book is to give a precise 
definition of the fermionic projector and to employ methods of 
hyperbolic differential equations for its detailed analysis. The 
fermionic projector makes it possible to formulate a new type of 
variational principles in space-time. The mathematical tools for the 
analysis of the corresponding Euler-Lagrange equations are developed. A 
particular variational principle is proposed which gives rise to an 
effective interaction showing many similarities to the interactions of 
the standard model.

The main chapters of the book are easily accessible for beginning 
graduate students in mathematics or physics. Several appendices provide 
supplementary material which will be useful to the experienced researcher.
\end{quote}}

\maketitle

\setcounter{page}{4}

\tableofcontents

\chapter*{Preface to the Second Online Edition}

\vspace*{-0.25em}

In the almost twelve years since this book was completed, the fermionic projector approach
evolved to what is known today as the theory of causal fermion systems.
There has been progress in several directions: the mathematical
setting was generalized, the mathematical methods were improved and enriched,
and the physical applications have been concretized and worked out in more detail.
The current status of the theory is presented in a coherent way in
the recent monograph~\cite{cfs}. An untechnical physical introduction is given in~\cite{dice2014}.

Due to these developments, parts of the present book are superseded by the
more recent research papers or the monograph~\cite{cfs}.
However, other parts of this book have not been developed further and are still up to date.
For some aspects not covered in~\cite{cfs}, the present book is still the best reference.
Furthermore, the present book is still of interest as being the first publication in which
the causal action principle was presented.
Indeed, comparing the presentation in the present book
to the later developments should give the reader a deeper understanding of why
certain constructions were modified and how they were improved.
In order to facilitate such a study, we now outline the developments which led from the present
book to the monograph~\cite{cfs}. In order not to change the original bibliography,
a list of references to more recent research papers is given at the end of this preface,
where numbers are used (whereas the original bibliography using letters is still
at the end of the book). Similar as in the first online edition, I took the opportunity
to correct a few typos. Also, I added a few footnotes 
beginning with ``{\textsf{Online version}:}''.
Apart from these changes, the online version coincides precisely
with the printed book in the AMS/IP series. In particular, all equation numbers
are the same.

Maybe the most important change in the mathematical setup was the
move from indefinite inner product spaces to Hilbert spaces, as we now explain in detail.
Clearly, the starting point of all my considerations was Dirac theory. On Dirac wave functions in
Minkowski space, one can introduce the two inner products
\begin{align}
( \Psi | \Phi) &= \int_{\sR^3} (\overline{\Psi} \gamma^0 \Phi)(t, \vec{x})\: d^3x \tag{1} \label{sprodMin} \\
\bra \Psi|\Phi \ket &= \int_{\sR^4} \overline{\Psi(x)} \Phi(x) \: d^4x \:. \tag{2} \label{stipMin}
\end{align}
The first inner product~\eqref{sprodMin} is positive definite and thus defines a scalar product.
For solutions of the Dirac equation, it is time independent due to current conservation,
making the solution space of the Dirac equation to a Hilbert space
(more generally, the scalar product can be computed by integrating the normal component
of the Dirac current over any Cauchy surface).
The inner product~\eqref{stipMin}, on the other hand,
is indefinite. It is well-defined and covariant even on wave functions which do not satisfy the Dirac equation,
giving rise to an indefinite inner product space (which can be given a Krein space structure).
It should be pointed out that the time integral in~\eqref{stipMin} in general diverges for solutions of the
Dirac equation, a problem which I always considered to be more of technical than of fundamental nature
(this technical problem can be resolved for example
by working as in~\eqref{56}--\eqref{58} with a $\delta$-normalization in the mass
parameter or by making use of the mass oscillation property as introduced in~\cite{infinite}).

The fermionic projector approach is based on the belief that on a microscopic scale (like the Planck scale),
space-time should not be modeled by Minkowski space but should have
a different, possibly discrete structure. Consequently, the Dirac equation in Minkowski space
should not be considered as being fundamental, but it should be replaced by equations of different type.
For such a more fundamental description, the scalar product~\eqref{sprodMin} is problematic, 
because it is not clear how the analog of an integral over a hypersurface should be defined,
and why this integral should be independent of the choice of the hypersurface.
The indefinite inner product~\eqref{stipMin}, however, can easily be generalized to 
for example a discrete space-time if one simply replaces the
integral in~\eqref{stipMin} by a sum over all space-time points.
Such considerations led me to regard the indefinite inner product~\eqref{stipMin} as being
more fundamental than~\eqref{sprodMin}. This is the reason why throughout this book,
we work preferably with indefinite inner product spaces. In particular, the structure of
``discrete space-time'' is introduced on an underlying indefinite inner product space
(see~\S\ref{psec13}).

My views changed gradually over the past few years. The first input which triggered this process
was obtained when developing the existence theory for the causal action principle.
While working on this problem in the simplest setting of a finite number of
space-time points~\cite{discrete}, it became clear that in
order to ensure the existence of minimizers, one needs to assume that the 
image of the fermionic projector~$P$ is a {\em{negative definite}} subspace
of the indefinite inner product space~$(H, \bra .|. \ket)$. The fact that~$P$ has
a definite image makes it possible to introduce a Hilbert space~$(\H, \langle .|. \rangle_\H)$
by setting~$\langle .|. \rangle_\H = -\bra .| P\, . \ket$ and dividing out the null space.
This construction, which was first given in~\cite[Section~1.2.2]{rrev}, gave
an underlying Hilbert space structure. However, at this time, the connection of the corresponding
scalar product to integrals over hypersurfaces as in~\eqref{sprodMin} remained obscure.

From the mathematical point of view, having an underlying Hilbert space structure has the major
benefit that functional analytic methods in Hilbert spaces become applicable.
When thinking about how to apply these methods, it became clear that also measure-theoretic
methods are useful. This led me to generalize the mathematical setting such as to allow
for the description of not only discrete, but also continuous space-times.
This setting was first introduced in~\cite{continuum} when working out the existence theory.
This analysis also clarified which constraints one must impose in order to obtain
a mathematically well-posed variational problem.

The constructions in~\cite{continuum} also inspired the notion of the {\em{universal measure}},
as we now outline. When working out the existence theory, it became clear that instead of using the
kernel of the fermionic projector, the causal action principle can be formulated equivalently
in terms of the local correlation operators~$F(x)$ which relate the Hilbert space scalar product
to the spin scalar product by
\[ \langle \psi | F(x) \phi \rangle_\H = -\Sl \psi(x) | \phi(x) \Sr_x \:. \]
In this formulation, the only a-priori structure of space-time is that of being a measure space~$(M, \mu)$.
The local correlation operators give rise to a mapping
\[ F \::\: M \rightarrow {\mathscr{F}} \:,\quad x \mapsto F(x) \:, \]
where~${\mathscr{F}}$ is the subset of finite rank operators on~$\H$ which are symmetric and 
(counting multiplicities) have at most~$2N$ positive and at most~$2N$ negative eigenvalues
(where~$N$ denotes the number of sectors).
Then, instead of working with the measure~$\mu$, the causal action can be expressed
in terms of the push-forward measure~$\rho = F_* \mu$, being a measure on~${\mathscr{F}}$
(defined by~$\rho(\Omega) = \mu(F^{-1}(\Omega))$).
As a consequence, it seems natural to leave out the measure space~$(M, \mu)$ and to work instead
directly with the measure~$\rho$ on~${\mathscr{F}}$, referred to as the universal measure.
We remark that working with~$(M, \mu)$ has the potential benefit that it is possible to
prescribe properties of the measure~$\rho$. In particular, if~$\mu$ is a discrete measure,
then so is~$\rho$ (for details see~\cite[Section~1.2]{continuum}). However, the analysis of the
causal action principle in~\cite{support} suggests
that minimizing measures are always discrete, even if one varies over all regular Borel measures
(which may have discrete and continuous components).
With this in mind, it seems unnecessary to arrange the discreteness of the measure~$\rho$
by starting with a discrete measure space~$(M, \mu)$.
Then the measure space~$(M, \mu)$ becomes obsolete.
These considerations led me to the conviction that one should work with the universal measure~$\rho$,
which should be varied within the class of all regular Borel measures. Working with general regular Borel measures
also has the advantage that it becomes possible to take convex combinations of universal measures,
which seems essential for getting the connection to second-quantized bosonic fields
(see the notions of decoherent replicas of space-time and of microscopic mixing of wave functions
in~\cite{qft} and~\cite{qftlimit}).

Combining all the above results led to the framework of {\em{causal fermion systems}}, where
a physical system is described by a Hilbert space~$(\H, \langle .|. \rangle_\H)$ and
the universal measure~$\rho$ on~${\mathscr{F}}$. This framework was first introduced in~\cite{rrev}.
Subsequently, the analytic, geometric and topological structures encoded in a causal fermion system
were worked out systematically; for an overview see~\cite[Chapter~1]{cfs}.

From the conceptual point of view, the setting of causal fermion systems
and the notion of the universal measure considerably changed both the
role of the causal action principle and the concept of what space-time is.
Namely, in the causal action principle in this book, one varies the fermionic projector~$P$ in
a given discrete space-time.
In the setting of causal fermion systems, however, one varies instead the universal measure~$\rho$,
being a measure on linear operators on an abstract Hilbert space.
In the latter formulation, there is no space-time to begin with. On the contrary, space-time is
introduced later as the support of the universal measure.
In this way, the causal action principle evolved from a variational principle for wave functions
in space-time to a variational principle for space-time itself as well as for all structures therein.

In order to complete the summary of the conceptual modifications, we remark that
the connection between the scalar product~$\langle .|. \rangle_\H$ and surface integrals as in~\eqref{sprodMin},
which had been obscure for quite a while, was finally clarified when working out Noether-like theorems
for causal variational principles~\cite{noether}. Namely, surface integrals now have a proper generalization to
causal fermion systems in terms of so-called {\em{surface layer integrals}}.
It was shown that the symmetry of the causal action under unitary transformations acting on~$\mathscr{F}$
gives rise to conserved charges which can be expressed by surface layer integrals.
For Dirac sea configurations, these conserved charges coincide with the surface
integrals~\eqref{sprodMin}.

Another major development concerns the description of {\em{neutrinos}}.
In order to explain how these developments came about, we first note that
in this book, neutrinos are modelled as left-handed massless Dirac particles (see~\S\ref{esec1}).
This has the benefit that the neutrinos drop out of the closed chain due to
chiral cancellations (see~\S\ref{esec21} and~\S\ref{esec22}).
When writing this book, I liked chiral cancellations, and I even regarded them as
a possible explanation for the fact that neutrinos appear only with one chirality.
As a side remark, I would like to mention that
I was never concerned about experimental observations which indicate that
neutrinos do have a rest mass, because I felt that these experiments are too indirect
for making a clear case. Namely, measurements only tell us that there are
fewer neutrinos on earth than expected from the number of neutrinos generated in fusion processes
in the sun. The conventional explanation for this seeming disappearance
of solar neutrinos is via neutrino oscillations, making it
necessary to consider massive neutrinos. However, it always seemed to me that
there could be other explanations for the lack of neutrinos on earth
(for example, a modification of the weak interaction or other, yet unknown fundamental forces),
in which case the neutrinos could well be massless.

My motivation for departing from massless neutrinos was not related to experimental evidence, but
had to do with problems of mathematical consistency. Namely, I noticed that left-handed neutrinos
do not give rise to stable minimizers of the causal action (see~\cite[Section~4.2]{cfs}).
This general result led me to incorporate right-handed neutrino components, and to
explain the fact that only the left-handed component is observed by the postulate that the
regularization breaks the chiral symmetry. This procedure cured the mathematical consistency
problems and had the desired side effect that neutrinos could have a rest mass, in agreement with
neutrino oscillations.

We now comment on other developments which are of more technical nature.
These developments were mainly triggered by minor errors or shortcomings in the present book.
First, Andreas Grotz noticed when working on his master thesis in 2007 that the
normalization conditions for the fermionic projector as given in~\eqref{eq:2a1} and~\eqref{eq:2a2}
are in general violated to higher order in perturbation theory. This error was corrected in~\cite{grotz}
by a rescaling procedure.
This construction showed that there are two different perturbation expansions: with and
without rescaling. The deeper meaning of these two expansions became clearer later when
working out different normalizations of the fermionic projector.
This study was initiated by the quest for a non-perturbative construction of the fermionic projector,
as was carried out in globally hyperbolic space-times in~\cite{finite, infinite}.
It turned out that in space-times of finite lifetime, one cannot work with the $\delta$-normalization in the mass
parameter as used in~\eqref{56}--\eqref{58} (the ``mass normalization''). Instead, a proper normalization
is obtained by using a scalar product~$(.|.)$ which is represented similar to~\eqref{sprodMin}
by an integral over a spacelike hypersurface (the ``spatial normalization'').
As worked out in detail in~\cite{norm} with a convenient contour integral method,
the causal perturbation expansion without rescaling realizes the spatial normalization condition,
whereas the rescaling procedure in~\cite{grotz} gives rise to the mass normalization.
The constructions in curved space-time in~\cite{finite, infinite} as well as the general connection between the
scalar product~$( .|. )$ and the surface layer integrals in~\cite{noether} showed that
the physically correct and mathematically consistent normalization condition is
the spatial normalization condition. With this in mind, the combinatorics of the causal perturbation
expansion in this book is indeed correct, but the resulting fermionic projector does not satisfy the mass
but the spatial normalization condition.

Clearly, the analysis of the continuum limit in Chapters~\ref{esec3}--\ref{esec5}
is superseded by the much more detailed analysis in~\cite[Chapters~3-5]{cfs}.
A major change concerns the treatment of the logarithmic singularities on the light cone,
as we now point out. In the present book, some of the contributions involving logarithms are
arranged to vanish by imposing that the regularization should satisfy the relation~\eqref{e:3C}.
I tried for quite a while to construct an example of a regularization which realizes this relation,
until I finally realized that there is no such regularization, for the following reason: \\[-0.5em]

\textsc{Lemma~I.}
There is no regularization which satisfies the condition~\eqref{e:3C}.

\Proof The linear combination of monomials~$M$ in~\eqref{e:3B} involves a factor~$T^{(1)}_{[2]}$,
which has a logarithmic pole on the light cone (see~\eqref{Tldef}, \eqref{Tadef}
and~\eqref{l:3.1}). Restricting attention to the corresponding
contribution~$\sim \log|\vec{\xi}|$, we have
\[ M \asymp -\frac{1}{16 \pi^3}\:
T^{(-1)}_{[0]}\: \overline{T^{(-1)}_{[0]}\: T^{(0)}_{[0]}}\:\log|\vec{\xi}| \:. \]
As a consequence,
\begin{align*}
(M & - \overline{M})\: \overline{T^{(0)}_{[0]}}^{-1} =
-\frac{\log|\vec{\xi}|}{16 \pi^3}\: \frac{\big| T^{(-1)}_{[0]} \big|^2}{\overline{T^{(0)}_{[0]}}}
\big( \overline{T^{(0)}_{[0]}} - T^{(0)}_{[0]} \big)  \\
&= -\frac{\log|\vec{\xi}|}{16 \pi^3}\: \bigg| \frac{T^{(-1)}_{[0]}}{T^{(0)}_{[0]}} \bigg|^2
\Big( \big| T^{(0)}_{[0]} \big|^2 - \big( T^{(0)}_{[0]} \big)^2 \Big)
= -\frac{\log|\vec{\xi}|}{8 \pi^3}\: \bigg| \frac{T^{(-1)}_{[0]}}{T^{(0)}_{[0]}} \bigg|^2
\, \Big(\text{Im} \,T^{(0)}_{[0]} \Big)^2 \leq 0 \:.
\end{align*}
Since this expression has a fixed sign, it vanishes in a weak evaluation on the light cone
only if it vanishes identically to the required degree.
According to~\eqref{l:3.1}, the function~$\text{Im}\, T^{(0)}_{[0]}$ is a regularization
of the distribution~$-i \pi \delta (\xi^2) \:\varepsilon (\xi^0)/(8 \pi^3)$ on the scale~$\varepsilon$.
Hence on the light cone it is of the order~$\varepsilon^{-1}$. This gives the claim.
\QED
This no-go result led me to reconsider the whole procedure of the continuum limit.
At the same time, I tried to avoid imposing relations between the regularization parameters,
which I never felt comfortable with because I wanted the continuum limit to work for at
least a generic class of regularizations. Resolving this important issue took
me a lot of time and effort. My considerations eventually led to the method of compensating the
logarithmic poles by a {\em{microlocal chiral transformation}}.
These construction as well as many preliminary considerations are given in~\cite[Section~3.7]{cfs}.

Finally, I would like to make a few comments on each chapter of the book.
Chapters~\ref{secintro}--\ref{secpfp} are still up to date, except for the generalizations
and modifications mentioned above. Compared to the presentation in~\cite{cfs},
I see the benefit that these chapters might be easier to read and might convey
a more intuitive picture of the underlying physical ideas.
Chapter~\ref{psec2} is still the best reference for the general derivation of the formalism of the
continuum limit. In~\cite[Chapter~2]{cfs} I merely explained the regularization effects in examples
and gave an overview of the methods and results in Chapter~\ref{psec2}, but without repeating
the detailed constructions. Chapter~\ref{esec2} is still the only reference where the
form of the causal action is motivated and derived step by step. Also, the
notion of state stability is introduced in detail, thus providing the basis for the
later analysis in~\cite{reg, vacstab}. As already mentioned above, the analysis in
Chapters~\ref{esec3}--\ref{secegg} is outdated. I recommend the reader to study
instead~\cite[Chapters~3--5]{cfs}. The Appendices are still valuable. I added a few
footnotes which point to later improvements and further developments.
%
\\[1.5em]
\hspace*{1cm} \hfill Felix Finster, Regensburg, August 2016 \\[0.5cm]


\centerline{\large{\bf{References}}}

\vspace*{.5em}

\newcommand\oldchapter{}
\let\oldchapter=\chapter
\renewcommand{\chapter}[2]{}

\providecommand{\bysame}{\leavevmode\hbox to3em{\hrulefill}\thinspace}
\providecommand{\MR}{\relax\ifhmode\unskip\space\fi MR }
\providecommand{\MRhref}[2]{%
  \href{http://www.ams.org/mathscinet-getitem?mr=#1}{#2}
}
\providecommand{\href}[2]{#2}

\let\chapter=\oldchapter

\include{pfp1}
\mainmatter
\setcounter{page}{123}
\chapter{The Euler-Lagrange Equations in the Vacuum}
\setcounter{equation}{0} \label{esec2} In this chapter we discuss a
general class of equations of discrete space-time in the vacuum, for
a fermionic projector which is modeled according to the
configuration of the fermions in the standard model. Our goal is to
motivate and explain the model variational principle introduced
in~{\S}\ref{psec15} in more detail and to give an overview of other
actions which might be of physical interest. The basic structure of
the action will be obtained by considering the continuum limit
({\S}\ref{esec21}--{\S}\ref{esec23}), whereas the detailed form of
our model variational principle will be motivated by a consideration
which also uses the behavior of the fermionic projector away from
the light cone ({\S}\ref{esec24}).

\section{The Fermion Configuration of the Standard Model}
\setcounter{equation}{0} \label{esec1}
Guided by the configuration of the leptons and quarks in the standard model,
we want to introduce a continuum fermionic projector which seems
appropriate for the formulation of a realistic physical model.
We proceed in several steps and begin for simplicity with the first
generation of elementary particles, i.e.\ with the quarks $d$, $u$
and the leptons $e$, $\nu_e$. The simplest way to incorporate these
particles into the fermionic projector as defined in~{\S}\ref{jsec3} is
to take the direct sum of the corresponding Dirac seas,
\begin{equation}
P^{\mbox{\scriptsize{sea}}}
\;=\; \bigoplus_{a=1}^4 X_a \:\frac{1}{2} \left(p_{m_a} - k_{m_a} \right)
    \label{e:1a}
\end{equation}
with $m_1=m_d$, $m_2=m_u$, $m_3=m_e$, $m_4=0$ and $X_1=X_2=X_3=\1$,
$X_4=\chi_L$. The spin dimension in~(\ref{e:1a}) is $(8,8)$. Interpreting
isometries of the spin scalar product as local gauge transformations
(see~{\S}\ref{psec11}), the gauge group is $U(8,8)$.
Clearly, the ordering of the Dirac seas in the direct sum in~(\ref{e:1a}) is a
pure convention.  Nevertheless, our choice entails no loss of generality
because any other ordering can be obtained from~(\ref{e:1a}) by a suitable
global gauge transformation.

In the standard model, the quarks come in three ``colors,'' with an underlying
$SU(3)$ symmetry.  We can build in this symmetry here by taking three identical
copies of each quark Dirac sea.
This leads us to consider instead of~(\ref{e:1a}) the fermionic projector
\begin{equation}
    P^{\mbox{\scriptsize{sea}}} \;=\; \bigoplus_{a=1}^N X_a \:\frac{1}{2} \left(p_{m_a} - k_{m_a} \right)    \label{e:1k}
\end{equation}
with $N=8$ and $m_1=m_2=m_3=m_d$, $m_4=m_5=m_6=m_u$, $m_7=m_e$, $m_8=0$,
and $X_1=\cdots=X_7=\1$, $X_8=\chi_L$. Now the spin dimension is
$(16,16)$, and the gauge group is $U(16,16)$.

Let us now consider the realistic situation of three generations.
Grouping the elementary particles according to their lepton number
and isospin, we get the four families $(d, s, b)$, $(u, c, t)$,
$(\nu_e, \nu_\mu, \nu_\tau)$ and $(e, \mu, \tau)$. In the
standard model, the particles within each family couple to the gauge
fields in the same way. In order to also arrange this here, we take
the (ordinary) sum of these Dirac seas. Thus we define the
{\em{fermionic projector of the vacuum}}\index{fermionic projector!of
  the vacuum} by
\begin{equation}
P(x,y) \;=\; \bigoplus_{a=1}^N \:\sum_{\alpha=1}^3 X_a
\:\frac{1}{2} \left(p_{m_{a\alpha}} - k_{m_{a\alpha}} \right)
\label{e:1i}
\end{equation}
with $N=8$, $X_1=\cdots=X_7=\1$ and $X_8=\chi_L$; furthermore
$m_{11}=m_{21}=m_{31}=m_d$, $m_{12}=m_{22}=m_{32}=m_s$,
$m_{13}=m_{23}=m_{33}=m_b$, $m_{41}=m_{51}=m_{61}=m_u$,
\ldots, $m_{71}=m_e$, $m_{72}=m_\mu$,
$m_{73}=m_\tau$, and $m_{81}=m_{82}=m_{83}=0$.
We refer to the direct summands in~(\ref{e:1i}) as {\em{sectors}}\index{sector}.
The spin dimension in~(\ref{e:1i}) is again $(16,16)$.

The fermionic projector of the vacuum~(\ref{e:1i}) fits into the
general framework of~{\S}\ref{jsec3} (it is a special case of~(\ref{vs1})
obtained by setting~$g(a)=3$). Thus the interaction can be
introduced exactly as in~{\S}\ref{jsec3} by defining the
auxiliary fermionic projector~(\ref{vafp1}), inserting bosonic
potentials~${\mathcal{B}}$ into the auxiliary Dirac equation~(\ref{b})
and finally taking the partial trace~(\ref{part}).
We point out that our only free parameters are the nine masses of
the elementary leptons and quarks.  The operator ${\mathcal{B}}$
which describes the interaction must satisfy the causality
compatibility condition~(\ref{89}).  But apart
from this mathematical consistency condition, the operator
${\mathcal{B}}$ is arbitrary.  Thus in contrast to the
standard model, we do not put in the structure of the fundamental
interactions here, i.e.\ we do not specify the gauge groups, the
coupling of the gauge fields to the fermions, the coupling
constants, the CKM matrix, the Higgs mechanism, the masses of the
$W$- and $Z$-bosons, etc.  The reason is that in our description,
the physical interaction is to be determined and described
by our variational principle in discrete space-time.

\section{The General Two-Point Action}
\setcounter{equation}{0} \label{esec20}
The model variational principle in~{\S}\ref{psec15} was formulated via
a two-point action (\ref{p:35}, \ref{p:44}, \ref{p:45}).
In the remainder of this chapter, we shall consider the general two-point action
\begin{equation}
    S \;=\; \sum_{x,y \in M} {\mathcal{L}}[P(x,y)\:P(y,x)]
    \label{e:2a}
\end{equation}
more systematically and study for which Lagrangians ${\mathcal{L}}$
the corresponding Euler-Lagrange (EL) equations are satisfied in the vacuum (a
problem arising for actions other than two-point actions is
discussed in Remark~\ref{remark31}). Let us derive the EL
equations corresponding to~(\ref{e:2a}). We set \label{A_xy}
\begin{equation}
    A_{xy} \;=\; P(x,y)\:P(y,x)
    \label{e:2Adef}
\end{equation}
and for simplicity often omit the subscript ``$_{xy}$'' in what
follows. In a gauge, $A$ is represented by a $4N \times 4N$
matrix, with $N=8$ for the fermion configuration of the standard
model~(\ref{e:1i}).  We write the matrix components with Greek
indices, $A=(A^\alpha_\beta)_{\alpha, \beta=1,\ldots,4N}$.  The
Lagrangian in~(\ref{e:2a}) is a functional on $4N \times 4N$
matrices.  Denoting its gradient by ${\mathcal{M}}$, \label{mathcalM}
\[ {\mathcal{M}}[A] \;=\; ({\mathcal{M}}[A]^\alpha_\beta)_{\alpha, \beta=1,\ldots,4N}
\spc {\mbox{with}}\spc {\mathcal{M}}[A]^\alpha_\beta \;=\;
\frac{\partial {\mathcal{L}}[A]}{\partial A^\beta_\alpha} \:, \]
the variation of ${\mathcal{L}}$ is given by
\begin{equation}
    \delta {\mathcal{L}}[A] \;=\; \sum_{\alpha, \beta=1}^{4N}
    {\mathcal{M}}[A]^\alpha_\beta \:\delta A^\beta_\alpha \;=\; \Tr \left(
    {\mathcal{M}}[A] \:\delta A \right) \:,
    \label{e:2x}
\end{equation} \label{Tr2}
where ``$\Tr$'' denotes the trace of $4N \times 4N$ matrices. Summing over $x$ and $y$ yields the variation of the action,
\[ \delta S \;=\; \sum_{x,y \in M} \delta {\mathcal{L}}[A_{xy}]
\;=\; \sum_{x,y \in M} \Tr \left( {\mathcal{M}}[A_{xy}] \:\delta
A_{xy} \right) \:. \] We substitute the identity
\begin{equation}
    \delta A_{xy} \;=\; \delta P(x,y) \:P(y,x) \:+\: P(x,y) \:\delta P(y,x)
    \label{e:2D}
\end{equation}
and use the symmetry $x \leftrightarrow y$ as well as the fact that the
trace is cyclic to obtain
\begin{equation}
\delta S \;=\; 4 \: \sum_{x,y \in M} \Tr \left( Q(x,y) \:\delta P(y,x) \right)
    \label{e:2bold}
\end{equation}
with
\begin{equation}
Q(x,y) \;=\; \frac{1}{4} \left( {\mathcal{M}}[A_{xy}]\:P(x,y)
\:+\: P(x,y) \:{\mathcal{M}}[A_{yx}] \right) \:.
\label{526a}
\end{equation}
This equation can be simplified, in the same spirit as the
transformation from (\ref{p:49prel}) to~(\ref{p:49}) for our model
variational principle.
\begin{Lemma} \label{lemmaMtrans}
In the above setting,
\[ {\mathcal{M}}[A_{xy}]\: P(x,y) \;=\; P(x,y)\: {\mathcal{M}}[A_{yx}]\:. \]
\end{Lemma}
{\Proof} We consider for fixed~$x,y \in M$ variations of the general form
\[ \delta P(x,y) \;=\; C \:,\spc \delta P(y,x) \;=\; C^* \]
with~$C$ any $4N \times 4N$-matrix. Then, using~(\ref{e:2x}, \ref{e:2D})
and cyclically commuting the factors inside the trace, we obtain
\[ \delta {\mathcal{L}}[A_{xy}] \;=\;
\Tr \Big( {\mathcal{M}}[A_{xy}]\: P(x,y)\: C^* \:+\:
P(y,x)\: {\mathcal{M}}[A_{xy}]\: C \Big) . \]
Since the Lagrangian is symmetric~(\ref{symmetry}), this is equal to
\[ \delta {\mathcal{L}}[A_{yx}] \;=\;
\Tr \Big( {\mathcal{M}}[A_{yx}]\: P(y,x)\: C \:+\:
P(x,y)\: {\mathcal{M}}[A_{yx}]\: C^* \Big) . \]
Subtracting these two equations, we get
\begin{eqnarray*}
\lefteqn{ \Tr \Big( \left( {\mathcal{M}}[A_{xy}]\: P(x,y) -
P(x,y)\: {\mathcal{M}}[A_{yx}] \right) C^* \Big) } \\
&=& \Tr \Big( \left( {\mathcal{M}}[A_{yx}]\: P(y,x)
- P(y,x)\: {\mathcal{M}}[A_{xy}] \right) C \Big) \:.
\end{eqnarray*}
Changing the phase of~$C$ according to~$C \to e^{i \varphi} C$,
$\varphi \in [0, 2 \pi)$, one sees that both sides of the equation vanish
separately, and thus
\[ \Tr \Big( \left( {\mathcal{M}}[A_{xy}]\: P(x,y) -
P(x,y)\: {\mathcal{M}}[A_{yx}] \right) C^* \Big) \;=\; 0\:. \]
Since~$C$ is arbitrary, the claim follows.
\QED
Using this lemma, we can simplify~(\ref{526a}) to
\begin{equation}
Q(x,y) \;=\; \frac{1}{2} \:{\mathcal{M}}[A_{xy}]\:P(x,y) \:.
\label{e:2c}
\end{equation}

We can also write~(\ref{e:2bold}) in the compact form
\begin{equation}
\delta S \;=\; 4\: \tr (Q \:\delta P) \:, \label{e:2b}
\end{equation}
where $Q$ is the operator on $H$ with kernel~(\ref{e:2c}).
Exactly as in~{\S}\ref{psec15} we consider unitary variations of
$P$ (\ref{p:50a}) with finite support, i.e.
\[ \delta P \;=\; i\: [B, P] \:, \]
where $B$ is a Hermitian operator of finite rank. Substituting
into~(\ref{e:2b}) and cyclically commuting the operators in the trace yields that
\[ \delta S \;=\; 4i\: \tr (Q \:[B,P]) \;=\; 4i\: \tr ([P,Q] \:B) \:. \]
Since $B$ is arbitrary, we conclude that
\begin{equation}
    [P,Q] \;=\; 0 \label{e:2d}
\end{equation}
with $Q$ according to~(\ref{e:2c}). These are the EL
equations\index{Euler-Lagrange (EL) equations}. \label{ELequations}

\section{The Spectral Decomposition of $P(x,y)\:P(y,x)$}
\setcounter{equation}{0} \label{esec21}
As outlined in~{\S}\ref{psec15} in a model example, the EL
equations~(\ref{e:2d}) can be analyzed using the spectral decomposition of the matrix $A$, (\ref{p:46}). On the other hand, we saw in Chapter~\ref{psec2}
that~$A$ should be looked at in an expansion about the light
cone. We shall now combine these methods and compute the eigenvalues and
spectral projectors of $A$ using the general formalism of the continuum limit introduced in~{\S}\ref{psec26}.

Since the fermionic projector of the vacuum~(\ref{e:1i}) is a direct sum,
we can study the eight sectors separately.
We first consider the {\em{neutrino sector}} $n=8$, i.e.
\[ P(x,y) \;=\; \sum_{\alpha=1}^3 \chi_L \:\frac{1}{2} \left(
p_{m_\alpha} - k_{m_\alpha} \right)
\spc {\mbox{with}} \spc m_\alpha=0. \]
Assuming that the regularized Dirac seas have a vector-scalar
structure~(\ref{p:2f2}) and regularizing as explained
after~(\ref{p:D0}), the regularized fermionic projector, which with a
slight abuse of notation we denote again by $P(x,y)$, is of the form
\begin{equation}
    P(x,y) \;=\; \chi_L \:g_j(x,y) \:\gamma^j \label{e:2u}
\end{equation}
with suitable functions $g_j$. Since $P$ is Hermitian, $P(y,x)$ is given by
\[ P(y,x) \;=\; P(x,y)^* \;=\;
\chi_L \:\overline{g_j(x,y)} \:\gamma^j \:. \]
Omitting the arguments $(x,y)$ of the functions $g_j$, we obtain
for the $4 \times 4$ matrix $A$
\begin{equation}
    A \;=\; \chi_L \:\gslsh \;\chi_L \:\overline{\gslsh} \;=\;
    \chi_L\: \chi_R \:\gslsh\:\overline{\gslsh} \;=\; 0 \:.
    \label{e:26a}
\end{equation}
Hence in the neutrino sector, $A_{xy}$ is identically equal to zero. We refer to
cancellations like in~(\ref{e:26a}), which come about because the neutrino
sector contains only left-handed particles, as {\em{chiral
    cancellations}}\index{chiral cancellation}.

Next we consider the {\em{massive sectors}} $n=1,\ldots,7$ in~(\ref{e:1i}), i.e.
\begin{equation}
    P(x,y) \;=\; \sum_{\alpha=1}^3
    \frac{1}{2} \left(p_{m_\alpha} - k_{m_\alpha} \right) \:. \label{e:2n}
\end{equation}
Again assuming that the regularized Dirac seas have a vector-scalar
structure, the regularized fermionic projector is
\begin{equation}
    P(x,y) \;=\; g_j(x,y) \:\gamma^j + h(x,y) \label{e:2g}
\end{equation}
with suitable functions $g_j$ and $h$. Using that $P$ is Hermitian, we obtain
\begin{equation}
P(y,x) \;=\; \overline{g_j(x,y)} \:\gamma^j + \overline{h(x,y)} \:. \label{e:2h}
\end{equation}
Again omitting the arguments $(x,y)$, we obtain for the $4 \times 4$ matrix $A_{xy}$
\begin{equation}
    A \;=\; \gslsh \:\overline{\gslsh} \:+\: h \:\overline{\gslsh} \:+\:
    \gslsh \:\overline{h} \:+\: h \overline{h} \:. \label{e:2nn}
\end{equation}
It is useful to decompose $A$ in the form
\[ A \;=\; A_1 \:+\: A_2 \:+\: \mu \]
with
\[ A_1 \;=\; \frac{1}{2} \:[\gslsh, \overline{\gslsh}] \:, \spc
A_2 \;=\; h \:\overline{\gslsh} \:+\: \gslsh \:\overline{h} \:,\spc
\mu \;=\; g \overline{g} \:+\: h \overline{h} \]
and $g \overline{g} \equiv g_j \:\overline{g^j}$. Then the matrices~$A_1$
and $A_2$ anti-commute, and thus
\begin{equation}
(A-\mu)^2 \;=\; A_1^2 + A_2^2 \;=\; (g \overline{g})^2 \:-\: g^2
\:\overline{g}^2 \:+\: (g \overline{h} + h \overline{g})^2 \:.
    \label{e:2d2}
\end{equation}
The right side of~(\ref{e:2d2}) is a multiple of the identity matrix, and
so~(\ref{e:2d2}) is a quadratic equation for $A$.
The roots $\lambda_\pm$ of this equation, \label{lambda_pm}
\begin{equation}
\lambda_\pm \;=\; g \overline{g} \:+\: h \overline{h} \:\pm\:
\sqrt{(g \overline{g})^2 \:-\: g^2\:\overline{g}^2 \:+\:
(g \overline{h} + h \overline{g})^2 } \:,
    \label{e:2e}
\end{equation}
are the zeros of the characteristic polynomial of $A$. However, we
must be careful about associating eigenspaces to $\lambda_\pm$ because $A$
need not be diagonalizable. Let us first consider the case that the two
eigenvalues in~(\ref{e:2e}) are distinct. If we assume that $A$ is
diagonalizable, then $\lambda_\pm$ are the two eigenvalues of $A$, and the
corresponding spectral projectors $F_\pm$ are computed to be
\label{F_pm}
\begin{eqnarray}
F_\pm & = & \frac{\1}{2} \:\pm\: \frac{1}{\lambda_+ - \lambda_-} \left(
A \:-\: \frac{1}{2}\:(\lambda_++\lambda_-)\:\1 \right) \label{e:2f1}  \\
& = & \frac{\1}{2} \:\pm\:\frac{\frac{1}{2} \:[\gslsh, \overline{\gslsh}] \:+\:
h \overline{\gslsh} \:+\: \gslsh \overline{h}}{2 \:
\sqrt{(g \overline{g})^2 \:-\: g^2\:\overline{g}^2 \:+\:
(g \overline{h} + h \overline{g})^2 } } \:.
    \label{e:2f}
\end{eqnarray}
The explicit formula~(\ref{e:2f}) even implies that $A$ is diagonalizable.
Namely, a short calculation yields that
\[ A \:F_\pm \;=\; \lambda_\pm \:F_\pm \spc{\mbox{and}}\spc
F_+ + F_- \;=\; \1 \:, \]
proving that the images of $F_+$ and $F_-$ are indeed eigenspaces of
$A$ which span $\C^4$.  Moreover, a short computation using~(\ref{e:2g})
and~(\ref{e:2f}) yields that
\begin{equation}
F_\pm \:P(x,y) \;=\; \frac{\gslsh+h}{2} \:\pm\:
\frac{\gslsh \:(g \overline{g} + h \overline{h}) \:-\:
\overline{\gslsh} \:(g^2 - h^2) \:+\: (g \overline{g}) \:h \:+\: g^2 \:\overline{h}
}{2 \: \sqrt{(g \overline{g})^2 \:-\: g^2\:\overline{g}^2 \:+\:
(g \overline{h} + h \overline{g})^2 } } \:.
    \label{e:2l}
\end{equation}
Writing out for clarity the dependence on $x$ and $y$, the spectral
decomposition of $A$ is
\begin{equation}
    A_{xy} \;=\; \sum_{s=\pm} \lambda_s^{xy}\: F_s^{xy} \:.
    \label{e:2m}
\end{equation}
The following lemma relates the spectral representation of $A_{xy}$ to that
of $A_{yx}$; it can be regarded as a special case of Lemma~\ref{lemmaFBBF}.
\begin{Lemma} \label{elemma21}
\begin{eqnarray}
\lambda_\pm^{xy} & = & \lambda_\mp^{yx} \label{e:2j}  \\
F_\pm^{xy}\: P(x,y) & = & P(x,y) \:F_\mp^{yx} \:,\spc
\label{e:2k}
\end{eqnarray}
\end{Lemma}
{\Proof}
According to~(\ref{e:2g}, \ref{e:2h}), $A_{yx}$ is obtained from
$A_{xy}$ by the transformations $g_j \leftrightarrow \overline{g_j}$,
$h \leftrightarrow \overline{h}$.
The eigenvalues~(\ref{e:2e}) are invariant under these
transformations. Our convention for labeling the eigenvalues is~(\ref{e:2j}).
Using this convention, we obtain from~(\ref{e:2f1}) that
\begin{eqnarray*}
F_\pm^{xy}\:P(x,y) & = & \frac{1}{2}\:P(x,y) \pm
\frac{1}{\lambda^{xy}_+ - \lambda^{xy}_-} \left(A_{xy} \:P(x,y) \:-\:
\frac{1}{2}\:(\lambda^{xy}_++\lambda^{xy}_-)\: P(x,y) \right) \\
P(x,y) \:F_\mp^{yx} & = & \frac{1}{2}\:P(x,y) \pm
\frac{1}{\lambda^{xy}_+ - \lambda^{xy}_-} \left(P(x,y)\: A_{yx} \:-\:
\frac{1}{2}\:(\lambda^{xy}_++\lambda^{xy}_-)\: P(x,y) \right) .
\end{eqnarray*}
The identity $P(x,y) \:A_{yx} = P(x,y) \:P(y,x) \:P(x,y) = A_{xy} \:P(x,y)$
yields~(\ref{e:2k}).
\QED
If the eigenvalues in~(\ref{e:2e}) are equal, the matrix $A$ need not be
diagonalizable (namely, the right side of~(\ref{e:2d2}) may be zero
without~(\ref{e:2nn}) being a multiple of the identity matrix). Since such
degenerate cases can be treated by taking the limits $\lambda_+ - \lambda_-
\to 0$ in the spectral representation~(\ref{e:2m}), we do not worry about them here.

Before going on, we point out that according to~(\ref{e:2e}), the $4 \times 4$
matrix $A$ has at most two distinct eigenvalues.  In order to understand better
how this degeneracy comes about, it is useful to consider the space $V$ of real
vectors which are orthogonal to $g_j$ and $\overline{g_j}$ (with respect to the
Minkowski metric),
\[ V \;=\; \{ v \:|\: v_j \:g^j = 0 = v_j \:\overline{g^j} \} \:. \]
Since we must satisfy two conditions in four dimensions, ${\mbox{dim }} V \geq 2$.
Furthermore, a short calculation using~(\ref{e:2nn}) shows that for every $v
\in V$,
\begin{equation}
[A,\: v_j \gamma^j \gamma^5] \;=\; 0 \:. \label{e:2s}
\end{equation}
Thus the eigenspaces of $A$ are invariant subspaces of the operators
$v_j \gamma^j \gamma^5$. In the case when the two
eigenvalues~(\ref{e:2e}) are distinct, the family of operators $(v_j \gamma^j
\gamma^5)_{v \in V}$ acts transitively on the two-dimensional eigenspaces of
$A$. Notice that the operators $v_j \gamma^j \gamma^5$ map left-handed spinors
into right-handed spinors and vice versa. Thus one may regard~(\ref{e:2s}) as
describing a symmetry between the left- and right-handed component of $A$. We
refer to the fact that $A$ has at most two distinct eigenvalues as the
{\em{chiral degeneracy}}\index{chiral degeneracy} of the massive sectors in the vacuum.

Our next step is to rewrite the spectral representation using the formalism
of~{\S}\ref{psec26}. Expanding in powers of $m$ and regularizing gives
for a Dirac sea the series
\begin{equation}
\sum_{n=0}^\infty \frac{1}{n!} \left( \frac{i \xi\slsh}{2}
\:T^{(n-1)}_{[2n]}(x,y) \:+\: T^{(n)}_{[2n+1]}(x,y) \right) \:.
    \label{e:2o}
\end{equation}
In composite expressions, one must carefully keep track that every factor $\xi$
is associated to a corresponding factor $T^{(n)}_{[p]}$.  In~{\S}\ref{psec26}
this was accomplished by putting a bracket around both factors.  In order to
have a more flexible notation, we here allow the two factors to be written
separately, but in this case the pairing is made manifest by adding an
index $^{(n)}$, and if necessary also a subscript $_{[r]}$, to the factors $\xi$. With this notation, the contraction rules~(\ref{p:D6}--\ref{p:D8}) can be written as
\begin{equation}
(\xi^{(n_1)}_{[r_1]})_j\: (\xi^{(n_2)}_{[r_2]})^j \;=\;
\frac{1}{2}\:(z^{(n_1)}_{[r_1]} + z^{(n_2)}_{[r_2]}) \:,\quad
(\xi^{(n_1)}_{[r_1]})_j\: (\overline{\xi^{(n_2)}_{[r_2]}})^j \;=\;
\frac{1}{2}\:(z^{(n_1)}_{[r_1]} + \overline{z^{(n_2)}_{[r_2]}})
\label{e:2p}
\end{equation}
(and similar for the complex conjugates), where we introduced factors
$z^{(n)}_{[r]}$ \label{z^(n)_[r]} which by definition combine with the corresponding factor
$T^{(n)}_{[r]}$ according to
\begin{equation}
z^{(n)}_{[r]} \:T^{(n)}_{[r]} \;=\; -4\:(n\:T^{(n+1)}_{[r]} +
T^{(n+2)}_{\{r\}}) \:+\: {\mbox{(smooth functions)}} \:. \label{e:24b}
\end{equation}
In our calculations, most separate factors $\xi$ and $z$ will be
associated to $T^{(-1)}_{[0]}$. Therefore, we shall in this case often omit the
indices, i.e.
\[ \xi \;\equiv\; \xi^{(-1)}_{[0]} \:,\spc
z \;\equiv\; z^{(-1)}_{[0]} \:. \]
We point out that the calculation rule~(\ref{e:24b}) is valid only modulo
smooth functions. This is because in Chapter~\ref{psec2}
we analyzed the effects of the ultraviolet regularization,
but disregarded the ``regularization'' for small momenta related to the logarithmic mass
problem. However, this is not a problem because the smooth contribution
in~(\ref{e:24b}) is easily determined from the behavior away from the light
cone, where the factors $T^{(n)}_\circ$ are known smooth functions and
$z^{(n)}_{[r]}=\xi^2$.

We remark that one must be a little bit careful when regularizing the sum
in~(\ref{e:2n}) because the regularization functions will in general be
different for each Dirac sea. This problem is handled most conveniently by
introducing new ``effective'' regularization functions for the sums of the
Dirac seas.
More precisely, in the integrands~(\ref{p:D1}--\ref{p:D3})
we make the following replacements,
\begin{equation}
\left. \begin{array}{rclcrcl}
h \:a^{\frac{p-1}{2}} & \to & \displaystyle \sum_{\alpha=1}^3 h_\alpha\:
a_\alpha^{\frac{p-1}{2}} &,\;\;\;\;\;\;&
h \:a^{\frac{p-1}{2}}\:b & \to & \displaystyle \sum_{\alpha=1}^3 h_\alpha\:
a_\alpha^{\frac{p-1}{2}} \:b_\alpha \\
g\:a^{\frac{p}{2}} & \to & \displaystyle \sum_{\alpha=1}^3 g_\alpha\:
a_\alpha^{\frac{p}{2}} &,\;\;\;\;\;\;&
g \:a^{\frac{p}{2}}\:b & \to & \displaystyle \sum_{\alpha=1}^3 g_\alpha\:
a_\alpha^{\frac{p}{2}} \:b_\alpha\:.
\end{array} \right\} \label{e:26z}
\end{equation}
As is easily verified, all calculation rules for simple fractions remain
valid also when different regularization functions are involved.  This implies
that the contraction rules~(\ref{e:2p}, \ref{e:24b}) are valid for the
sums of Dirac seas as well.

Using~(\ref{e:2o}) and the contraction rules, we can
expand the spectral decomposition around the singularities on the light cone.
Our expansion parameter is the {\em{degree on the light
    cone}}\index{degree!on the light cone}, also denoted by
``$\deg$''\label{deg}.  It is defined in accordance with~(\ref{p:DL}) by
\[ \deg (T^{(n)}_\circ) \;=\; 1-n \:,\spc \deg(z^{(n)})=-1\:, \]
and the degree of a function which is smooth and non-zero on the light cone
is set to zero. The degree of a product is obtained by adding the degrees of all
factors, and of a quotient by taking the difference of the degrees of the
numerator and denominator. The leading contribution to the
eigenvalues is computed as follows,
\begin{eqnarray*}
\lefteqn{ g \overline{g} + h \overline{h} \;=\; (\deg<3) } \\
&&+ \frac{1}{4}\left( (\xi_j\:
T^{(-1)}_{[0]}) (\overline{\xi^j\: T^{(-1)}_{[0]}}) \:+\:
(\xi_j\: T^{(-1)}_{[0]}) (\overline{\xi^j\: T^{(0)}_{[2]}}) \:+\:
(\xi_j\: T^{(0)}_{[2]}) (\overline{\xi^j\: T^{(-1)}_{[0]}}) \right) \\
&=& \frac{1}{8}\:(z+\overline{z})\:T^{(-1)}_{[0]}\:
\overline{T^{(-1)}_{[0]}} \:+\: (\deg<3) \\
&=& \frac{1}{2}\left(T^{(0)}_{[0]}\: \overline{T^{(-1)}_{[0]}} +
T^{(-1)}_{[0]} \:\overline{T^{(0)}_{[0]}} \right) \:+\: (\deg<3) 
\end{eqnarray*}
\begin{eqnarray*}
(g \overline{g})^2 - g^2\: \overline{g}^2 &=& \frac{1}{16}\left(
(\xi_j\: T^{(-1)}_{[0]}) (\overline{\xi^j\: T^{(-1)}_{[0]}}) \right)^2 \\
&&-\frac{1}{16} \:(\xi\: T^{(-1)}_{[0]})^2\: (\overline{\xi\: T^{(-1)}_{[0]}})^2
\:+\: (\deg<6) \\
&=& \frac{1}{4} \left( T^{(0)}_{[0]} \:\overline{T^{(-1)}_{[0]}}
+ T^{(-1)}_{[0]} \:\overline{T^{(0)}_{[0]}} \right)^2 \\
&&-T^{(-1)}_{[0]}\: T^{(0)}_{[0]}\: \overline{T^{(-1)}_{[0]}\:T^{(0)}_{[0]}}
\:+\: (\deg<6) \\
&=& \frac{1}{4} \left(  T^{(0)}_{[0]} \:\overline{T^{(-1)}_{[0]}}
- T^{(-1)}_{[0]} \:\overline{T^{(0)}_{[0]}} \right)^2 \:+\: (\deg<6) \\
(g \overline{h} + h \overline{g})^2 &=& \frac{1}{4} \left( (i \xi
\:T^{(-1)}_{[0]}) \:\overline{T^{(0)}_{[1]}} + T^{(0)}_{[1]}\:
(\overline{i  \xi \:T^{(-1)}_{[0]}}) \right)^2 \:+\: (\deg<6) \\
&=& (\deg<6)\:,
\end{eqnarray*}
and thus
\begin{eqnarray}
\lambda_\pm &=& \frac{1}{2} \left( T^{(0)}_{[0]}\: \overline{T^{(-1)}_{[0]}}
+ T^{(-1)}_{[0]} \:\overline{T^{(0)}_{[0]}} \right)
\:+\: (\deg<3) \nonumber \\
&&\pm \frac{1}{2}\:\sqrt{\left(T^{(0)}_{[0]}\: \overline{T^{(-1)}_{[0]}}
- T^{(-1)}_{[0]} \:\overline{T^{(0)}_{[0]}} \right)^2
\:+\: (\deg<6)} \nonumber \\
&=& \frac{1}{2} \left( T^{(0)}_{[0]}\: \overline{T^{(-1)}_{[0]}}
+ T^{(-1)}_{[0]} \:\overline{T^{(0)}_{[0]}} \right)
\:+\: (\deg<3) \nonumber \\
&&\pm \frac{1}{2} \left(T^{(0)}_{[0]}\: \overline{T^{(-1)}_{[0]}}
- T^{(-1)}_{[0]} \:\overline{T^{(0)}_{[0]}} \right) \pm
\frac{(\deg<6)}{4 \left(T^{(0)}_{[0]}\: \overline{T^{(-1)}_{[0]}}
- T^{(-1)}_{[0]} \:\overline{T^{(0)}_{[0]}} \right)} \:+\:
\cdots \nonumber \\
&=& \left\{ \begin{array}{cc}
T^{(0)}_{[0]}\: \overline{T^{(-1)}_{[0]}} & {\mbox{for ``$+$''}} \\[.4em]
T^{(-1)}_{[0]} \:\overline{T^{(0)}_{[0]}} & {\mbox{for ``$-$''}} \end{array}
\right. \:+\: (\deg < 3). \label{e:2qa}
\end{eqnarray}
The spectral projectors are calculated similarly,
\begin{eqnarray}
F_\pm &=& \frac{\1}{2} \pm
\frac{\frac{1}{8} \:[\xi\slsh \:T^{(-1)}_{[0]},\: \overline{\xi\slsh
\:T^{(-1)}_{[0]}}] \:-\: \frac{i}{2} \left(T^{(0)}_{[1]}\:
(\overline{\xi\slsh T^{(-1)}_{[0]}}) \:-\: (\xi\slsh T^{(-1)}_{[0]})
\:\overline{T^{(0)}_{[1]}} \right)}
{T^{(0)}_{[0]}\: \overline{T^{(-1)}_{[0]}}
- T^{(-1)}_{[0]} \:\overline{T^{(0)}_{[0]}}} \nonumber \\
&&+ (\deg<0) \label{e:2r} \\
\lefteqn{ \hspace*{-0.5cm}
F_\pm\: P(x,y) \;=\; \frac{i}{4}\:(\xi\slsh\:T^{(-1)}_{[0]}) \:+\:
(\deg < 2) } \nonumber \\
&&\hspace*{-1cm} \pm \frac{i}{4}\: \frac{(\xi\slsh \:T^{(-1)}_{[0]}) (T^{(0)}_{[0]}\:
\overline{T^{(-1)}_{[0]}} + T^{(-1)}_{[0]}\:
\overline{T^{(0)}_{[0]}} ) \:-\:
2\:(\overline{\xi\slsh \:T^{(-1)}_{[0]}})\: T^{(-1)}_{[0]}\:
T^{(0)}_{[0]}} {T^{(0)}_{[0]}\: \overline{T^{(-1)}_{[0]}}
- T^{(-1)}_{[0]} \: \overline{T^{(0)}_{[0]}}}.\spc \label{e:2z}
\end{eqnarray}
The last expression contains inner factors $\xi$.  In situations when these
factors are not contracted to other inner factors in a composite expression, we
can treat them as outer factors. Then (\ref{e:2z}) simplifies to
\begin{equation}
F_\pm \:P(x,y) \;=\; \left\{ \begin{array}{cc}
0 & {\mbox{for ``$+$''}} \\
\frac{i}{2}\:\xi\slsh\:T^{(-1)}_{[0]} & {\mbox{for ``$-$''}}
\end{array} \right. \:+\: (\deg <2)\:. \label{e:2s2}
\end{equation}
By expanding, one can compute the eigenvalues and spectral projectors also to
lower degree on the light cone.  We do not want to enter the details of this
calculation here because in this chapter we only need that the lower degrees
involve the masses of the Dirac seas.  This is illustrated by the following
expansion of the eigenvalues,
\begin{eqnarray}
\lambda_\pm &=& \frac{1}{4} \: \times\: \left\{ \begin{array}{cc}
(z\: T^{(-1)}_{[0]})\: \overline{T^{(-1)}_{[0]}} \:+\:
(z\: T^{(0)}_{[2]})\: \overline{T^{(-1)}_{[0]}} \:+\:
(z\: T^{(-1)}_{[0]})\: \overline{T^{(0)}_{[2]}} & {\mbox{for ``$+$''}} \\[.5em]
T^{(-1)}_{[0]}\: (\overline{z\: T^{(-1)}_{[0]}}) \:+\:
T^{(-1)}_{[0]}\: (\overline{z\: T^{(0)}_{[2]}}) \:+\:
T^{(0)}_{[2]}\: (\overline{z\: T^{(-1)}_{[0]}}) & {\mbox{for ``$-$''}} \end{array} \right. \nonumber \\[.1em]
&&+\: T^{(0)}_{[1]}\: \overline{T^{(0)}_{[1]}}
\:\mp\:\frac{T^{(0)}_{[1]}\: \overline{T^{(-1)}_{[0]}}
-T^{(-1)}_{[0]} \:\overline{T^{(0)}_{[1]}}}
{ T^{(0)}_{[0]}\: \overline{T^{(-1)}_{[0]}}
- T^{(-1)}_{[0]} \:\overline{T^{(0)}_{[0]}}} \:
(T^{(0)}_{[1]}\: \overline{T^{(0)}_{[0]}}
-T^{(0)}_{[0]} \:\overline{T^{(0)}_{[1]}}) \nonumber \\
&&+\: (\deg < 2) \:. \label{e:2q}
\end{eqnarray}
Similarly, the contributions to degree $<2$ involve even higher powers of
the masses.

Let us specify how we can give the above spectral decomposition of $A_{xy}$
a mathematical meaning.  \label{mmean}
A priori, our formulas for $\lambda_\pm$ and
$F_\pm$ are only formal expressions because the formalism of the continuum limit
applies to simple fractions, but
dividing by $(T^{(0)}_{[0]}\: \overline{T^{(-1)}_{[0]}} - T^{(-1)}_{[0]}
\:\overline{T^{(0)}_{[0]}})$ is not a well-defined operation.  In order to
make mathematical sense of the spectral decomposition and
in order to ensure at the same
time that the EL equations have a well-defined continuum limit, we
shall only consider Lagrangians for which all expressions obtained by
substituting the spectral representation of $A$ into the
EL equations are linear combinations of simple fractions.
Under this assumption, working with
the spectral representation of $A_{xy}$ can be regarded merely as a convenient
formalism for handling the EL equations, the latter being
well-defined according to~{\S}\ref{psec26}.  Having
Lagrangians of this type in mind, we can treat $A$ in the massive sectors as a
diagonalizable matrix with two distinct eigenvalues $\lambda_\pm$ and
corresponding spectral projectors $F_\pm$.

The explicit formulas~(\ref{e:2qa}, \ref{e:2r}) show that the
eigenvalues of $A$ are to leading degree not real, but appear in {\em{complex
conjugate pairs}}\index{eigenvalues!in complex conjugate pairs}, i.e.
\begin{equation}
\overline{\lambda_+} \;=\; \lambda_- \spc{\mbox{and}}\spc
F_+^* \;=\; F_- \:.
    \label{e:2t}
\end{equation}
If one considers perturbations of these eigenvalues by taking into account the
contributions of lower degree, $\lambda_+$ and $\lambda_-$ will clearly still
be complex. This implies that the relations~(\ref{e:2t}) remain valid
(see the argument after~(\ref{p:55}) and the example~(\ref{e:2q})).
We conclude that in our expansion about the singularities on the light cone,
the eigenvalues appear to every degree in complex conjugate pairs~(\ref{e:2t}).

We finally summarize the results obtained in the neutrino and massive sectors
and introduce a convenient notation for the eigenvalues and spectral
projectors of $A_{xy}$. We found that in the continuum limit,
$A$ can be treated as a diagonalizable matrix. We denote
the distinct eigenvalues of $A$ by $(\lambda_k)_{k=1,\ldots,K}$ and the
corresponding spectral projectors by $F_k$. Since $A$ vanishes in the neutrino
sector, zero is an eigenvalue of $A$ of multiplicity four; we choose
the numbering such that $\lambda_1=0$. Due to the chiral degeneracy,
all eigenspaces are at least two-dimensional. Furthermore, all non-zero
eigenvalues of $A$ are complex and appear in complex conjugate
pairs~(\ref{e:2t}). It is useful to also consider the eigenvalues counting
their multiplicities. We denote them by $(\lambda_\alpha)_{\alpha=1,\ldots,
4N}$ or also by $(\lambda_{ncs})_{n=1,\ldots,8,\: c=L\!/\!R,\: s=\pm}$, where
$n$ refers to the sectors and $c,s$ count the eigenvalues within each
sector. More precisely,
\begin{equation}
\lambda_{8a} \;=\; 0 \spc {\mbox{and}}\spc
\lambda_{nc \pm} = \lambda_\pm^{(n)} \:,\;\;\;
    n=1,\ldots,7 \label{e:2H}
\end{equation}
with $\lambda_\pm^{(n)}$ as given by~(\ref{e:2e}) or~(\ref{e:2q}), where
the index ``$^{(n)}$'' emphasizes that the eigenvalues $\lambda_\pm$
depend on the regularization functions in the corresponding sector.

\section{Strong Spectral Analysis of the Euler-Lagrange
  Equations}\index{strong spectral analysis}
\setcounter{equation}{0} \label{esec22}
In this section we shall derive conditions which
ensure that the EL equations~(\ref{e:2d}, \ref{e:2c}) are
satisfied in the vacuum and argue why we want to choose our
Lagrangian in such a way that these sufficient conditions are
fulfilled. Since the Lagrangian ${\mathcal{L}}[A]$ must be
independent of the matrix representation of $A$, it depends only
on the eigenvalues $\lambda_\alpha$,
\[ {\mathcal{L}}[A_{xy}] \;=\;
{\mathcal{L}}(\lambda^{xy}_1,\ldots,\lambda^{xy}_{4N})\:, \] and
furthermore
${\mathcal{L}}(\lambda^{xy}_1,\ldots,\lambda^{xy}_{4N})$ is
symmetric in its arguments. In preparation, we consider the
case when the eigenvalues of $A$ are non-degenerate. Then the
variation of the eigenvalues is given in first order perturbation
theory by $\delta \lambda_\alpha = \Tr (F_\alpha \:\delta A)$. Let
us assume that ${\mathcal{L}}$ depends smoothly on the
$\lambda_\alpha$, but is not necessarily holomorphic (in
particular, ${\mathcal{L}}$ is allowed to be a polynomial in
$|\lambda_\alpha|$). Then
\begin{eqnarray}
\delta {\mathcal{L}} & = & \sum_{\alpha=1}^{4N} \left( \frac{\partial
{\mathcal{L}}(\lambda)}{\partial \:{\mbox{Re}} \lambda_\alpha}\;
{\mbox{Re}}\: \Tr (F_\alpha \:\delta A) \:+\: \frac{\partial
{\mathcal{L}}(\lambda)}{\partial \:{\mbox{Im}} \lambda_\alpha}\;
{\mbox{Im}}\:
\Tr (F_\alpha \:\delta A) \right) \\
&=& {\mbox{Re}} \sum_{\alpha=1}^{4N} \frac{\partial
{\mathcal{L}}(\lambda)}{\partial \lambda_\alpha}\; \Tr (F_\alpha
\:\delta A) \:, \label{e:2y}
\end{eqnarray}
where we set
\begin{eqnarray*}
\frac{\partial {\mathcal{L}}(\lambda)}{\partial \:{\mbox{Re}}
\lambda_\alpha} &=& \lim_{\sR \ni \varepsilon \to 0}
\frac{1}{\varepsilon} \left(
{\mathcal{L}}(\lambda_1,\ldots,\lambda_{\alpha-1},
\lambda_\alpha+\varepsilon, \lambda_{\alpha+1}, \ldots,
\lambda_{4N}) -
{\mathcal{L}}(\lambda_1,\ldots,\lambda_{4N}) \right) \\
\frac{\partial {\mathcal{L}}(\lambda)}{\partial \:{\mbox{Im}}
\lambda_\alpha} &=& \lim_{\sR \ni \varepsilon \to 0}
\frac{1}{\varepsilon} \left(
{\mathcal{L}}(\lambda_1,\ldots,\lambda_{\alpha-1},
\lambda_\alpha+i \varepsilon, \lambda_{\alpha+1}, \ldots,
\lambda_{4N}) - {\mathcal{L}}(\lambda_1,\ldots,\lambda_{4N})
\right)
\end{eqnarray*}
and
\begin{equation}
\frac{\partial {\mathcal{L}}(\lambda)}{\partial \lambda_\alpha}
\;\equiv\; \frac{\partial {\mathcal{L}}(\lambda)}{\partial
\:{\mbox{Re}} \lambda_\alpha} \:-\: i\: \frac{\partial
{\mathcal{L}}(\lambda)}{\partial \:{\mbox{Im}} \lambda_\alpha} \:.
\label{e:2cd}
\end{equation}

If some of the $\lambda_\alpha$s coincide, we must apply
perturbation theory with degeneracies. One obtains in generalization
of~(\ref{e:2y}) that
\begin{equation}
\delta {\mathcal{L}} \;=\; {\mbox{Re}} \sum_{k=1}^{K} \left.  \frac{\partial
{\mathcal{L}}(\lambda)}{\partial \lambda_\alpha} \:\Tr (F_k\:
\delta A) \right|_{\lambda_\alpha=\lambda_k} \:.  \label{e:2w}
\end{equation}
Here our notation means that we choose the index $\alpha$ such
that $\lambda_\alpha=\lambda_k$.  Clearly, $\alpha$ is not unique
if $\lambda_k$ is a degenerate eigenvalue; in this case $\alpha$
can be chosen arbitrarily due to the symmetry of ${\mathcal{L}}$.
We also write~(\ref{e:2w}) in the shorter form
\begin{equation}
    \delta {\mathcal{L}} \;=\; {\mbox{Re}} \sum_{k=1}^K
    \frac{\partial {\mathcal{L}}(\lambda)}{\partial \lambda_k}\: \Tr (F_k\: \delta A) \:.
    \label{e:2ce}
\end{equation}
In this formula it is not necessary to take the real part. Namely,
as the Lagrangian is real and symmetric in its arguments, we know
that $\overline{{\mathcal{L}}(\lambda_1,\ldots,\lambda_{4N})}=
{\mathcal{L}}(\overline{\lambda_1},\ldots,\overline{\lambda_{4N}})$,
and thus, according to~(\ref{e:2cd}),
\[ \overline{\frac{\partial {\mathcal{L}}(\lambda_1,\ldots,\lambda_{4N})}
{\partial \lambda_k}} \;=\; \frac{\partial
{\mathcal{L}}(\overline{\lambda_1},\ldots,\overline{\lambda_{4N}})}
{\partial \overline{\lambda_k}}\:. \]
Using furthermore that the
eigenvalues of $A$ appear in complex conjugate pairs, (\ref{e:2t}),
one sees that the operator
\[ \sum_{k=1}^K \frac{\partial {\mathcal{L}}}{\partial \lambda_k}\:F_k \]
is Hermitian. Hence we can write the sum in~(\ref{e:2ce}) as the trace of
products of two Hermitian operators on $H$, being automatically real. We
conclude that
\begin{equation}
    \delta {\mathcal{L}} \;=\; \sum_{k=1}^K \frac{\partial {\mathcal{L}}(\lambda)}{\partial
    \lambda_k}\: \Tr (F_k\: \delta A) \:. \label{e:2cf}
\end{equation}

Comparing~(\ref{e:2cf}) with~(\ref{e:2x}) gives \label{mathcalM2}
\begin{equation}
    {\mathcal{M}}[A_{xy}] \;=\; \sum_{k=1}^{K_{xy}} \frac{\partial
    {\mathcal{L}}(\lambda)}{\partial \lambda^{xy}_k}\: F^{xy}_k \:.
    \label{e:39z}
\end{equation}
We substitute this identity into~(\ref{e:2c}) and apply
Lemma~\ref{elemma21} in each sector to obtain
\begin{equation}
Q(x,y) \;=\; \frac{1}{2} \:\sum_{k=1}^{K_{xy}} \frac{\partial
{\mathcal{L}}(\lambda^{xy})}{\partial \lambda^{xy}_k}\: F^{xy}_k\:
P(x,y) \;=\; \frac{1}{2} \:\sum_{k=1}^{K_{xy}} \frac{\partial
{\mathcal{L}}(\lambda^{xy})}{\partial \lambda^{xy}_k}\: P(x,y)\:
F^{yx}_k \:.
    \label{e:2v}
\end{equation}
Using these relations, we can write the EL
equations~(\ref{e:2d}) as
\begin{eqnarray}
\lefteqn{ \int d^4z \left( P(x,z)\: P(z,y) \:\sum_{k=1}^{K_{zy}}
\frac{\partial {\mathcal{L}}(\lambda^{zy})}{\partial
\lambda^{zy}_k}\: F^{yz}_k \right. } \\
&& \hspace*{1.5cm} \left. \:-\: \sum_{k=1}^{K_{xz}}
\frac{\partial {\mathcal{L}}(\lambda^{xz})}{\partial
\lambda^{xz}_k}\: F^{xz}_k \: P(x,z)\: P(z,y) \right) \;=\; 0 \:.
\label{e:2t2}
\end{eqnarray}
This equation splits into separate equations on the eight sectors.  In the
neutrino sector, we have according to~(\ref{e:2u}),
\begin{equation}
    P(x,z)\: P(z,y) \;=\; \chi_L \:\gslsh(x,z) \;\chi_L \:\gslsh(z,y) \;=\;
    \chi_L \:\chi_R\:\gslsh(x,z) \:\gslsh(z,y) \;=\; 0 \:.
    \label{e:2triv}
\end{equation}
Since both summands in~(\ref{e:2t2}) contain a
factor $P(x,z)\:P(z,y)$, the EL equations are trivially satisfied in the
neutrino sector due to chiral cancellations\index{chiral cancellation}. In the massive sectors, there are
no chiral cancellations.  As shown in Appendix~\ref{appD}, there are no
further cancellations in the commutator if generic perturbations of the physical
system are taken into account.  This means that~(\ref{e:2t2}) will be satisfied if and only if $Q$ vanishes in the massive
sectors, i.e.
\begin{equation}
\sum_{k=1}^{K_{xy}} \frac{\partial
{\mathcal{L}}(\lambda^{xy})}{\partial \lambda^{xy}_k}\: F^{xy}_k\:
P(x,y)\; X X^* \;=\; 0 \:. \label{e:2s3}
\end{equation}
As explained on page~\pageref{mmean}, we shall only consider Lagrangians for
which~(\ref{e:2s3}) is a linear combination of simple fractions.
Thus we can evaluate~(\ref{e:2s3})
weakly on the light cone and apply the integration-by-parts
rules. We will now explain why it is reasonable to demand that
(\ref{e:2s3}) should be valid even in the {\em strong} sense, i.e.\
without weak evaluation and the integration-by-parts rule. First, one should
keep in mind that the integration-by-parts rules are valid only to leading
order in $(l E_P)^{-1}$. As is worked out in~Appendix~\ref{pappB}, the
restriction to the leading order in $(l E_P)^{-1}$ is crucial
when perturbations by the bosonic fields are considered
(basically because the microscopic form of the bosonic potentials
is unknown). But in the vacuum, one can consider the higher orders in
$(l E_P)^{-1}$ as well (see the so-called regularization expansion
in~{\S}\ref{psec24} and~{\S}\ref{psec25}).
Therefore, it is natural to impose that in the vacuum
the EL equations should be satisfied to all orders in $(l E_P)^{-1}$.
Then the integration-by-parts rules do not apply, and weak evaluation becomes
equivalent to strong evaluation. A second argument in favor of a
strong analysis is that even if we restricted attention to the leading
order in $(l E_P)^{-1}$ and allowed for integrating by parts, this would
hardly simplify the equations~(\ref{e:2s3}), because the integration-by-parts
rules depend on the indices $n$ of the involved factors $T^{(n)}_\circ$ and
$\overline{T^{(n)}_\circ}$.  But the
relations between the simple fractions given by the integration-by-parts rules are
different to every degree, and thus the conditions~(\ref{e:2s3}) could be
satisfied only by imposing to every degree conditions on the regularization
parameters. It seems difficult to satisfy all these extra
conditions with our small number of regularization functions. Clearly, this
last argument does not rule out the possibility that there might be a
Lagrangian together with a special regularization such that~(\ref{e:2s3})
is satisfied to leading order in $(l E_P)^{-1}$ only after applying the
integration-by-parts rules. But such Lagrangians are certainly difficult to
handle, and we shall not consider them here.

For these reasons, we here restrict attention to Lagrangians for
which~(\ref{e:2s3}) is satisfied strongly. Then~(\ref{e:2s3})
simplifies to the conditions
\begin{equation}
\frac{\partial {\mathcal{L}}(\lambda^{xy})}{\partial
\lambda^{xy}_{ncs}} \;=\; 0 \spc {\mbox{for $n=1,\ldots,7$.}}
    \label{e:2A}
\end{equation}

\section{Motivation of the Lagrangian, the Mass Degeneracy
Assumption} \index{Lagrangian!polynomial} \index{action!polynomial}
\setcounter{equation}{0} \label{esec23}
Let us discuss the conditions~(\ref{e:2A}) in
concrete examples.  We begin with the class of Lagrangians which
are {\em{polynomial}} in the eigenvalues $\lambda_\alpha$ of $A$.
Since different powers of $\lambda_\alpha$ have a different degree
on the light cone, there cannot be cancellations between them.
Thus it suffices to consider polynomials which are
{\em{homogeneous of degree $h$}}, $h \geq 1$. Furthermore, as the
Lagrangian should be independent of the matrix representation of
$A$, it can be expressed in terms of traces of powers of $A$.
Thus we consider Lagrangians of the form
\begin{equation}
{\mathcal{L}}[A] \;=\; {\mathcal{P}}_h[A] \:, \label{e:2B}
\end{equation}
where ${\mathcal{P}}_l$ denotes a polynomial in $\Tr(A^p)$
homogeneous in $A$ of degree $l$, i.e.
\begin{eqnarray}
{\mathcal{P}}_l &=& \sum_n c_n \: R_{p_1} \:\cdots\:
R_{p_{\max(n)}}
\spc{\mbox{with}} \spc \sum_{j=1}^{\max(n)} p_j \;=\; l, \label{e:2H2} \\
R_p &=& \Tr (A^p) \;=\; \sum_{\alpha=1}^{4N} \lambda_\alpha^p
\end{eqnarray}
and coefficients $c_n$, which for simplicity we assume to be rational. For
example, the general Lagrangian of degree $h=3$ reads
\begin{equation}
{\mathcal{L}}[A] \;=\; c_1 \:R_3 \:+\: c_2 \: R_1\: R_2 \:+\: c_3
\:R_1\:R_1\:R_1
    \label{e:2C}
\end{equation}
with three coefficients $c_n \in \Q$ (only two of which are of
relevance because the normalization of ${\mathcal{L}}$ clearly has no
effect on the EL equations).  We assume that ${\mathcal{L}}$ is
{\em{non-trivial}} in the sense that at least one of the
coefficients $c_n$ in the definition~(\ref{e:2B}, \ref{e:2H2}) of
${\mathcal{L}}$ should be non-zero.  The EL equations
corresponding to~(\ref{e:2C}) can easily be computed using that
$\delta R_p = p \: \Tr (A^{p-1} \:\delta A)$ together
with~(\ref{e:2D}) and the fact that the trace is cyclic. The
resulting operator $Q$ in~(\ref{e:2d}) is of the form
\begin{equation}
Q(x,y) \;=\; \left[ {\mathcal{P}}_0 \:A^{h-1} \:+\:
{\mathcal{P}}_1\: A^{h-2} \:+\: \cdots \:+\: {\mathcal{P}}_{h-1}
\right] P(x,y) \label{e:2E}
\end{equation}
where ${\mathcal{P}}_l$ are homogeneous polynomials of the
form~(\ref{e:2H2}) (${\mathcal{P}}_0$ is a rational number). In
the example~(\ref{e:2C}),
\[ Q(x,y) \;=\; \left[ \frac{3}{2}\:c_1\: A^2 \:+\: c_2\: R_1\:A \:+\:
\frac{1}{2}\: (c_2 \:R_2 + 3 c_3\: R_1^2) \right] P(x,y) \:. \]
By substituting the regularized formulas of the light-cone expansion
into~(\ref{e:2E}), one sees that $Q(x,y)$ is to every degree on the light cone
a polynomial in $T^{(n)}_\circ$ and $\overline{T^{(n)}_\circ}$, well-defined
according to~{\S}\ref{psec26}. Thus for polynomial actions, our
spectral decomposition is not needed. But it is nevertheless
a convenient method for handling the otherwise rather complicated combinatorics
of the Dirac matrices.

For the polynomial Lagrangian~(\ref{e:2B}), the conditions~(\ref{e:2A}) become
\begin{equation}
{\mathcal{P}}_0 \:\lambda^{h-1} \:+\:
{\mathcal{P}}_1\:\lambda^{h-2} \:+\: \cdots \:+\:
{\mathcal{P}}_{h-1} \;=\; 0 \spc {\mbox{for
$\lambda=\lambda_{ncs}$, $n=1,\ldots,7$}} \label{e:2F}
\end{equation}
and the ${\mathcal{P}}_l$ as in~(\ref{e:2H2}). It is useful to
analyze these conditions algebraically as polynomial equations
with rational coefficients for the eigenvalues of $A$. To this
end, we need to introduce an abstract mathematical notion which
makes precise that, according to~(\ref{e:2H}), the eigenvalues
$\lambda_{ncs}$ have certain degeneracies, but that there are no
further relations between them. We say that the matrix $A$ has $n$
{\em{independent eigenvalues}}\index{eigenvalues!independent} if $A$ has $n$ distinct
eigenvalues, one of them being zero and the others being
algebraically independent. The following lemma shows that the
conditions~(\ref{e:2F}) can be fulfilled only if the degree of the
Lagrangian is sufficiently large.
\begin{Lemma} \label{lemma22}
For a non-trivial Lagrangian of the form~(\ref{e:2B}) which satisfies the
conditions~(\ref{e:2F}),
\begin{equation}
h \;\geq\; n \:, \label{e:2G}
\end{equation}
where $n$ denotes the number of independent eigenvalues of $A$.
\end{Lemma}
{\Proof} First suppose that the ${\mathcal{P}}_l$
in~(\ref{e:2F}) are not all zero. Then we can regard the left
side of~(\ref{e:2F}) as a polynomial in $\lambda$ of degree at most
$h-1$. According to~(\ref{e:2H}), the eigenvalues $\lambda_{8a}$
in the lepton sector all vanish. Thus the polynomial
in~(\ref{e:2F}) has at least $n-1$ distinct zeros, and thus its
degree must be at least $n-1$. This proves~(\ref{e:2G}).

It remains to consider the case when the coefficients
${\mathcal{P}}_l$ in~(\ref{e:2F}) all vanish. Since the Lagrangian
is non-trivial, at least one of the ${\mathcal{P}}_l$ is
non-trivial, we write ${\mathcal{P}}_l \not \equiv 0$. On the
other hand, ${\mathcal{P}}_l[A] = 0$ and furthermore $l \leq h-1$
from~(\ref{e:2F}). Hence to conclude the proof it suffices to show
that
\begin{equation}
{\mathcal{P}}_l \not \equiv 0 \;{\mbox{ and }}\; {\mathcal{P}}_l[A] = 0
\;\;\; \Longrightarrow\ \;\;\; l \geq n-1. \label{e:2I}
\end{equation}
To prove~(\ref{e:2I}), we proceed inductively in $n$. For $n=1$
there is nothing to show. Assume that~(\ref{e:2I}) holds for given
$n$ and any matrix $A$ with $n$ independent eigenvalues. Consider
a matrix $A$ with $n+1$ independent eigenvalues. A given
non-trivial polynomial ${\mathcal{P}}_{l+1}$ can be uniquely
decomposed in the form
\begin{equation}
{\mathcal{P}}_{l+1} \;=\; R_1 \:{\mathcal{P}}_l \:+\: R_2
\:{\mathcal{P}}_{l-1} \:+\:\cdots\:+\: R_{l+1}\:{\mathcal{P}}_0
\:, \label{e:2J}
\end{equation}
where the polynomials ${\mathcal{P}}_{l-k}$ contain only factors
$R_j$ with $j>k$. Since ${\mathcal{P}}_{l+1} \not \equiv 0$, at
least one of the factors ${\mathcal{P}}_{l-k}$ is non-trivial. Let
$k \geq 0$ be the smallest natural number such that
${\mathcal{P}}_{l-k} \not \equiv 0$. The functional
${\mathcal{P}}_{l+1}[A]$ is a homogeneous polynomial of degree
$l+1$ in the eigenvalues $\lambda_1,\ldots,\lambda_{n+1}$ of $A$.
We pick those contributions to this polynomial which are
homogeneous in $0 \neq \lambda_{n+1}$ of degree $n+1$. These
contributions all come from the summand $R_{k+1}
{\mathcal{P}}_{l-k}$ in~(\ref{e:2J}) because the summands to its
left are trivial and the summands to its right are composed only
of factors $R_l$ with $l>k+1$. Hence, apart from the prefactor
$\lambda^{k+1}_{n+1}$ and up to irrelevant combinatorial factors
for each of the monomials, these contributions coincide with the
polynomial ${\mathcal{P}}_{l-k}(\lambda_1,\ldots,\lambda_n)$
evaluated for a matrix with $n$ independent eigenvalues. We
conclude that if
${\mathcal{P}}_{l+1}(\lambda_1,\ldots,\lambda_{n+1})$ vanishes,
then ${\mathcal{P}}_{l-k}(\lambda_1,\ldots,\lambda_n)$ must also
be zero. The induction hypothesis yields that $l-k \geq n-1$ and
thus $l+1 \geq n$. \QED

We seek a Lagrangian which is as simple as possible. One strategy is to
consider the general polynomial Lagrangian~(\ref{e:2B}) and to choose the
degree $h$ as small as possible. According to Lemma~\ref{lemma22}, the degree cannot be
smaller than the number of independent eigenvalues of $A$. Thus if we treat the
eigenvalues as being algebraically independent in the sectors containing the
Dirac seas $(d,s,b)$, $(u,c,t)$, and $(e, \mu, \tau)$, then $h$ is bounded from
below by $h \geq 3 \times 2 +1 = 7$. Unfortunately, polynomials of degree $\geq
7$ involve many coefficients $c_n$ and are complicated. Therefore, it is
desirable to reduce the number of independent eigenvalues. Since the eigenvalues
depend on the masses and regularization functions
of the particles involved (see~(\ref{e:2q})), we
can reduce the number of distinct eigenvalues only by assuming that the massive
sectors are identical. The best we can do is to assume that
\begin{equation}
m_u = m_d = m_e \:,\spc m_c = m_s = m_\mu \:,\spc m_t = m_b = m_\tau
\:, \label{e:2K}
\end{equation}
and that the regularization functions in the massive sectors coincide.
Then the additional degeneracies in the massive sectors reduce the number of
distinct eigenvalues to three. If~(\ref{e:2K}) holds, the bound of
Lemma~\ref{lemma22} is even optimal. Namely, a simple calculation shows that the
polynomial Lagrangian of degree three~(\ref{e:2C}) with
\begin{equation}
c_1 \;=\; 14 \:,\spc c_2 \;=\; -\frac{3}{2} \:,\spc c_3 \;=\; \frac{1}{28}
    \label{e:2L}
\end{equation}
satisfies the conditions~(\ref{e:2A}). According to Lemma~\ref{lemma22}, a
degree $h<2$ would make it necessary to impose relations between $\lambda_+$
and $\lambda_-$, which is impossible in our formalism~(\ref{e:2qa}). We
conclude that~(\ref{e:2C}, \ref{e:2L}) is the polynomial Lagrangian of
minimal degree which satisfies the conditions~(\ref{e:2A}).
We refer to~(\ref{e:2K}) as the {\em{mass degeneracy
    assumption}}\index{mass degeneracy assumption}.  In order to
understand what this condition means physically, one should keep in mind
that~(\ref{e:2K}) gives conditions for the bare masses, which due to the
self-interaction are different from the effective masses (this is a bit similar
to the situation in the grand unified theories, where simple algebraic relations
between the bare quark and lepton masses are used with some
success~\cite{Ro}).

Another strategy for finding promising Lagrangians is to consider homogeneous
polynomials of higher degree, but which are of a particularly simple form. A good
example for such a Lagrangian is the determinant,
\begin{equation}
{\mathcal{L}}[A] \;=\; \det A \:. \label{e:2M}
\index{action!determinant}
\index{Lagrangian!determinant}
\end{equation}
By writing $\det A = \det(\1 - (\1-A))$, expanding in powers of
$(\1-A)$ and multiplying out, this Lagrangian can be brought into
the form~(\ref{e:2B}, \ref{e:2H2}) with $h=4N$. The
Lagrangian~(\ref{e:2M}) is appealing because of its simple form.
Furthermore, it has the nice property that whenever the
eigenvalues of $A$ appear in complex conjugate pairs~(\ref{e:2t}),
the product of these eigenvalues is positive, and thus
${\mathcal{L}} \geq 0$. Unfortunately, this Lagrangian has the following
drawback. The matrix $A$ vanishes identically in the neutrino
sector~(\ref{e:26a}), and so $A$ has a zero eigenvalue of
multiplicity four. As a consequence, ${\mathcal{L}}$ and its
variations vanish until perturbations of at least fourth order are
taken into account, making the analysis rather complicated. For
this reason, (\ref{e:2M}) does not seem the best Lagrangian for
developing our methods, and we shall not consider it here.

In the polynomial Lagrangian~(\ref{e:2C}, \ref{e:2L}) we did not use
that the eigenvalues of $A$ appear in complex conjugate pairs. This
fact can be exploited to construct an even simpler Lagrangian. Assume again
that the masses are degenerate~(\ref{e:2K}). Then the absolute values
$|\lambda_\alpha|$ of the eigenvalues of $A$ take only the two values $0$
and $|\lambda_+|=|\lambda_-|$, with multiplicities $4$ and $28$,
respectively. Thus if we consider homogeneous polynomials in
$|\lambda_\alpha|$, there is already a Lagrangian of degree two which
satisfies the conditions~(\ref{e:2A}), namely
\begin{equation}
    {\mathcal{L}} \;=\; \sum_{\alpha=1}^{4N} |\lambda_\alpha|^2 \:-\: \frac{1}{28}
    \left(\sum_{\alpha=1}^{4N} |\lambda_\alpha| \right)^2 \:.
    \label{e:2N}
\end{equation}
Using the notion of the spectral weight~(\ref{p:38}),
this Lagrangian can be written as
\begin{equation}
{\mathcal{L}}[A] \;=\; |A^2| \:-\: \frac{1}{28} \:|A|^2 \:.
    \label{e:2O}
    \index{Lagrangian!model}
    \index{variational principle|see{action}}
\end{equation}
The factor $1/28$ may be replaced by a Lagrange multiplier $\mu$,
\begin{equation}
{\mathcal{L}}[A] \;=\; |A^2| \:-\: \mu \:|A|^2 \:, \label{e:2P}
\end{equation}
because the value of $\mu=1/28$ is uniquely determined from the condition that
the EL equations should be satisfied in the vacuum.
The functional~(\ref{e:2P}) can be regarded as the effective
Lagrangian corresponding to the variational principle with
constraint~(\ref{p:44}, \ref{p:45}). We conclude that~(\ref{e:2O}) is
precisely our model variational principle introduced in~{\S}\ref{psec15}.

The above considerations give a motivation for our model
Lagrangian~(\ref{e:2O}) as well as for the mass degeneracy
assumption~(\ref{e:2K}). Also, it is nice that
many special properties of the fermionic projector of the
vacuum were used.  Namely, the EL equations corresponding
to~(\ref{e:2O}) are fulfilled only because of chiral cancellations in
the neutrino sector and the fact that the eigenvalues of
$A$ appear in complex conjugate pairs.
But unfortunately, our arguments so far do not determine the action
uniquely. In particular, variational principles formulated with the
spectral weight of higher powers of~$A$, like for example
\[ {\mathcal{L}}[A] \;=\; |A^4| \:-\: \frac{1}{28} \:|A^2|^2 \]
or, more generally,
\beq \label{63d}
{\mathcal{L}}[A] \;=\; |A^{2n}| \:-\: \frac{1}{28} \:|A^n|^2
\qquad (n \geq 1),
\eeq
are all satisfied in the vacuum for exactly the same reasons as~(\ref{e:2O}).
The next section gives an argument which distinguishes~(\ref{e:2O}) from
the other Lagrangians.

\section{Stability of the Vacuum}
\setcounter{equation}{0} \label{esec24}
In {\S}\ref{psec23} we argued that the pointwise product~$P(x,y)\: P(y,x)$
depends essentially on the unknown high-energy behavior of the fermionic
projector and is therefore undetermined. This argument, which led us to
a weak analysis near the light cone and was the starting point for
the formalism of the continuum limit in~{\S}\ref{psec26}, must clearly
be taken seriously if one wants to get information for general
regularizations.
On the other hand, the equations of discrete space-time (if analyzed
without taking the continuum limit) should yield constraints for
the regularization, and one might expect that for the special
regularizations which satisfy these constraints,  one can make statements on the fermionic projector even
pointwise. In particular, it is tempting to conjecture that away from
the light cone, where the fermionic projector of the continuum
is smooth (see~{\S}\ref{jsec5}),
the fermionic projector of discrete space-time should
be well-approximated pointwise by the fermionic projector of the continuum.
Since going rigorously beyond the continuum limit is difficult and requires
considerable work, we cannot prove this conjecture here.
But we can take it as an ad-hoc assumption
that away from the light cone (i.e.\ for $|s|,|l| \gg E_P^{-1}$), the physical fermionic projector should coincide to leading orders in~$s E_P^{-1}$
and~$l E_P^{-1}$ with the continuum fermionic projector. We refer to this
assumption that the fermionic projector is
{\em{macroscopic away from the light cone}}\index{macroscopic away
  from the light cone}.

In this section we will analyze the EL equations in the vacuum under
the assumption that the fermionic projector is macroscopic away from
the light cone. This will give us some insight into how causality
enters the EL equations. More importantly, a stability analysis of the
vacuum will make it possible to uniquely fix our variational
principle.  Our analysis here can be considered as being complementary
to the continuum limit: whereas in the continuum limit we restrict
attention to the singularities on the light cone, we here consider
only the behavior away from the light cone, where the fermionic
projector is smooth. This gives us smooth functions defined for $y\in
M\setminus L_x$, which however have poles when $y$ approaches the light
cone around $x$. In this way, we will again encounter singularities on
the light cone, but of different nature than those considered in the
continuum limit. Unfortunately, our method gives no information on the
behavior of these singularities, and therefore we must treat them with
an ad-hoc ``regularity assumption'' (see Def.~\ref{def612}). For this reason,
the  arguments given here should be considered only  as a first step
towards an analysis beyond the continuum limit.  The methods and
results of this section will not be needed later in this book.

We begin by deriving the spectral decomposition of the closed
chain away from the light cone.
Due to the direct sum structure of the fermionic projector, we can
again consider the sectors separately. In the neutrino sector, the
product~$P(x,y)\:P(y,x)$ vanishes identically due to chiral
cancellations~(\ref{e:26a}). Again assuming that the masses are
degenerate (see~{\S}\ref{esec23}), it remains to consider one massive
sector~(\ref{e:2n}) with three mass parameters~$m_\alpha$, $\alpha=1,2,3$.
Since we assume that the fermionic projector is macroscopic
away from the light cone, we do not need a regularization.
\begin{Lemma}\label{5.6.1}
In a massive sector, the closed chain~$A = P(x,y)\:P(y,x)$ has away
from the light cone the following spectral decompositions.
If~$y-x$ is spacelike,
\beq \label{61a}
A \;=\; \lambda\: \1 \spc {\mbox{with $\lambda \in \R$}}\:.
\eeq
If on the other hand~$y-x$ is timelike, $A$ has either the spectral
decomposition~(\ref{61a}) with $\lambda\geq 0$ or
\beq \label{62z}
A \;=\; \sum_{s=\pm} \lambda_s\: F_s
\eeq
with spectral projectors~$F_\pm$ on two-dimensional eigenspaces and
corresponding eigenvalues $\lambda_\pm$ which are real and positive,
\beq \label{61b}
\lambda_\pm \;\geq\; 0 \:.
\eeq
\end{Lemma}
{\Proof}
We write the fermionic projector (\ref{e:2n}) in the form~(\ref{e:2g}) with
\begin{eqnarray*}
g_j(x,y) &=& i \partial_j \:\frac{1}{2} \sum_{\alpha=1}^3 (P_{m_\alpha^2} -
K_{m_\alpha^2})(x,y) \\
h(x,y) &=& \frac{1}{2} \sum_{\alpha=1}^3 m_\alpha (P_{m_\alpha^2} -
K_{m_\alpha^2})(x,y) \:,
\end{eqnarray*}
where~$P_a$ and~$K_a$ denote the fundamental solutions of the Klein-Gordon
equation,
\begin{eqnarray}
P_a(x,y) &=& \int \frac{d^4p}{(2 \pi)^4}\: \delta(p^2-a)\: e^{-ip(x-y)}
\label{Padef} \\
K_a(x,y) &=& \int \frac{d^4p}{(2 \pi)^4}\: \epsilon(p^0)\:
\delta(p^2-a)\: e^{-ip(x-y)} \:. \label{Kdef}
\end{eqnarray}
Due to Lorentz symmetry, the vector field~$g$ is parallel
to~$y-x$. Thus we can write it as
\beq \label{grep}
g_j(x,y) \;=\; i(y-x)_j\: f(x,y)
\eeq
with a complex scalar distribution~$f$. Furthermore,
the distribution~$K_a$ is causal in the sense that
${\mbox{supp}}\: K_a(x,.) \subset J_x$. This can be seen by decomposing
it similar to~(\ref{32}) as
\[ K_a \;=\; \frac{1}{2 \pi i}\left( S^\lor_a - S^\land_a \right) , \]
where~$S^\lor$ and $S^\land$ are the causal Green's functions of the Klein-Gordon
equation (see~(\ref{l:11})). Alternatively, one can for any
spacelike vector~$y-x$ choose a reference frame where~$y-x$ points in
spatial direction, $y-x=(0, \vec{v})$. Then the Fourier integral~(\ref{Kdef}) can be written as
\[ K_a \;=\; \frac{1}{(2 \pi)^4} \int_{-\infty}^\infty d\omega\:
\epsilon(\omega)
\int_{\sR^3} d\vec{p} \; \delta(\omega^2 - |\vec{p}|^2-a)\: e^{-i \vec{p} \vec{v}} \:, \]
and a symmetry argument shows that the integrals over the upper
and lower mass shells cancel each other.

If $y-x$ is spacelike, $K_a$ drops out of our formulas
due to causality,
\[ g_j(x,y) \;=\; i \partial_j \:\frac{1}{2} \sum_{\alpha=1}^3 P_{m_\alpha^2}(x,y)
\:,\spc
h(x,y) \;=\; \frac{1}{2} \sum_{\alpha=1}^3 m_\alpha P_{m_\alpha^2}(x,y) \: .\]
Performing in (\ref{Padef}) the transformation~$p \to -p$, we find
that~$P_a$ has even parity, $P_a(y,x)=P_a(x,y)$.
As a consequence,
\[ \overline{g(x,y)} \;=\; g(y,x) \;=\; -g(x,y)\:,\spc
\overline{h(x,y)} \;=\; h(y,x) \;=\; h(x,y)\:. \]
Plugging these relations into~(\ref{e:2nn}), we obtain
\[ A \;=\; \gslsh \:\overline{\gslsh} + h \:\overline{h} \;=\;
\left( \Bra g, \overline{g} \Ket + |h|^2 \right) \1 \:, \]
where in the last step we used~(\ref{grep}). This proves (\ref{61a}).

If~$y-x$ is timelike, the situation is more complicated because
both~$P_a$ and~$K_a$ contribute. Using the
representation~(\ref{e:2g}, \ref{grep}), we obtain
\[ A \;=\; i \xi \slsh \left(\overline{f} h + f \overline{h} \right)
+ \xi^2\:|f|^2\:\1 + |h|^2\:\1 \:. \]
Away from the light cone, the matrix~$\xi\slsh$ has the two real
eigenvalues~$\pm \sqrt{\xi^2}$, each with
multiplicity two. Hence the eigenvalues of~$A$ are given by
\[ \lambda_{\pm} \;=\;
\pm \sqrt{\xi^2} \left(\overline{f} h + f \overline{h} \right)
+ \xi^2\:|f|^2  + |h|^2
\;=\; \left| \sqrt{\xi^2} f \pm h \right|^2 \;\geq\; 0 \:. \]

\vspace*{-.6cm} \QED

This lemma has important consequences. We first recall that the
conditions (\ref{e:2A}) obtained from a strong analysis near the light
cone led us to only consider Lagrangians for which~${\mathcal{M}}$
vanishes if~$A$ has in the massive sector two eigenvalues appearing
as complex conjugate pairs. According to~(\ref{61a}), this last
condition is satisfied if~$y-x$ is space-like. This means
that for all variational principles discussed in~{\S}\ref{esec23},
the matrix~${\mathcal{M}}_{xy}$ vanishes for space-like~$y-x$.
According to~(\ref{e:2c}), the same is true for~$Q(x,y)$.
In other words, $Q$ is supported inside the (closed) light cone.
This remarkable property can be understood as a manifestation of some kind of
``causality'' in the EL equations. However, we point out that this
property is no longer true in an interacting system (see Chapter~\ref{esec3}).

Using that~$Q$ is supported inside the light cone, we can in what
follows restrict attention to {\em{timelike}} $y-x$. For convenience,
we will also use~(\ref{62z}) in the case that $A$ is a multiple of the
identity matrix~(\ref{61a}); we then simply set $\lambda_+ =\lambda_-
=\lambda$. The main statement of Lemma~\ref{5.6.1} for timelike $y-x$ is
that, according to~(\ref{61b}), ~$A$ is a {\em{positive}} matrix.
This means that the
concept of the spectral weight, which was important in~{\S}\ref{esec23} (and which will also be crucial in
Chapter~\ref{esec3}), becomes trivial in the vacuum.
The spectral weight~(\ref{p:38}) reduces to the ordinary trace, and so our
Lagrangians~(\ref{63d}) simplify away from the light cone
to polynomial Lagrangians.

Suppose that $P$ is a {\em{stable minimum}} of the action,
in the sense that it is impossible to decrease the action by a
variation of $P$,
\beq \label{61z}
S[P+\delta P] \;\geq\; S[P]\quad\mbox{for all variations~$\delta P$}.
\eeq
In order to derive necessary conditions for stability, we shall
consider this inequality for special variations. Our idea is to vary
$P$ by changing the momentum and spin orientation of individual
fermionic states. In order to have discrete states, we consider the
system as in {\S}\ref{jsec6} in finite $3$-volume. According to the
replacement rule (\ref{eq:A}), the fermionic projector of the vacuum
then becomes
\begin{eqnarray}
P(x,y) &=& \int^\infty_{-\infty} \frac{dk^0}{2\pi} \frac{1}{V}
\sum_{k\in L^3} \hat{P}(k) e^{-ik(x-y)} \label{62aa} \\
\hat{P}(k) &=& \sum^3_{\alpha=1} (k\slsh + m_\alpha)\:
\delta(k^2-m^2_\alpha)\: \theta (-k^0)\:.
\end{eqnarray}
Since we want to preserve the vector-scalar structure, we take both
fermionic states for any $k$ on one of the mass shells and bring them
into states of momentum $q$ which are not on the mass shells,
\beq \label{569a}
\delta P \;=\; -c\: (k\slsh +m)\: e^{-ik(x-y)} + \tilde{c}\: (l\slsh
+\tilde{m})\: e^{-iq(x-y)}
\eeq
with $m\in\{m_1,m_2,m_3\} \not\ni \sqrt{q^2}$ and $k^2=m^2$, $k^0<0$,
$l^2=\tilde{m}^2$. Here~$c$ and~$\tilde{c}$ are normalization
constants; carrying out the~$k^0$-integral in (\ref{62aa}) one sees
that
\beq \label{62c}
c \;=\; \frac{1}{4\pi k^0} \frac{1}{V}\:.
\eeq
We must make sure that we do not change the normalization of the
fermionic states. First of all, this means that we must preserve the
sign of the inner product of the states, and this implies that $l$
must be on the lower mass shell, $l^0<0$. Furthermore, the phase
factors $e^{-ik(x-y)}$ and $e^{-iq(x-y)}$ in (\ref{569a})
drop out of normalization integrals and thus
have no influence on the normalization, but the normalization constants
must be kept fixed. This means that $cm=\tilde{c}\tilde{m}$, and by
rescaling $\tilde{c}$ and $\tilde{m}$ we can arrange that $\tilde{c} = c$
and $\tilde{m} = m$. We conclude that the variation of $P$ must be of
the form
\beq \label{62d}
\delta P = -c\: (k\slsh +m)\: e^{-ik(x-y)} + c\:(l\slsh +m)\: e^{-iq(x-y)}
\eeq
with
\beq \label{62e}
m\in\{m_1, \ldots,m_g \} \not\ni \sqrt{q^2}\,,\quad k^2=l^2=m^2\,,\quad
k^0,l^0<0\:.
\eeq
This variation preserves the normalization of the fermionic states, as
one sees easily from the fact that it can be realized by a unitary
transformation $U$ (\ref{p:50a}) ($U$ picks the states of momentum
$k$, multiplies them by the phase factor $e^{-i(q-k)x}$ and finally
``Lorentz boosts'' the spinors with a constant unitary transformation as
considered in Lemma~\ref{Dirtrans}). We point out that the variation
(\ref{62d}, \ref{62e}) is discrete (and not a continuous family of variations
$\delta P(\tau)$). But due to the factor $V^{-1}$ in (\ref{62c}), we
can make $\delta P$ arbitrarily small by making the $3$-volume
sufficiently large. Therefore, it is admissible to treat
$\delta P$ perturbatively.
The variation of the action (\ref{e:2b}) is (up to an irrelevant
normalization constant) computed to be
\beq \label{62g}
\delta S \;=\; - \Tr (\hat{Q}(k) (k\slsh +m)) + \Tr (\hat{Q}(q) (l\slsh +m))\:,\eeq
where $\hat{Q}$ is the Fourier transform of $Q$, \label{hatQ(p)}
\[ \hat{Q}(p) \;=\; \int Q(\xi)\: e^{-ip \xi}\; d^4\xi\:. \]
Evaluating the stability inequality (\ref{61z}) gives rise to the
notion of state stability. We first give the definition and explain it
afterwards. We denote the {\em{mass cone}}\index{mass cone} by
\[ \mathcal{C} \;=\; \{p\mid p^2>0\} \]
and label the {\em{upper}}\index{mass cone!upper} and
{\em{lower mass cone}}\index{mass cone!lower} by superscripts
$\lor$ and $\land$, respectively, \label{mathcalC} \label{mathcalClor}
\label{mathcalCland}
\begin{equation} \label{calcdef}
\mathcal{C}^\lor \;=\; \{p\mid p^2>0\,\,\mbox{and}\,\, p^0>0\}\,,\spc
\mathcal{C}^\land \;=\; \{p \mid p^2>0\,\,\mbox{and}\,\, p^0<0\} \,.
\end{equation}

\begin{Def} \label{def611}\index{state stability}
  The fermionic projector of the vacuum is called {\bf{state stable}}
  if the corresponding operator $\hat{Q}(p)$ is well-defined in the
  lower mass cone $\mathcal{C}^\land$ and can be written as \beq
  \label{62f} \hat{Q} (p) \;=\; a\:\frac{p\slsh}{|p|} + b \eeq with
  continuous real functions $a$ and $b$ on $\mathcal{C}^\land$ having
  the following properties:
\begin{itemize}
\item[(i)] $a$ and $b$ are Lorentz invariant,
\[ a \;=\; a(p^2)\:,\spc b \;=\; b(p^2) \:. \]
\item[(ii)] $a$ is non-negative.
\item[(iii)] The function $a+b$ is minimal on the mass shells,
\[ (a+b)(m^2_\alpha) \;=\; \inf_{q \in {\mathcal{C}}^\land} (a+b)(q^2)
\quad\mbox{for~$\alpha=1,2,3$}. \]
\end{itemize}
\end{Def}

It is natural to assume that $Q$ is Lorentz invariant because $P$ has
this property too. Thus the main point of (\ref{62f}) is that
$\hat{Q}$ is finite and continuous inside the lower mass cone. This is
needed because otherwise the term $\Tr(\hat{Q}(k)(k\slsh +m))$ in
(\ref{62g}) would be ill-defined (strictly speaking,  we only need
that $\hat{Q}$ is finite on the lower mass shells, but this seems
impossible to arrange without $\hat{Q}$ being well-defined and
continuous inside the whole lower mass cone).
Using (\ref{62f}), we obtain for any $q\in\mathcal{C}^\land$,
\beq \label{62h}
\frac{1}{4}\: \Tr(\hat{Q}(q)(l\slsh +m)) \;=\; a\:\frac{\langle
  q,l\rangle}{|q|} + b\:m\,.
\eeq
Since the vectors $q$ and $l$ are both in the lower mass cone, their
inner product $\langle q,l\rangle$ is positive,
and it can be made arbitrarily large by choosing
$|l^0| \gg 1$. Hence if $a$ were negative, (\ref{62h}) would not be
bounded from below, leading to an instability. This explains (ii). If
$a$ is non-negative,
\[ \frac{1}{4}\: \inf_l \:\Tr (\hat{Q}(q)(l\slsh +m))
\;=\; a(q) \:m + b(q)\:m\,. \]
Comparing this with the first term in (\ref{62g}) gives (iii).

The question arises whether the conditions of Def.~\ref{def611}, which
are clearly necessary for stability, are also sufficient.
The fact that we vary pairs of fermions keeping the vector-scalar
structure is indeed no restriction because if one varies one fermionic
state, the additional pseudoscalar, axial and bilinear contributions
(see~(\ref{eq:C10}) for details) drop out when taking similar to (\ref{62h})
the trace with $\hat{Q}$. Also, at least if $a$ is strictly positive
and if the function $a+b$ has no minima away from the mass shells, we
find that the variation~(\ref{62g}) really increases the action, and
we obtain stability. Hence the main restriction of state stability is
that we consider stability only within the class of homogenous
fermionic projectors. The second restriction is that Def.~\ref{def611}
involves no condition for $\hat{Q}(p)$ if $p$ is outside the lower
mass cone. This is because we want to allow for the possibility that
$\hat{Q}(q)$ is infinite for $q\notin \mathcal{C}^\land$, in such a way
that the expression~(\ref{62h}) equals $+\infty$. Treating such
infinities would make it necessary to introduce an ultraviolet
regularization and to analyze the divergences as the regularization is
removed. Since this would go far beyond the treatment in this section,
we here simply disregard those $q$ for which ultraviolet divergences
might appear.

\begin{figure}[tb]
\begin{center}
\scalebox{0.9}
{\includegraphics{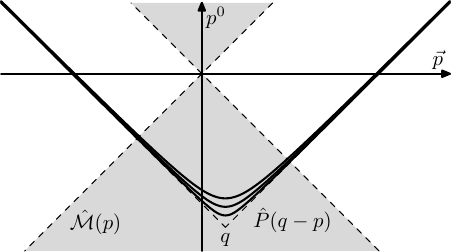}}
\caption{The convolution~${\hat{\mathcal{M}}} * \hat{P}$.}
\label{fig4}
\end{center}
\end{figure}

We come to the analysis of state stability. First, we rewrite $Q$ using
the spectral representation of $A$. Our starting point is~(\ref{e:2C}),
\beq \label{62j}
Q(x,y) \;=\; \frac{1}{2}\: {\mathcal{M}}(x,y)P(x,y)\:.
\eeq
Away from the light cone, we can apply Lemma~\ref{5.6.1} and write
$\mathcal{M}(x,y)$ as
\beq \label{62k}
\mathcal{M}(x,y) \;=\; \begin{cases}
\displaystyle \sum\limits_{s=\pm} \frac{\partial\mathcal{L}}{\partial\lambda_s} F_s
&\quad \mbox{if $y-x$ is timelike}\\
\qquad 0&\quad\mbox{if $y-x$ is spacelike}\,.
\end{cases}
\eeq
The important point for us is that the integral kernel of $Q$ has a
product structure (\ref{62j}). This product in position space can also
be expressed as a convolution in momentum space,
\beq \label{62m}
\hat{Q} (q) \;=\; \frac{1}{2} \:\int \frac{d^4p}{(2\pi)^4}\:
\hat{\mathcal{M}} (p) \:\hat{P}(q-p)\:.
\eeq
As is illustrated in Figure~\ref{fig4}, the integration range is
unbounded.
Therefore, the existence of the integral as well as its
value might depend sensitively on an ultraviolet regularization. Since
in this section we work with the unregularized fermionic projector, we
simply impose that (\ref{62m}) should hold even after the ultraviolet
regularization has been removed.

\begin{Def}\label{def612}
The fermionic projector satisfies the assumption of a {\bf
  distributional $\mathcal{M}P$-product}\index{distributional $\mathcal{M}P$-product} if for every $q$ for which
  $\hat{Q}(q)$ exists, the convolution integral (\ref{62m}) is
  well-defined without regularization and coincides with
  $\hat{Q}(q)$.
\end{Def}

We point out that the above assumption is not merely a technical
simplification, but it makes a highly non-trivial statement on the
high-energy behavior of the physical fermionic projector. Ultimately,
it needs to be justified by an analysis of the variational principle
with ultraviolet regularization. At this stage, it can at least be
understood from the fact that the assumption of a distributional
$\mathcal{M}P$-product makes our stability analysis robust to
regularization details. This means physically that the system should
also be stable under ``microscopic'' perturbations of the fermionic
projector on the Planck scale.

In what follows we will assume that the conditions in Def.~\ref{def611}
and Def.~\ref{def612} are satisfied. Then the convolution integral
(\ref{62m}) is well-defined for all $q\in\mathcal{C}^\land$. This
implies that $\hat{\mathcal{M}}(p)$ must be a well-defined distribution on
the set $\overline{\mathcal{C}^\land}\cup\{p\mid p^2<0\}$. Furthermore,
since both $\hat{P}$ and $\hat{Q}$ are Lorentz invariant, we may
assume that $\hat{\mathcal{M}}$ is also Lorentz invariant (otherwise
we could replace $\hat{\mathcal{M}}$ by its Lorentz invariant part,
and (\ref{62m}) would remain true). But if $\mathcal{M}(x,y)$ is
Lorentz invariant, it involves no bilinear contribution, and its
vector component is a multiple of $y-x$.
According to~(\ref{62k}), the matrix~$\mathcal{M}(x,y)$ commutes
with~$A$. This implies that, for all $x,y$ for which~$\mathcal{M}(x,y)
\neq 0$, the matrix~$A$ involves no bilinear contribution.
Since the bilinear part of~$A$ is given by the commutator of~$P(x,y)$
and $P(y,x)$, we find that~$P(x,y)$ and~$P(y,x)$ commute.
It follows that~$A_{xy} = A_{yx}$ for all~$x,y$
for which~$\mathcal{M}(x,y) \neq 0$. As a
consequence, $\mathcal{M}(x,y) = \mathcal{M}(y,x)$, and thus
\beq \label{62n}
\hat{\mathcal{M}}(-p) \;=\; \hat{\mathcal{M}}(p) \:.
\eeq
This identity yields that $\hat{\mathcal{M}}$ is a well-defined
distribution even on $\mathcal{C}^\lor$. We thus come to the following
conclusion.

\begin{Prp} \label{prp6.1.4}
If the fermionic projector is state stable and the
$\mathcal{M}P$-product is distributional, then $\hat{\mathcal{M}}(p)$
is a Lorentz invariant distribution of even parity (\ref{62n}).
\end{Prp}
This proposition poses a strong constraint for the ultraviolet
regularization of the fermionic projector. It is a difficult question
whether there are regularizations which satisfy this constraint. But at
least we can say that the result of Proposition~\ref{prp6.1.4} does not
immediately lead to inconsistencies, as the following lemma shows.

\begin{Lemma} \label{612}
For all actions considered in~{\S}\ref{esec23}, there is a
distribution $\tilde{\mathcal{M}}$ on $M$ which coincides with
$\mathcal{M}$ away from the light cone,
\[\tilde{\mathcal{M}}(y-x) \;=\; \mathcal{M}(x,y)\qquad\mbox{for all}\,\,
x,y\in M\,\,\mbox{with}\,\, (y-x)^2\neq 0\,.\]
\end{Lemma}
{\Proof} An explicit calculation using (\ref{62k}) and the
  representation of $P(x,y)$ with Bessel functions shows that for all
  actions considered in~{\S}\ref{esec23}, $\mathcal{M}(x,y)$ is for
  $\xi\equiv y-x$ inside the upper light cone a smooth function
  with the following properties. It can be written as
\[ \mathcal{M}(\xi) \;=\; i\Pdd_{\xi} f(\xi^2) + g(\xi^2) \spc (\xi\in I^\land)\]
with complex-valued functions $f,g\in C^\infty(\mathbb{R}^+)$, which
  have at most polynomial growth as $\xi^2\to\infty$ and at
  most a polynomial singularity on the light cone, i.e.
\[ |f(\xi^2)|+|g(\xi^2| \;\leq\; c\left(\xi^{2n}  +\frac{1}{\xi^{2n}} \right)\]
for suitable constants $c>0$, $n\in\mathbb{N}$.

Setting $z=\xi^2$, the Laplacian of a Lorentz invariant
function $h(\xi^2)$ is computed to be
\beq \label{63d2}
\Box h(z) \;=\; 4z h''(z) + 8h'(z) = \frac{4}{z}(z^2h'(z))'\,.
\eeq
This allows us to invert the Laplacian explicitly,
\beq \label{63e}
\Box^{-1} h(z) \;=\; \frac{1}{4}\int^z_{z_0} \frac{d\tau}{\tau^2} \int^\tau_{\tau_0} \sigma \:h(\sigma) \:d\sigma
\eeq
with two free constants $\tau_0$ and $z_0$. We choose $z_0=1$ and set
$\tau_0=0$ if $\int^1_0 \sigma|h(\sigma)|d\sigma <\infty$ and $\tau_0=1$ otherwise.
Then applying $\Box^{-1}$ decreases the order of the pole on the light cone
by one. Hence the functions
\[ F^\lor (\xi) \;=\; \Box^{-n-1} f(\xi^2)\:,\spc
G^\lor(\xi) \;=\; \Box^{-n-1}g(\xi^2) \]
are smooth functions on $I^\lor$ which are locally bounded near the light cone
and have at most polynomial growth as $\xi^2 \to \infty$.
Repeating the above construction inside the lower light cone gives
rise to functions $F^\land$ and $G^\land$ on $I^\land$.  Extending
these functions by zero outside their respective domains of
definition, the functions
\[ F \;=\; F^\lor + F^\land\:, \spc G \;=\; G^\lor + G^\land \]
are regular distributions on $M$. By construction, the distribution
\[ \tilde{\mathcal{M}} \;=\; i\Pdd \Box^{n+1} F + \Box^{n+1} G \]
coincides inside the light cone with $\mathcal{M}$ and vanishes
outside the light cone.  \QED

The next lemma states a few important properties of the distribution
$\hat{\mathcal{M}}$.
\begin{Lemma} \label{6.1.3}
Under the assumptions of Proposition~\ref{prp6.1.4}, the vector component of
$\hat{\mathcal{M}}$ is supported inside the mass cone $\{p\mid
p^2\geq 0\}$. If the scalar component of $\mathcal{M}$ has a
nonvanishing contribution with the following asymptotics as $I
\ni  \xi \to 0$,
\beq \label{63aa}
\mathcal{M}(\xi) \;\sim\; \xi^{2r} \log^s
\xi^2\quad\mbox{with}\quad r,s\in\mathbb{Z}, \,\, s\geq 1\,,
\eeq
then its Fourier transform satisfies outside the mass cone for some
$c>0$ the bound
\beq \label{63b}
|\hat{\mathcal{M}}(p)| \;\geq\; \frac{1}{c} (-p^2)^{-2-r}\quad{\mbox{if~$p^2<-c$}}\,.
\eeq
\end{Lemma}
{\Proof} The statement for the vector component immediately follows from a
symmetry argument: Due to Lorentz invariance, the vector component of
$\hat{\mathcal{M}}(p)$ can be written as $p\slsh f(p)$ with a scalar
distribution $f$. Since $\hat{\mathcal{M}}$ has even parity
(\ref{62n}), it follows that $f(-p)=-f(p)$. Again using Lorentz
invariance, we conclude that $f(p)$ vanishes if $p$ is outside the
mass cone.

For the scalar component of $\hat{\mathcal{M}}$, the above symmetry
argument does not apply, and (as can easily be verified by an explicit
calculation) there is indeed a contribution outside the mass
cone. More specifically, in the case $r\geq 0$ we can rescale the
Fourier transform by $\xi\to \tau^{-1}\xi$,
\[ \int \xi^{2r} \log^s (\xi^2)\: e^{i\tau p\xi} \:d^4\xi \;=\; \tau^{-4-2r}
\int \xi^{2r}(\log\xi^2-\log\tau^2)^s\: e^{ip\xi} \:d^4\xi\,, \]
and obtain that (\ref{63aa}) gives rise to a contribution to
$\hat{\mathcal{M}}$ with the following asymptotics as $p^2\to-\infty$,
\beq \label{63c1}
\hat{\mathcal{M}}(p) \;\sim\; (-p^2)^{-2-r}\: (\log^s(-p^2)+\mbox{l.o.t.})\spc
(r\geq 0)\,,
\eeq
where `l.o.t.' denotes lower order in $\log(-p^2)$.
If $r<0$, the singularity on the light cone cannot be treated with
this scaling argument. But we can nevertheless compute the Fourier
transform by iteratively applying the operator $\Box^{-1}_p$,
\[ (-\Box_p)^{2r} \int \log^s(\xi^2)\: e^{ip\xi} d^4\xi \;=\; \int \xi^{2r}
\log^s(\xi^2)\:e^{ip\xi}d^4\xi\,. \]
Using (\ref{63e}) and (\ref{63c1}), we obtain a
contribution to $\hat{\mathcal{M}}(p)$ with the following asymptotics
as $p^2\to -\infty$,
\beq \label{63c2}
\hat{\mathcal{M}}(p) \;\sim\; (-p^2)^{-2-r}\: (\log^{s+1}(\xi^2) +
\mbox{l.o.t.}) \spc(r<0)
\eeq
(here we do not need to worry about the integration constants in
(\ref{63e}) because these correspond to contributions localized on the
light cone, which will be considered separately below, see
(\ref{63f})).
The asymptotic formulas (\ref{63c1}, \ref{63c2}) explain the estimate
(\ref{63b}). However, the proof is not yet finished because the singular
part of $\mathcal{M}$ which is localized on light cone might give rise to a
contribution to $\hat{\mathcal{M}}$ which cancels (\ref{63c1}) or (\ref{63c2}). Thus it
  remains to show that all contributions to $\mathcal{M}$ which are
  localized on the light cone have an asymptotics different from
  (\ref{63c1}, \ref{63c2}) and thus cannot compensate these contributions.

A scalar Lorentz invariant distribution $h$ which is localized on the
light cone satisfies for some $n>0$ the relation
$\xi^{2n}h(\xi)=0$. Hence its Fourier transform is a
distributional solution of the equation
\[ \Box^n \hat{h}(p) \;=\; 0\,. \]
In the case $n=1$, we see from (\ref{63d2}) that away from the mass
cone, $\hat{h}(p)$ must be a linear combination of the functions $1$
and $p^{-2}$. By iteratively applying (\ref{63e}) one finds that for
general $n$, $\hat{h}$ is of the form
\beq \label{63f}
\frac{1}{p^2} \,\, \mbox{(polynomial in $p^2$)}\,\,+ \log(-p^2)\,\,
\mbox{(polynomial in $p^2$)}
\eeq
The asymptotics of these terms as $p^2 \to -\infty$ is clearly different from that
in~(\ref{63c1}, \ref{63c2}).
\QED

We are now ready to prove the main result of this section.
\begin{Thm} \label{thm6.1.4}
Consider the action corresponding to a Lagrangian of type
(\ref{63d}). Assume that the fermionic projector of the vacuum is
state stable and that the $\mathcal{M}P$-product is distributional
(see Def.~\ref{def611} and Def.~\ref{def612}). Then $n=1$.
\end{Thm}
{\Proof} Since here we consider only one sector, we need
to change the normalization of the spectral weight in (\ref{63d}),
\[ \mathcal{L} [A] = |A^{2n}| -\frac{1}{4} |A^n|^2\,. \]
According to Lemma~\ref{5.6.1}, $\mathcal{L}$ vanishes identically for
spacelike $y-x$, whereas for timelike $y-x$ we may replace the
spectral weight by an ordinary trace. Hence away from the light cone and
up to an irrelevant constant factor~$2n$,
\beq \label{63g}
\mathcal{M} \;=\; \left[ A^n-\frac{1}{4}\Tr(A^n) \right] A^{n-1}\,.
\eeq
According to Proposition~\ref{prp6.1.4}, $\hat{\mathcal{M}}$ is a Lorentz
invariant distribution of even parity. Following Def.~\ref{def611} and
Def.~\ref{def612}, the convolution integral (\ref{62m}) should be
well-defined for any $q$ inside the lower mass cone. According to
Lemma~\ref{6.1.3}, the vector component of $\hat{\mathcal{M}}$ is
supported inside the closed mass cone. Thus for the corresponding
contribution to (\ref{62m}), the integration range is indeed compact
(see Figure~\ref{fig4}), and so (\ref{62m}) is well-defined in the
distributional sense.

It remains to consider the scalar component of
$\hat{\mathcal{M}}$. This requires a more detailed analysis. We again
write the fermionic projector in the form (\ref{e:2g}, \ref{grep}),
\[ P(x,y) \;=\; i\xi\slsh f + h\,.\]
Using the representation (\ref{l:3.1}) of the distribution $T_{m^2} =
\frac{1}{2}(P_{m^2}-K_{m^2})$ in position space, we can write $f$ and
$g$ in the upper light cone as
\begin{eqnarray*}
f &=& \frac{c_1}{\xi^4} + \frac{c_2}{\xi^2} +
\alpha\: (\log \xi^2+i\pi)+\beta\\
g &=& \frac{c_3}{\xi^2} +\gamma\: (\log \xi^2+i\pi)
+\delta
\end{eqnarray*}
with constants $c_j\neq 0$ and smooth real function
$\alpha,\beta,\gamma,\delta$ with $\gamma_0:=\gamma(0)\neq 0$.
A short calculation yields
\begin{eqnarray*}
A &=& \xi^2|f|^2+|h|^2-\xi\slsh
\mbox{Im}(f\overline{h})\\
&=& \gamma^2_0 \log^2\xi^2 + \mbox{(lower orders in
  $\log\xi^2$)} \:+\:
  \xi\slsh \left(\frac{1}{\xi^4}+ \mbox{(higher
    orders in $\xi^2$)}\right)\,.
\end{eqnarray*}
The important point is that the vector component of $A$ involves no
logarithms, whereas the scalar component involves terms
$\sim\log^2\xi^2$. Substituting this expansion of $A$ into (\ref{63g}),
we obtain for the scalar component of $\mathcal{M}$
sums of expressions of the general form
$\xi^{2r}\log^s\xi^2$. We select those terms for which
$s$ is maximal, and out of these terms for which $r$ is
minimal. We get a contribution only when both the square bracket and
the factor $A^{n-1}$ in~(\ref{63g}) contain an odd number of Dirac matrices,
and a short computation yields the asymptotics
\beq \label{64a}
{\mathcal{M}} \;\sim\; (n-1)\: \xi^{-6} \log^{4n-6} \xi^2\,.
\eeq
If $n=1$, the scalar component of $\hat{\mathcal{M}}$ vanishes, and we
get no further conditions. However if $n>1$, Lemma~\ref{6.1.3} yields
that $\hat{\mathcal{M}}(p)$ is outside the mass cone for large $p$
bounded away from zero by
\[ |\hat{\mathcal{M}}(p)| \;\geq\;  -\frac{p^2}{c}\,.\]
As a consequence, the convolution integral (\ref{62m}) diverges. \QED

We close this section with a few remarks. First, we can now discuss
the stability of the vacuum for the polynomial actions
(\ref{e:2B}). \index{action!polynomial}
The strong analysis on the light cone in~{\S}\ref{esec23}
forced us to only consider polynomial Lagrangians which vanish
identically if $A$ has two independent eigenvalues. According to Lemma~\ref{5.6.1}, this implies that $Q$ vanishes identically away from the
light cone, and so the above stability analysis becomes
trivial. Indeed, the following general consideration shows that for
polynomial actions, the vacuum is {\em{not stable}} in the strict sense:
Under the only assumption that $A$ has vector-scalar structure, the
matrix $A$ has at most two independent eigenvalues (see (\ref{e:2m})
and Lemma~\ref{elemma21}), and so the Lagrangian vanishes
identically. Thus the variational principle is trivial; it gives no
conditions on the structure of the fermionic projector of the vacuum.

It might seem confusing that in the above analysis, the operator
$Q(\xi)$ had poles on the light cone (see (\ref{64a}, \ref{62j})),
although it vanishes identically in the formalism of the continuum limit.
This can be understood from the fact that on the light cone, the matrix
$A$ has a
pair of {\em complex} conjugated eigenvalues (see (\ref{e:2t})),
whereas away from the light cone its eigenvalues are in general
distinct {\em real} eigenvalues (see Lemma~\ref{5.6.1}). If a matrix
has non-real eigenvalues, this property remains valid if the
matrix is slightly perturbed. Therefore, treating the eigenvalues of
$A$ perturbatively, it is impossible to get from the region on the
light cone to the region away from the light cone. This is the reason
why the result of Lemma~\ref{5.6.1} cannot be obtained in the
formalism of the continuum limit. The transition between the
asymptotic regions near and away from the light cone can be studied
only by analyzing the EL equations with regularization in detail, and
this goes beyond the scope of this book. However, we point out that
the order of the pole in (\ref{64a}) is lower than the order of all
singularities which we will study in the continuum limit. This
justifies that we can neglect (\ref{64a}) in what follows; taking into
account (\ref{64a}) would have no effect on any of the results in this
book.

The above argument shows that from all Lagrangians mentioned in~{\S}\ref{esec23}, only for~(\ref{e:2O}) the vacuum can be stable (in the
sense made precise in Theorem~\ref{thm6.1.4}). The important question arises
whether the above necessary conditions are also sufficient, i.e.\ if
for the Lagrangian (\ref{e:2O}) the vacuum is indeed state
stable. This question cannot be answered here; in any case a more
detailed analysis would yield additional constraints for the mass
parameters $m_\alpha$ and the regularization.
But at least we can say that for the Lagrangian (\ref{e:2O}), the
vacuum has very nice properties which point towards stability:
First of all, $Q$ vanishes in the continuum limit. But it does not vanish
identically away from the light cone, and thus the EL equations are non-trivial
in the vacuum. Furthermore, $Q(\xi)$ is {\em supported in
the light cone}, giving an interesting (although not yet fully
understood) link to causality. The crucial point for Theorem~\ref{thm6.1.4} is that the {\em scalar component} of $\mathcal{M}$ {\em
  vanishes} (as is obvious from (\ref{63g})), and therefore
$\hat{\mathcal{M}}$ is {\em supported inside the mass cone}
(see Lemma~\ref{6.1.3}).
The last property implies via a simple support argument that, for $q$
inside the lower mass cone, the convolution integral (\ref{62m}) is
finite (see Figure~\ref{fig4}). This means that the stability conditions (ii)
and (iii) of Def.~\ref{def611} (which we did not consider here) could
be analyzed without any assumption on the regularization. However, for
$q$ outside the lower mass cone, the convolution integral (\ref{62m})
will in general diverge (see again Figure~\ref{fig4}), indicating that the
vacuum could be stable even under perturbations where fermionic states
with momenta outside the lower mass cone are occupied.

\chapter{The Dynamical Gauge Group}
\setcounter{equation}{0} \label{esec3}
We now begin the analysis of the continuum limit
of the EL equations with interaction. In order to work in a concrete example,
we shall analyze our model variational principle of~{\S}\ref{psec15}.
But the methods as well as many of the results carry over to other
variational principles, as will be discussed in the Remarks at the
end of Chapter~\ref{esec3} and at the end of Chapter~\ref{esec4}.
We again consider the fermionic projector of the standard
model~{\S}\ref{esec1}. For the bosonic potentials in the corresponding
auxiliary Dirac equation~(\ref{b}) we make the ansatz
\begin{equation}
{\mathcal{B}} \;=\; C\!\slsh + \gamma^5 \: E\!\slsh + \Phi + i
\gamma^5 \:\Xi
    \label{e:3a}
\end{equation}
with a vector potential $C$, an axial potential $E$ and
scalar/pseudoscalar potentials $\phi$ and $\Xi$, which again in
component notation ${\mathcal{B}}={\mathcal{B}}^{(a \alpha)}_{(b
\beta)}$ we assume to be of the form
\begin{equation}
C \;=\; C^a_b\: \delta^\alpha_\beta \:,\;\;\;\;\;\;
E \;=\; E^a_b\: \delta^\alpha_\beta \:,\;\;\;\;\;\;
\phi \;=\; \phi^{(a \alpha)}_{(b \beta)} \:,\;\;\;\;\;\;
\Xi \;=\; \Xi^{(a \alpha)}_{(b \beta)} \:.    \label{e:3c}
\end{equation}
Exactly as in~{\S}\ref{jsec5} it is convenient to introduce the chiral
potentials\index{potential!chiral} \label{A_L/R} \label{Y_L/R}
\begin{equation}
    A_{L\!/\!R} \;=\; C \pm E \:,
    \label{e:3b}
\end{equation}
and to define the {\em{dynamical mass matrices}}\index{mass matrix!dynamical} by
\begin{equation}
m Y_{L\!/\!R} \;=\; m Y - \phi \mp i \Xi \:. \label{e:3dyn}
\end{equation}
Then the auxiliary Dirac equation takes the form
\begin{equation}
\left( i \Pdd + \chi_L (\Aslsh_R - mY_R) + \chi_R (\Aslsh_L - mY_L) \right)
P(x,y) \;=\; 0\:.    \label{e:3dir}
\end{equation}
Clearly, the potentials in~(\ref{e:3a}) must be causality
compatible~(\ref{89}).
We assume in what follows that this condition is satisfied, and we
will specify what it means in the course of our analysis.

Let us briefly discuss the ansatz~(\ref{e:3a}). The vector and axial potentials
in~(\ref{e:3a}) have a similar form as the gauge potentials in the standard
model. Indeed, when combined with the chiral potentials~(\ref{e:3b}), they can be
regarded as the gauge potentials corresponding to the gauge group $U(8)_L
\times U(8)_R$. This so-called {\em{chiral gauge group}} includes the
gauge group of the standard model\index{gauge group!chiral}.
At every space-time point, it has a natural
representation as a pair of $8 \times 8$ matrices acting on the sectors; we
will work in this representation throughout. Compared to the most general
ansatz for the chiral potentials, the only restriction
in~(\ref{e:3b}, \ref{e:3c}) is that the chiral potentials are the same for the three
generations. This can be justified from the behavior of the fermionic
projector under generalized gauge transformations, as will be explained in
Remark~\ref{rem32} below. The scalar potentials in~(\ref{e:3a}) do not
appear in the standard model, but as we shall see, they will play an important
role in our description of the interaction (here and
in what follows, we omit the word ``pseudo'' and by a ``scalar potential''
mean a scalar or a pseudoscalar potential).
We point out that we do not consider a gravitational field. The
reason is that here we want to restrict attention to the
interactions of the standard model. But since the principle of the fermionic
projector respects the equivalence principle, one could clearly include a
gravitational field; we plan to do so in the future. Compared to a general
multiplication operator, (\ref{e:3a}) does not contain bilinear potentials
(i.e.\ potentials of the form $H_{jk} \sigma^{jk}$ with
$\sigma^{jk}= \frac{i}{2} [\gamma^j, \gamma^k]$). Clearly, bilinear
potentials do not appear in the standard model, but it is not obvious why they should be
irrelevant in our description. Nevertheless, we omit bilinear potentials
in order to keep the analysis as simple as possible. To summarize, (\ref{e:3a})
is certainly not the most general ansatz which is worth studying. But since
the potentials in~(\ref{e:3a}) are considerably more general than the gauge
potentials in the standard model, it seems reasonable to take~(\ref{e:3a}) as
the starting point for our analysis.

\section{The Euler-Lagrange Equations to Highest Degree on the Light Cone}
\markboth{6. THE DYNAMICAL GAUGE GROUP}
{6.1. THE EL EQUATIONS TO HIGHEST DEGREE ON THE LIGHT CONE}
\setcounter{equation}{0} \label{esec31}
We come to the detailed calculations.
We again work with the spectral decomposition of $A_{xy}$ and
proceed degree by degree on the light cone. In this section we
consider the highest degree. Then the fermionic projector is
influenced only by the chiral potentials (and not by the scalar
potentials or the particle states), and the chiral potentials
merely describe local phase transformations of the fermionic
projector. More precisely, truncating all contributions of degree
$<2$ and denoting this ``truncated fermionic projector'' by
$P_0(x,y)$, we have (see~{\S}\ref{jsec5} and Appendix~\ref{pappLC})\footnote{{\textsf{Online version}:}
Here the factor~$g^2$ obtained by summing over the generations when forming
the partial trace (see~\eqref{pt}) is absorbed into the definition of the factors~$T^{(-1)}_{[0]}$.
This differs from the convention used in the book~\cite{cfs}
(listed in the references in the preface to the second online edition),
where for clarity the factors~$g$ which count the number of generations
are always written out.}
\begin{equation}
P_0(x,y) \;=\; \left( \chi_L \:X_L\: \intL_x^y + \chi_R \:X_R \intR_x^y \right)
\frac{i}{2}\:\xi\slsh\: T^{(-1)}_{[0]}(x,y)\:,    \label{e:3d}
\end{equation}
where we used for the ordered exponentials the short notation
\begin{equation}
\intc_x^y \;=\; {\mbox{Pexp}} \left( -i \int_0^1 A_c^j(\tau y + (1-\tau)x)\: (y-x)_j \:
d\tau \right) \label{e:3h}
\end{equation}
with $c=L$ or $R$. We also truncate the matrix $A_{xy}$ by setting
\[ A_0(x,y) \;=\; P_0(x,y) \:P_0(y,x)\:. \]
It follows from~(\ref{e:3d}) that
\begin{equation}
A_0 \;=\; \left\{ \chi_L \:X_L \:\intL_x^y \intR_y^x X_R \:+\:
\chi_R \:X_R \:\intR_x^y \intL_y^x X_L \right\} \:\frac{1}{4}\:
(\xi\slsh\:T^{(-1)}_{[0]})(\overline{\xi\slsh\:T^{(-1)}_{[0]}}) \:.
    \label{e:3e}
\end{equation}
We can assume that the matrix inside the curly brackets is
diagonalizable; indeed, this is the generic situation, and the general case
immediately follows from it by approximation. The matrix $A_0$ is invariant on the
left- and right-handed spinors. If considered on one of these invariant
subspaces, the curly brackets depend only on the sector indices
$a,b=1,\ldots,8$, whereas the factors to their right involve only Dirac matrices.
This allows us to factor the spectral decomposition of $A_0$ as follows. We
first diagonalize the phase factor, i.e. \label{W_c} \label{nu_nc}
\label{I_nc}
\begin{equation}
W_c \;\equiv\; X_c \;\intc_x^y \intcb_y^x \!X_{\overline{c}} \;=\;
\sum_{n=1}^8 \nu_{nc}\:I_{nc} \label{e:3f}
\end{equation}
with eigenvalues $\nu_{nc}$ (counting multiplicities) and corresponding
spectral projectors $I_{nc}$, where $\overline{c}$ is defined by
$\overline{L}=R$ and $\overline{R}=L$. The matrices $W_L$ and $W_R$ are
obtained from each other by taking their adjoint. Thus we can arrange that the
same holds for their spectral decompositions,
\begin{equation}
\overline{\nu_{nc}} \;=\; \nu_{n \overline{c}}\:,\spc
I_{nc}^* \;=\; I_{n \overline{c}}\:.    \label{e:3ff}
\end{equation}
The spectral representation of the second term in~(\ref{e:3e}) is
computed exactly as described in~{\S}\ref{esec21}. More
precisely, it is obtained from~(\ref{e:2e}, \ref{e:2f}) by
setting $h$ to zero, i.e.\ similar to~(\ref{e:2qa}, \ref{e:2r}),
\begin{equation}
\frac{1}{4}\:(\xi\slsh\:T^{(-1)}_{[0]})(\overline{\xi\slsh\:T^{(-1)}_{[0]}})
\;=\; \sum_{s=\pm} \lambda_s \:F_s    \label{e:3g}
\end{equation}
with
\begin{eqnarray}
\lambda_s &=& \frac{1}{4}\: T^{(-1)}_{[0]} \:\overline{T^{(-1)}_{[0]}}
\:\times\: \left\{ \begin{array}{cl} z & {\mbox{if $s=+$}} \\
\overline{z} & {\mbox{if $s=-$}} \end{array} \right. \label{e:3.4} \\
F_s &=& \frac{1}{z-\overline{z}} \:\times\: \left\{
\begin{array}{cl} \xi\slsh \overline{\xi\slsh} - \overline{z} &
{\mbox{if $s=+$}} \\
-\xi\slsh \overline{\xi\slsh}+z & {\mbox{if $s=-$ .}} \end{array} \right.
\label{e:3.5}
\end{eqnarray}
Combining~(\ref{e:3f}) and~(\ref{e:3g}) gives
\begin{equation}
A_0 \;=\; \sum_{n=1}^8 \:\sum_{c=L,R} \:\sum_{s=\pm}
\lambda_{ncs}\:F_{ncs} \label{e:3spec}
\end{equation}
with
\begin{equation}
\lambda_{ncs} \;=\; \nu_{nc} \:\lambda_s \:,\spc F_{ncs} \;=\; \chi_c\:
I_{nc}\: F_s \:.    \label{e:3k}
\end{equation}

It might be surprising at first sight that, although $A_0$ clearly is a
gauge-invariant expression, the phase shifts described by the ordered
exponentials in~(\ref{e:3d}) do not drop out in~(\ref{e:3e}). Let us explain
in detail how this comes about. We first recall that under gauge
transformations, the truncated fermionic projector transforms like
\begin{equation}
P_0(x,y) \;\longrightarrow\; U(x)\: P_0(x,y)\: U(y)^{-1}\:,    \label{e:3l}
\end{equation}
where $U$ is unitary with respect to the spin scalar product, $U(x)^*=U(x)^{-1}$.
These $U(2N, 2N)$ gauge transformations correspond to a local symmetry of the
system, which is related to the freedom in choosing a local basis for the
spinors (see~{\S}\ref{psec11}). When forming the closed chain, the gauge
transformations at $y$ drop out,
\[ P_0(x,y)\: P_0(y,x) \;\longrightarrow\; U(x)\: P_0(x,y)\:P_0(y,x)\:
U(x)^{-1}\:. \]
In order to see the relation between the phase transformations in~(\ref{e:3d})
and the above gauge transformations, it is useful to consider the situation when
the chiral potentials have the form of pure gauge potentials, i.e.
\begin{equation}
A_c^j \;=\; i V_c\:(\partial^j V_c^{-1})
    \label{e:3j}
\end{equation}
with unitary operators $V_L, V_R \in U(8)$. Then the ordered
exponential~(\ref{e:3h}) reduces to a product of unitary transformations at the
two end points,
\[ \intc_x^y \;=\; V_c(x) \:V_c(y)^{-1}\:. \]
Using that the potentials are causality compatible, (\ref{e:3d}) becomes
\begin{equation}
P_0(x,y) \;=\; \sum_{c=L,R} \chi_c\: V_c(x) \:X_c \left( \frac{i}{2}\:\xi\slsh
\: T^{(-1)}_{[0]}(x,y) \right) V_c(y)^{-1}\:.    \label{e:3i}
\end{equation}
Hence the left- and right-handed components of $P_0$ are transformed
independently by $V_L$ and $V_R$, respectively. In order to write these
transformations in a form similar to~(\ref{e:3l}), we combine $V_L$ and $V_R$
into one operator $V$,
\[ V \;=\; \chi_L\:V_L + \chi_R\:V_R \:. \]
The effect of the chiral potentials in~(\ref{e:3i}) is then described by the
transformation
\[ P_0(x,y) \;\longrightarrow\; V(x) \:P_0(x,y)\: V(y)^* \:, \]
and thus the closed chain transforms according to
\begin{equation}
P_0(x,y)\: P_0(y,x) \;\longrightarrow\; V(x) \:P_0(x,y)\: V(y)^*\: V(y)\:
P_0(y,x)\: V(x)^* \:.    \label{e:3k2}
\end{equation}
The point is that the transformation $V$ is in general {\em{not}}
unitary, because
\[ V^* \;=\; \chi_R\: V_L^{-1} + \chi_L V_R^{-1}\;
\stackrel{\mbox{\tiny{in general}}}{\neq}\; \chi_L\: V_L^{-1} +
\chi_R V_R^{-1} \;=\; V^{-1}\:. \] More precisely, $V$ is unitary
if and only if $V_L=V_R$ at every space-time point. According
to~(\ref{e:3j}), this implies the condition $A_L \equiv A_R$.
From~(\ref{e:3b}) we conclude that $V$ is unitary if and only if
the axial potentials $E$ in~(\ref{e:3a}) are identically equal to
zero. This means that only the subgroup $U(8) \subset U(8)_L
\times U(R)_R$ of the chiral gauge group, which gives rise to the
vector potential $C$ in~(\ref{e:3a}), describes local unitary
transformations of the fermionic projector and thus corresponds to
a local gauge symmetry in the sense of~{\S}\ref{psec11}. We
refer to this subgroup of the chiral gauge group as the {\em{free
gauge group}} \label{3fgg}\index{gauge group!free} ${\mathcal{F}}$; it can be identified
with a subgroup of the gauge group, ${\mathcal{F}} \subset U(2N,
2N)$ (we remark for clarity that the other degrees of freedom of
the gauge group $U(2N, 2N)$ are related to the gravitational
field~{\S}\ref{isec5} and are thus not considered here). The axial
potentials, however, describe local transformations which are not
unitary and thus cannot be identified with gauge transformations
in the sense of~{\S}\ref{psec11}. These non-unitary
transformations do not correspond to an underlying local symmetry
of the system. The interpretation of these results is that the
\label{3sgg} {\em{chiral gauge group is spontaneously
    broken}},\index{spontaneous symmetry breaking}
and only its subgroup ${\mathcal{F}}$ corresponds to an unbroken local
symmetry of the system.

A simple way to understand why the chiral gauge group is spontaneously broken
is that axial potentials describe relative phase shifts between the left- and
right-handed components of the fermionic projector. Such relative phases do not
drop out when we form composite expressions, as one sees in~(\ref{e:3k2}) or,
more generally, in~(\ref{e:3e}). By imposing that the relative phases be zero
in all composite expressions, we can distinguish those systems in which the
axial potentials vanish identically. In this way, one can fix the gauge up to
global chiral gauge transformations (i.e.\ transformations of the
form~(\ref{e:3i}) with constant matrices $V_c$) and up to local free gauge
transformations. Since this gauge fixing argument makes use of the phases which
appear in $P_0(x,y)$, one may regard the chiral gauge symmetry as being
spontaneously broken by the fermionic projector.

The spontaneous breaking of the chiral gauge symmetry by the fermionic
projector has, at least on the qualitative level, some similarity to the Higgs
mechanism in the standard model. We recall that in the Higgs
mechanism\index{Higgs mechanism} one
arranges by a suitable quartic potential in the classical Lagrangian that the
Higgs field $\Phi$ has a non-trivial ground state, i.e. $\Phi \neq 0$ in the
vacuum. The Higgs field is acted upon by a local gauge group. Since $\Phi \neq
0$, one can, by prescribing the phase of $\Phi$, fix the gauge globally. This
shows that the local gauge symmetry is spontaneously broken by the Higgs field,
a fact which can then be used to give the gauge bosons mass.
In our setting, we also have in the vacuum a non-trivial object, namely the
fermionic projector, which is composed of the Dirac seas corresponding to the
leptons and quarks. Thus our situation is indeed quite similar to the Higgs
mechanism, if one only keeps in mind that the role of the Higgs field in our
description is played by the fermionic projector of the vacuum. Clearly, this
analogy does not carry over to the mathematical details. But also in our
description, the spontaneous symmetry breaking makes it possible that
undifferentiated gauge potentials enter the EL-equations, giving the hope that the corresponding gauge bosons might be massive.

Since the chiral gauge symmetry is spontaneously broken, we cannot expect that
the EL equations admit chiral potentials corresponding to the whole group
$U(8)_L \times U(8)_R$. In order to quantify which restrictions for the
chiral potentials we get, we must work in a more general setting and introduce a
suitable mathematical notation. Contributions to the fermionic projector which
involve the phases of the chiral potentials, but not the gauge fields, currents,
or scalar potentials, are called {\em{gauge terms}}\index{gauge term}. Likewise, we refer
to the contributions of the gauge terms to a composite expression in the
fermionic projector as the gauge terms in the respective expression. The
simplest examples for gauge terms are~(\ref{e:3d}) or~(\ref{e:3e}), but we
shall encounter gauge terms to lower degree on the light cone as well.

\begin{Def} \label{def1}
A subgroup ${\mathcal{G}}$ of the chiral gauge group is called a
{\bf{dynamical gauge group}}\index{gauge group!dynamical} if the gauge terms of the potentials
corresponding to ${\mathcal{G}}$ vanish in the EL equations. The
subgroup ${\mathcal{G}} \cap {\mathcal{F}}$ is the {\bf{free
dynamical gauge group}}\index{gauge group!free dynamical}.
\end{Def}
Clearly, this definition does not give a unique dynamical gauge group. In
particular, every subgroup of a dynamical gauge group is again a dynamical gauge
group. Since we want to choose the dynamical gauge group as large as possible,
we will always restrict attention to dynamical gauge groups which are
{\em{maximal}} in the sense that they are not contained in a larger dynamical
gauge group.

We first analyze the gauge
term~(\ref{e:3d}) in the EL equations corresponding
to our variational principle~(\ref{e:2O}). We only consider the highest
degree on the light cone, which is degree~5. This gives the following
result.
\begin{Thm} \label{thm31}
The eigenvalues $\nu_{nc}$ of $W_c$ must satisfy the conditions
\begin{equation}
    \nu_{8c} = 0 \quad {\mbox{and}} \quad |\nu_{nc}| = |\nu_{n'c'}| {\mbox{ for
    $n,n'=1,\ldots,7$ and $c,c'=L,R$}}.
    \label{e:3n}
\end{equation}
The dynamical gauge group ${\mathcal{G}}$ is restricted by
\begin{equation}
    {\mathcal{G}} \subset (U(7) \times U(1))_L \times (U(7) \times U(1))_R \:,
    \label{e:3gg}
\end{equation}
where the $U(7)$ are unitary matrices acting on the seven massive sectors, and
the $U(1)$ act on the neutrino sector.

If conversely the conditions~(\ref{e:3n}) or~(\ref{e:3gg}) are satisfied,
then the EL equations are satisfied to degree $5$ on the light cone.
\end{Thm}
It is easy to see that the conditions in the above theorem are sufficient for
the EL equations to be satisfied, and this consideration also gives an idea of
how these conditions come about.  Namely, suppose that~(\ref{e:3gg}) holds.
Then the dynamical gauge potentials are invariant on the massive sectors as well
as on the neutrino sector.  Using a block matrix notation where the first
component refers to the massive sectors and the second component to the neutrino
sector, we see from~(\ref{e:3f}) that the matrices $W_L$ and $W_R$ have the
form
\begin{equation}
W_L \;=\; \left( \begin{array}{cc} U & 0 \\ 0 & 0 \end{array} \right) ,\spc
W_R \;=\; \left( \begin{array}{cc} U^* & 0 \\ 0 & 0 \end{array} \right)
    \label{e:3J}
\end{equation}
with $U$ a unitary $7 \times 7$ matrix.  Hence their eigenvalues $\nu_{nc}$
satisfy the conditions~(\ref{e:3n}). Using~(\ref{e:39z}), the gradient
of the Lagrangian is computed to be
\begin{equation}
{\mathcal{M}}[A] \;=\; 2 \sum_{k=1}^K \left\{ \left( |\lambda_k|
- \frac{1}{28}\: |A| \right) \frac{\overline{\lambda_k}}{|\lambda_k|} \right\}
F_k\:.
    \label{e:3br}
\end{equation}
We saw in~{\S}\ref{esec23} that the curly brackets vanish in
the vacuum. If~(\ref{e:3n}) is satisfied, the gauge terms change
the eigenvalues $\lambda_{ncs}$ only by a phase~(\ref{e:3k}).
Since these phases drop out when absolute values are taken, the
curly brackets in~(\ref{e:3br}) are zero even with interaction (to
the highest degree on the light cone). According to~(\ref{e:2c}),
this implies  that $Q$ vanishes, and so the EL equations are
satisfied.

It is more difficult to show that the conditions~(\ref{e:3n})
and~(\ref{e:3gg}) are also necessary. We give the proof in detail. \\[.5em]
{\em{Proof of Theorem~\ref{thm31}. }}
Using the argument given after the statement of the theorem, it remains to show
that the conditions~(\ref{e:3n}) and~(\ref{e:3gg}) are necessary.
Substituting the spectral decomposition of~$A_0$ into~(\ref{e:2C}),
we obtain, in analogy to~(\ref{e:2v}), the following representation
for~$Q$,
\[ Q(x,y) \;=\; \frac{1}{2}\: \sum_{n,c,s} \frac{\partial {\mathcal{L}}(\lambda^{xy})}{
\partial \lambda^{xy}_{ncs}}\: F^{xy}_{ncs}\: P_0(x,y) \:+\: (\deg < 5). \]
Computing the Euler-Lagrange equations similar to~(\ref{e:2t2}) and keeping
track of the chiral cancellations, we obtain in analogy
to~(\ref{e:2s3}) the equation
\[ \sum_{n,c,s} \frac{\partial {\mathcal{L}}(\lambda^{xy})}{
\partial \lambda^{xy}_{ncs}}\: F^{xy}_{ncs}\: P_0(x,y) \: \intcb_y^z
X_{\overline{c}} \:+\: (\deg < 5) \;=\; 0\:, \]
and by multiplying from the right by the macroscopic unitary matrix
$\intcb_z^y\!$, we can arrange that $z=y$.
We substitute~(\ref{e:3d}) as well as the right of~(\ref{e:3k}) and
apply~(\ref{e:2s2}) to obtain
\[ \sum_{nc} \frac{\partial {\mathcal{L}}(\lambda)}{\partial \lambda_{nc-}}\: \chi_c\:
I_{nc} \left\{ X_c \:\intc_x^y \intcb_y^x X_{\overline{c}} \right\} \left(
\frac{i}{2}\: \xi\slsh\: T^{(-1)}_{[0]} \right) \:+\: (\deg < 5) \;=\; 0 \:. \]
The curly brackets coincide with the matrix $W_c$, and since
$I_{nc}$ is a spectral projector of this matrix, we simply get a scalar factor
$\nu_{nc}$. Furthermore, we use the particular form of our
Lagrangian~(\ref{e:2O}) as well as the first equation
in~(\ref{e:3k}). This gives
\begin{equation}
i \sum_{n,c} \left( |\nu_{nc}| - \frac{1}{14}\: \sum_{n', c'} |\nu_{n'c'}|
\right) \: \chi_c\:\xi\slsh\: I_{nc}\:
\overline{\lambda_-}\: T^{(-1)}_{[0]} \;=\; 0 \:. \label{e:3q}
\end{equation}
Using~(\ref{e:2qa}), the non-smooth factors are a monomial of degree five,
\begin{equation}
\overline{\lambda_-}\: T^{(-1)}_{[0]} \;=\;
T^{(0)}_{[0]}\:T^{(-1)}_{[0]}\: \overline{T^{(-1)}_{[0]}}\:. \label{e:3mon11}
\end{equation}
We cannot assume that this monomial is equal to zero. Namely, the
fermionic projector differs to highest degree on the light cone
from the fermionic projector of the vacuum only by macroscopic
phase factors (this is guaranteed by the gauge invariance of the
regularized causal perturbation expansion, see Appendix~\ref{pappB}).
Therefore, exactly as explained for the vacuum in~{\S}\ref{esec22}, we can evaluate~(\ref{e:3q}) even strongly.
In particular, we can consider the regularization expansion
of~(\ref{e:3q}) (see~{\S}\ref{psec24}, {\S}\ref{psec25}). This means that, in
order to set the monomial~(\ref{e:3mon11}) in~(\ref{e:3q}) equal
to zero, we would have to impose an infinite number of
regularization conditions. We conclude that in order to
satisfy~(\ref{e:3q}), we must assume that the macroscopic
prefactor vanishes. Using that the spectral projectors $I_{nc}$
are linearly independent, we get the conditions
\[ \left( |\nu_{nc}| - \frac{1}{14}\: \sum_{n', c'} |\nu_{n'c'}|
\right) |\nu_{nc}|^2 \;=\; 0 \spc {\mbox{for all $n, c$}}. \]
This implies that the absolute values of $14$ of the eigenvalues $\nu_{nc}$ must
coincide, and that the remaining two eigenvalues must be zero. The matrix $W_L$
contains a factor $X_R$ and is thus singular of rank one. Choosing the numbering
such that $\nu_{L8}=0$, it follows from~(\ref{e:3ff}) that also $\nu_{R8}=0$.
Hence the two zero eigenvalues are those for $n=8$. This shows
that~(\ref{e:3n}) is a necessary condition.

Next we will show that~(\ref{e:3n}) implies the constraint for the dynamical
gauge group~(\ref{e:3gg}). We introduce for fixed $x$ and $y$ the abbreviations
\begin{equation}
    U \;=\; \intL_x^y \intR_y^x \spc {\mbox{and}} \spc T \;=\; X_R\:.
    \label{e:3s}
\end{equation}
We consider $U=(U^a_b)$ and $T=(T^a_b)$ as matrices on $\C^8$ endowed with the
standard Euclidean scalar product $\langle .,.  \rangle$.  Then $U$ is unitary, and
$T$ is a projector of rank $7$.  According to~(\ref{e:3ff}), the
conditions~(\ref{e:3n}) tell us that the matrix $UT$ must have $7$ eigenvalues
on the unit circle and one zero eigenvalue. For a vector $u$ in the kernel of
$UT$,
\[ 0 \;=\; \langle UT u,\: UT u \rangle \;=\; \langle Tu,\: Tu \rangle \:, \]
where we used in the last step that $U$ is unitary. Thus $u$ is also in the
kernel of $T$. On the other hand, if $u$ is an eigenvector of $UT$
corresponding to an eigenvalue on the unit circle,
\[ |u|^2 \;=\; \langle UTu,\: UTu \rangle \;=\; \langle Tu,\: Tu \rangle \:. \]
Since for a projector, $|Tu| < |u|$ unless $u$ is in the image of $T$, it follows
that $u$ is also an eigenvector of $T$, of eigenvalue one. We conclude that
every eigenvalue of $UT$ is also an eigenvalue of $T$, or equivalently that
\begin{equation}
[UT,\: T] \;=\; 0\:.    \label{e:3r}
\end{equation}

Let us analyze what this commutator condition means for the chiral
potentials. We already know from the causality compatibility condition that
\begin{equation}
[A_R, T] \;=\; 0 \:.    \label{e:3t}
\end{equation}
Hence substituting the definition of $U$, (\ref{e:3s}), into~(\ref{e:3r}) and
using that the resulting condition must hold for all $x$ and $y$, we obtain that
$[A_L T, T] = 0$. Subtracting the adjoint and using that $T$ is idempotent gives
the stronger statement
\begin{equation}
    [A_L, T] \;=\; 0 \:. \label{e:3u}
\end{equation}
From~(\ref{e:3t}) and~(\ref{e:3u}) we conclude that the chiral potentials must
be block diagonal in the sense that $(A_c)^a_b=0$ if $a=8$ and $b \neq 8$ or
vice versa. Such chiral potentials correspond precisely to the gauge group
in~(\ref{e:3gg}).
\QED

\begin{Remark} \label{rem31} \em
We point out that the subgroup $U(1)_L \times U(1)_R$ of the gauge group
in~(\ref{e:3gg}), which acts on the neutrino sector, is not uniquely determined
and could be replaced by any other subgroup which contains $U(1)_L$. This can
immediately be understood from the fact that the neutrino sector contains only
left-handed particles, and thus the form of the potential $A_R$, which acts on
the right-handed component, has no significance. To make this argument
rigorous, we consider the Dirac equation~(\ref{e:3dir}). Since $P=XP$, we may
replace the chiral potentials $A_c$ in~(\ref{e:3dir}) by $A_c X_c$, and this
indeed makes the component of $A_R$ in the neutrino sector equal to zero,
showing that this component is of no relevance. We conclude that we may
arbitrarily change the subgroup $U(1)_R$ in~(\ref{e:3gg}); e.g.\ we could
replace~(\ref{e:3gg}) by
\begin{equation}
(U(7) \times U(1))_L \times U(7)_R \spc {\mbox{or}} \spc
(U(7)_L \times U(7)_R) \times U(1) \:. \label{e:3E}
\end{equation}
We do not write out this obvious arbitrariness in what follows; instead
we will simply give the gauge groups in the most convenient form. \em
\end{Remark}

\section{The Gauge Terms in the Euler-Lagrange Equations}
\setcounter{equation}{0}
We come to the analysis of the EL equations to the next lower degree $4$
on the light cone. According to the formulas of the light-cone
expansion in Appendix~\ref{pappLC}, the structure of the fermionic projector
to the next lower degree is considerably more complicated than~(\ref{e:3d}), because in addition to gauge terms, there are also contributions involving the chiral fields and
currents as well as the scalar potentials. Fortunately, the following
general argument allows us to distinguish these different types of contributions
in the EL equations\label{3gapl}.
In the formulas for $P(x,y)$, the gauge terms always
involve ordered exponentials of the chiral potentials, integrated along the
line segment $\overline{xy}= \{\alpha x + (1-\alpha) y,\: 0 \leq \alpha \leq 1\}$.
We refer to such contributions as {\em{line contributions}}.
\index{line contribution}%
The fields, currents and scalar potentials, however, are in the light-cone
expansion evaluated at individual points, namely either at the end points $x,
y$ or at an intermediate point $z \in \overline{xy}$; we call the corresponding
contributions to the light-cone expansion {\em{point
    contributions}}\index{point contribution}. In the
case of an evaluation at an intermediate point $z$, the point contribution
clearly involves an integral over $z$ along $\overline{xy}$. But in contrast to
the line contribution, where the chiral potentials at different points enter
the ordered exponential in a nonlinear way, the line integrals in a point
contribution simply takes averages of the potentials, fields, or currents along
the line segment. For example by expanding the ordered exponential in a Dyson
series (see Def.~\ref{2.5.4}) and considering the higher
order terms, one sees immediately that the line and point contributions are
independent in the EL equations in the sense that the EL equations must be
satisfied separately by the line and point contributions.
Moreover, we can distinguish point contributions in the EL equations, provided
that their configuration of the tensor indices is different. Therefore, the point
contributions involving the scalar potentials, the chiral fields and
the currents are independent in the EL equations as well.

Using the above arguments, we can study the gauge terms and the contributions
involving the scalar potentials, the gauge fields and the currents
separately. In the remainder of this section, we consider
only the gauge terms. Thus we restrict attention to chiral potentials, i.e.
instead of~(\ref{e:3dir}) we consider the Dirac equation
\[ (i \Pdd - mY + \chi_L\: \Aslsh_R + \chi_R\: \Aslsh_L)\: P(x,y) \;=\; 0 \]
with $Y$ a fixed matrix. We will return to the general Dirac
equation~(\ref{e:3dir}) in Chapter~\ref{esec4}.

\begin{Def} \label{def32}
We introduce for $p \in \{4,\ldots,7\}$ the groups $B_p, F_p \subset U(8)$ by
\begin{eqnarray}
B_p &=& \{ \:\underbrace{g \oplus \cdots \oplus g}_{p
{\mbox{\scriptsize{ summands}}}}
\oplus \underbrace{g^{-1} \oplus \cdots \oplus g^{-1}}_{7-p
{\mbox{\scriptsize{ summands}}}} {\mbox{ with $g \in U(1)$}} \} \\
F_p &=& U(p) \times U(7-p) \times U(1) \label{e:3f1}
\end{eqnarray}
and define corresponding subgroups ${\mathcal{B}}_p$ \label{mathcalB_p} and
${\mathcal{F}}_p$ \label{mathcalF_p} of the dynamical gauge group by
\begin{equation}
{\mathcal{B}}_p \;=\; B_p \times \1 \;\subset\; U(8)_L \times
U(8)_R \:,\spc {\mathcal{F}}_p \;=\; \{ {\mbox{$(g,g)$ with $g \in
F_p$}} \} \;\subset\; {\mathcal{F}} \:. \label{e:3f2}
\end{equation}
Their product
\begin{equation}
{\mathcal{G}}_p \;=\; {\mathcal{B}}_p \cdot {\mathcal{F}}_p
\;\equiv\; \{ {\mbox{$bf$ with $b \in {\mathcal{B}}_p$, $f \in
{\mathcal{F}}_p$}} \} \;\subset\; U(8)_L \times U(8)_R
    \label{e:3A}
\end{equation}
is called the {\bf{${\mbox{\em{p}}}^{\mbox{\scriptsize{th}}}$
dynamical gauge group}}\index{gauge group!dynamical}.
\end{Def}
In block matrix notation, the elements of $B_p$ and $F_p$ can be written as
\begin{equation}
\left( \begin{array}{ccc} z\: \1_p & 0 & 0 \\ 0 & z^{-1}\: \1_q & 0 \\
0 & 0 & 0 \end{array} \right) \spc {\mbox{and}} \spc
\left( \begin{array}{ccc} g & 0 & 0 \\ 0 & h & 0 \\
0 & 0 & l \end{array} \right) \:,
    \label{e:3N}
\end{equation}
respectively, where the first component refers to the first $p$
sectors, the second component to the next $q \equiv 7-p$ sectors
and the last component to the neutrino sector. Here $z, l \in
U(1)$, $g \in U(p)$ and $h \in U(q)$. Clearly, $B_p$ and
${\mathcal{B}}_p$ are group isomorphic to $U(1)$. Notice that
${\mathcal{B}}_p$ acts only the left-handed component. The group
${\mathcal{F}}_p$ transforms the left- and right-handed components
in the same way, and so its corresponding gauge potentials are
vector potentials. The groups ${\mathcal{B}}_p$ and
${\mathcal{F}}_p$ commute, and this ensures that their
product~(\ref{e:3A}) is again a group. It is easy to verify that
${\mathcal{F}}_p$ is indeed the largest subgroup of
${\mathcal{F}}$ which commutes with ${\mathcal{B}}_p$,
\[ {\mathcal{F}}_p \;=\; \left\{ f \in {\mathcal{F}} \;|\; bfb^{-1} = f {\mbox{ for all }} b \in
{\mathcal{B}}_p \right\} \:. \]

We introduce an abbreviation for the linear combination of monomials,
\begin{equation}
M \;\equiv\; T^{(0)}_{[1]}\: T^{(0)}_{[1]}\: \overline{T^{(-1)}_{[0]}\:
T^{(0)}_{[0]}} \:-\: T^{(-1)}_{[0]}\: T^{(1)}_{[2]}\:
\overline{T^{(-1)}_{[0]}\: T^{(0)}_{[0]}} \:.    \label{e:3B}
\end{equation}
\begin{Thm} \label{thm32}
There are precisely the following possibilities for the choice of the
dynamical gauge group.
\begin{itemize}
\item[(1)] Without assuming any relations between the basic fractions, the
dynamical gauge group must be contained in the free gauge group,
\begin{equation}
{\mathcal{G}} \;\subset\; {\mathcal{F}}_0 \;=\; U(7) \times U(1)
\:.
    \label{e:3F}
\end{equation}
\item[(2)] If we allow for one relation between the basic fractions, the
dynamical gauge group is (possibly after a global gauge transformation)
restricted by
\begin{equation}
{\mathcal{G}} \;\subset\; {\mathcal{G}}_p \spc {\mbox{for some $p
\in \{4,\ldots,7\}$}}.    \label{e:3I}
\end{equation}
In this case, the relation between the basic fractions is\footnote{{\textsf{Online
version}:} As shown in Lemma~I in the preface to the second online edition,
there is no regularization which realizes this relation.}
\begin{equation}
(M - \overline{M})\: \overline{T^{(0)}_{[0]}}^{-1}
\;=\; 0    \label{e:3C}
\end{equation}
with $M$ according to~(\ref{e:3B}).
\item[(3)] If we allow for two relations between the basic fractions, we get no
constraints for the dynamical gauge group besides those of Theorem~\ref{thm31}.
The two relations between the basic fractions are~(\ref{e:3C}) and
\begin{equation}
M \: \overline{T^{(0)}_{[0]}}^{-1}
\;=\; 0 \:.   \label{e:3M}
\end{equation}
\end{itemize}
In each of these cases, the gauge terms vanish in the EL equations to degree
$4$ on the light cone.
\end{Thm}
{\Proof}
We first bring the EL equations to degree~$4$ into a more explicit form.
Theorem~\ref{thm31} implies that the variation of our Lagrangian vanishes to
highest degree on the light cone,
\begin{equation}
\frac{\partial {\mathcal{L}}(\lambda)}{\partial \lambda_{ncs}}
\:+\: (\deg < 3) \;=\; 0 \:. \label{e:3van}
\end{equation}
According to~(\ref{e:39z}, \ref{e:2c}), all
contributions to the EL equations vanish for which the variation of the
Lagrangian is considered to highest degree (even if the spectral projectors
or the factors $P(x,y)$ are expanded to lower degree). This means that we only
need to compute the Lagrangian to the next lower degree, whereas it
suffices to take into account both the spectral projectors and the factors
$P(x,y)$ to highest degree.

Since the Lagrangian is a function of the eigenvalues only, our task is to calculate
the contribution to the eigenvalues to the next lower degree two, denote by
$\Delta \lambda_{ncs}$. This calculation is carried out in a more general
context in Appendix~\ref{appC} (see Theorems~\ref{thmB1} and~\ref{thmB2}).
Specializing the obtained results gives
\[ \Delta \lambda_{8cs} \;=\; 0 \:, \]
whereas for $n=1,\ldots,7$,
\begin{eqnarray}
\Delta \lambda^{xy}_{nc-} & = & T^{(0)}_{[1]}\: \overline{T^{(0)}_{[1]}} -
T^{(-1)}_{[0]}\: \overline{T^{(1)}_{[2]}} \:+\:
\nu_{nc} \left( T^{(0)}_{[2]}\: \overline{T^{(0)}_{[0]}} + T^{(-1)}_{[0]}\:
\overline{T^{(1)}_{[2]}} \right)\spc \label{e:3v1} \\
&&+\frac{T^{(0)}_{[1]}\: \overline{T^{(-1)}_{[0]}} - T^{(-1)}_{[0]}\:
\overline{T^{(0)}_{[1]}}}{\overline{\lambda^{xy}_{nc-}} - \lambda^{xy}_{nc-}}
\left( \nu_{nc}\; T^{(0)}_{[1]}\: \overline{T^{(0)}_{[0]}}\:-\:
\overline{\nu_{nc}}\; T^{(0)}_{[0]}\: \overline{T^{(0)}_{[1]}} \right)\spc
    \label{e:3v2} \\
\Delta \lambda^{yx}_{nc+} &=& \overline{\Delta \lambda^{xy}_{n \overline{c}-}}
\end{eqnarray}
(here $\lambda^{xy}_{nc-}$ denotes the eigenvalues of $A_0$, (\ref{e:3k})).
The EL equations take again the form~(\ref{e:2s3}).
Substituting the asymptotic formula to highest
degree~(\ref{e:2s2}) and expanding our Lagrangian~(\ref{e:2O}) shows that the
EL equations become to degree $4$,
\[ \sum_{nc} \Delta \left( |\lambda_{nc-}| - \frac{1}{28}\: \sum_{n', c',
s'} |\lambda_{n' c' s'}| \right) \frac{\overline{\lambda_-}}{|\lambda_-|}\:
|\nu_{nc}|^2 \chi_c\: I_{nc}\: (i \xi\slsh\: T^{(-1)}_{[0]}) \;=\; 0\:. \]
Since the eigenvalues appear in complex conjugate pairs, we may replace the
sum over $s'$ by a factor two and set $s'=-$. Also, the non-vanishing
macroscopic factor $\xi\slsh$ can be omitted. Furthermore, we use that the
spectral projectors $I_{nc}$ are macroscopic and linearly independent, and
that $|\nu_{nc}|^2$ vanishes for $n=8$ and is equal to one otherwise, (\ref{e:3n}).
We thus obtain that the EL equations to degree $4$ are equivalent to the
conditions that for all $n=1,\ldots,7$,
\[ \left[ {\mbox{Re}}\left( \overline{\lambda_{nc-}}\: \Delta \lambda_{nc-}\right)
- \frac{1}{14} \sum_{n', c'} {\mbox{Re}} \left( \overline{\lambda_{n'c' -}}\:
\Delta \lambda_{n'c' -} \right) \right] \frac{1}{\overline{T^{(0)}_{[0]}}}
\;=\; 0 \:. \]
It is more convenient to write this condition in the form that the expression
\[ H_{nc} \;\equiv\; 2\: {\mbox{Re}} \left( \overline{\lambda_{nc-}}\:
\Delta \lambda_{nc-}\right) \overline{T^{(0)}_{[0]}}^{-1} \]
should be independent of $n$ and $c$,
\begin{equation}
H_{nc} \;=\; H_{n'c'} \spc {\mbox{for all $n,n'=1,\ldots,7$ and $c,c'=L,R$}}.
    \label{e:3z}
\end{equation}
Next we compute $H_{nc}$ by substituting the formulas for $\lambda_{nc-}$
and $\Delta \lambda_{nc-}$, (\ref{e:2qa}) and (\ref{e:3v1}, \ref{e:3v2}).
Since the last summand of $\Delta \lambda_{nc-}$, (\ref{e:3v2}) is imaginary,
we can use that for $\alpha \in i\R$,
\[ 2\: {\mbox{Re}} \left( \overline{\lambda_{nc-}}\:\alpha \right) \;=\;
\alpha\: (\overline{\lambda_{nc-}} - \lambda_{nc-}) \:, \]
and the denominator in~(\ref{e:3v2}) drops out. We thus obtain
\begin{equation}
H_{nc} \;=\; \left( \nu_{nc}\:M + \overline{\nu_{nc}}\: \overline{M} + L +
\overline{L} \right) \: \overline{T^{(0)}_{[0]}}^{-1}\:,    \label{e:3w}
\end{equation}
where $M$ is the linear combination of monomials~(\ref{e:3B}), and $L$
is given by
\begin{equation}
L \;\equiv\; T^{(0)}_{[0]}\: \overline{T^{(-1)}_{[0]}} \left( T^{(0)}_{[2]}\:
\overline{T^{(0)}_{[0]}} + T^{(-1)}_{[0]}\: \overline{T^{(1)}_{[2]}} \right)\:.
    \label{e:3y}
\end{equation}
We conclude that the EL equations to degree $4$ are equivalent to the
conditions (\ref{e:3z}) with $H_{nc}$ given by~(\ref{e:3w}, \ref{e:3B},
\ref{e:3y}).

Let us analyze what the conditions~(\ref{e:3z}) mean. First of all, the
contributions to~(\ref{e:3w}) which involve $L$ or $\overline{L}$ are clearly
independent of $n, c$ and thus drop out in~(\ref{e:3z}). In the case $n'=n$
and $c'=\overline{c}$, we can in~(\ref{e:3z}) apply the first part
of~(\ref{e:3ff}) to obtain the necessary conditions
\begin{equation}
(\nu_{nc} - \overline{\nu_{nc}})\:(M-\overline{M})\:
\overline{T^{(0)}_{[0]}}^{-1} \;=\; 0\:.    \label{e:3G}
\end{equation}
If we assume no relations between the basic fractions, this implies that
$\nu_{nc}=\overline{\nu_{nc}}$, and thus
\begin{equation}
    \nu_{nc} \;=\; \pm 1 \spc {\mbox{for all $n=1,\ldots,7$ and $c=L,R$}}.
    \label{e:3D}
\end{equation}
For $x=y$, the matrix $W_c$ becomes $W_c=X_c X_{\overline{c}}$, and thus the
eigenvalues in~(\ref{e:3D}) are equal to one. Since these eigenvalues depend
smoothly on $x$ and $y$, we conclude that $\nu_{nc}=1$ for all $x$ and $y$.
This means in the block matrix representation for $W_c$, (\ref{e:3J}), that the
unitary matrix $U$ is equal to the identity. Thus, according to~(\ref{e:3f}),
\begin{equation}
X_c \:\intc_x^y \intcb_y^x X_{\overline{c}} \;=\; W_c \;=\;
X_c X_{\overline{c}} \:.    \label{e:3K}
\end{equation}
Differentiating with respect to $y$ and setting $y=x$ gives
\[ X_c\; (A_c - A_{\overline{c}}) \: X_{\overline{c}} \;=\; 0\:. \]
Hence the left- and right-handed potentials must coincide on the massive sectors.
Using the argument in Remark~\ref{rem31}, we can arrange the same in
the neutrino sector. This gives the dynamical gauge group in~(\ref{e:3F}).
Conversely, if~(\ref{e:3F}) is satisfied, then the matrices $W_c$ are of the
form~(\ref{e:3K}). It follows that $\nu_{nc}=1$ for all $n=1,\ldots,7$ and
$c=L,R$, and thus~(\ref{e:3z}) holds.

We next consider the case when we allow for relations between the basic fractions.
The only way to avoid the conditions~(\ref{e:3D}) (which lead to the dynamical
gauge group~(\ref{e:3F})) is to assume that the factor $M-\overline{M}$
in~(\ref{e:3G}) vanishes. This gives precisely the relation~(\ref{e:3C}).
If~(\ref{e:3C}) holds, $H_{nc}$ simplifies to
\begin{equation}
H_{nc} \;=\; 2\:{\mbox{Re}} \left(\nu_{nc} \right)\: M\:+\: L + \overline{L} \:.
    \label{e:3H}
\end{equation}
If we assume no further relations between the basic fractions, the
conditions~(\ref{e:3z}) are equivalent to
\begin{equation}
{\mbox{Re}} \left( \nu_{nc} \right) \;=\; {\mbox{Re}} \left( \nu_{n'c'} \right)
\spc {\mbox{for all $n=1,\ldots,7$ and $c=L,R$}}.
    \label{e:3L}
\end{equation}
The only way to avoid these conditions is to impose in addition
that~(\ref{e:3M}) holds. If this is done, all terms involving $\nu_{nc}$ or
$\overline{\nu_{nc}}$ drop out in~(\ref{e:3w}), and~(\ref{e:3z}) is satisfied.

It remains to show that the conditions~(\ref{e:3L}) are equivalent
to~(\ref{e:3I}). Again using the argument in Remark~\ref{rem31},
it is obvious that the dynamical gauge group has the required form
on the neutrino sector, and thus we can in what follows restrict
attention to the seven massive sectors. Then ${\mathcal{G}}$ is a
subgroup of $U(7)_L \times U(7)_R$, and the matrices $W_c$ are
unitary and have the spectral representation
\begin{equation}
    W_c(x,y) \;=\; \intc_x^y \intcb_y^x \;=\; \sum_{n=1}^7 \nu_{nc}\: I_{nc}\:.
    \label{e:3O}
\end{equation}
Suppose that~(\ref{e:3I}) is satisfied. Then
\[ {\mathcal{G}} \;\ni\; \left( \:\intL_x^y,\: \intR_y^x \right) \;=\; (bf, f) \]
with $b \in {\mathcal{B}}_p$ and $f \in {\mathcal{F}}_p$, and thus
\[ W_c \;=\; bf \:f^{-1} \;=\; b \:. \]
As one sees immediately from~(\ref{e:3N}), the eigenvalues $\nu_{nc}$ of $b$
are equal to $z$ and $\overline{z}$ with $z \in U(1) \subset \C$. Thus the
conditions~(\ref{e:3L}) are satisfied.

Suppose conversely that the conditions~(\ref{e:3L}) hold. We denote the Lie
algebra of the dynamical gauge group by $\g$; it is a subalgebra
of $su(7) \oplus su(7)$. Let $\pi$ be the projection onto the axial part,
\begin{equation}
    \pi \;:\; g \rightarrow su(7) \;:\; (A_L, A_R) \mapsto A_L - A_R \:.
    \label{e:3P2}
\end{equation}
Its image $\pi(\g)$ is a subspace of $su(7)$ (but it is in general no subalgebra).
For the first part of our argument, we consider the situation ``locally'' for
$x$ near $y$. Expanding the ordered exponentials in~(\ref{e:3O}) in a power
series around $x$ yields according to~(\ref{e:3h}) that
\begin{equation}
W_{L\!/\!R}(x+\varepsilon u, x) \;=\; \1 \:\pm\: i \varepsilon
(A^L_j(x) - A^R_j(x))\: u^j \:+\: O(\varepsilon^2)\:.
    \label{e:3P}
\end{equation}
Since the gauge potentials at $x$ can be chosen freely with values in the
dynamical gauge algebra, the term $A \equiv (A^L_j(x) - A^R_j(x)) \:u^j$ can take
any value in $\pi(\g)$. The eigenvalues of~(\ref{e:3f}) have the expansion
$\nu_{nc} = 1 \pm i \varepsilon \lambda_n + o(\varepsilon)$, where $\lambda_n$
are the eigenvalues of $A$. We conclude from~(\ref{e:3L}) that
\begin{equation}
\sigma(A) \;=\; \{ \pm \lambda {\mbox{ with }} \lambda=\lambda(A) \in \R \}
\spc {\mbox{for all $A \in \pi(\g)$}}.
    \label{e:3R}
\end{equation}
We can assume in what follows that $\pi(\g)$ is non-trivial, $\pi(\g) \neq 0$,
because otherwise according to~(\ref{e:3P2}) the dynamical gauge potentials are
pure vector potentials, giving rise to~(\ref{e:3F}).

Next we consider the eigenvalues of $W^{xy}_c$ ``globally'' for $y$ far from
$x$. Expanding the ordered exponentials in~(\ref{e:3O}) along the line
$x+\varepsilon \xi$, $\xi \equiv y-x$ gives
\begin{eqnarray}
\lefteqn{ W_L(x+\varepsilon \xi, y) \;=\; W_L(x,y) } \nonumber \\
&&\:+\: i \varepsilon \left( A^L_j(x)\:
W_L(x,y) - W_L(x,y)\: A^R_j(y) \right) \xi^j \:+\: O(\varepsilon^2)\:.
    \label{e:3U}
\end{eqnarray}
It would be nicer to have the potentials $A^L$ and $A^R$ on the same side of the
factor $W_L$. Therefore, we perform a unitary transformation with
$U_\varepsilon = \1 - i \varepsilon A^L_j + O(\varepsilon^2)$ to obtain
\begin{equation}
U_\varepsilon\: W_L(x+\varepsilon \xi, y)\: U_\varepsilon^{-1} \;=\;
W_L(x,y) \:+\: i \varepsilon \:W_L(x,y) \: A \:+\: O(\varepsilon^2)\:,
    \label{e:3Q}
\end{equation}
where we set $A \equiv (A^L_j - A^R_j) \:\xi^j$. Let us analyze
what~(\ref{e:3L}) and our information on $A$, (\ref{e:3R}), tell us about
the form of $W_L$; for simplicity, we work rather elementary with matrices. As
explained before~(\ref{e:3R}), we are free to choose $A \in \pi(\g)$; we fix
any $A \neq 0$. We diagonalize the matrix $W_L$ for given $x$ and $y$. This
gives according to~(\ref{e:3L}),
\begin{equation}
W_L(x,y) \;=\; \left( \begin{array}{cc} z\: \1_p & 0 \\ 0 & \overline{z}\:
\1_q \end{array} \right)    \label{e:3T}
\end{equation}
with $z \in U(1)$, where we used a block matrix notation similar to that
in~(\ref{e:3N}) and again set $q=7-p$. We can without loss of generality assume
that $p \in \{4,5,6\}$.
We fist consider the case $z \neq \overline{z}$. Computing the eigenvalues
of~(\ref{e:3Q}) in first order perturbation theory, the conditions~(\ref{e:3L})
yield that $A$ must be of the form
\[ A \;=\; \left( \begin{array}{cc} \lambda\: \1_p & C^* \\ C & -\lambda\: \1_q
\end{array} \right) \]
with a $q \times p$ matrix $C$ and $\lambda \in \R$. By changing the basis on
the eigenspaces of $W_L(x,y)$, we can even arrange that
\begin{equation}
A \;=\; \left( \begin{array}{ccc} \lambda\: \1_p & C^* & 0 \\
C & -\lambda\: \1_p & 0 \\ 0 & 0 & - \lambda\: \1_{7-2p}
    \end{array} \right)    \label{e:3S}
\end{equation}
with a $p \times p$ matrix $C$. Thus $-\lambda$ is an eigenvalue of $A$.
According to~(\ref{e:3R}), the eigenvalues of $A$ are precisely $\pm \lambda$.
Since $A \neq 0$, we know too that $\lambda \neq 0$. It is a general
result on self-adjoint matrices that if the expectation value of a unit vector
coincides with the largest eigenvalue of the matrix, then this vector must be an
eigenvector. Applied to~(\ref{e:3S}), this result shows that the submatrix $C$
is zero. Thus
\[ A \;=\; \left( \begin{array}{cc} \lambda\: \1_p & 0 \\ 0 & -\lambda\: \1_q
\end{array} \right) \spc {\mbox{with $p \in \{4,5,6\}$, $\lambda \neq 0$}}. \]
This means that $W_L$ and $A$ have the same eigenspaces. Repeating the above
construction for general $x$ and $y$ while keeping $A$ fixed, one sees that the
matrices $W_L(x,y)$ all have the same eigenspaces as $A$ (and this is trivially
true even when $W_L$ degenerates to a multiple of the identity matrix). This
shows that in our basis, (\ref{e:3T}) holds even for all $x$ and $y$.
In the case $z=\overline{z}$ for our original matrix $W_L$ (chosen
before~(\ref{e:3T})), $W_L$ is a multiple of the identity matrix. If this is
true for all $x$ and $y$, then~(\ref{e:3T}) holds for $p=0$. Otherwise, we
choose $x$ and $y$ such that $W_L(x,y)$ is not a multiple of the identity
matrix and repeat the above argument. We conclude that for some $p \in
\{4,\ldots,7\}$ and possibly after a global gauge transformation, the matrix
$W_L$ has the form~(\ref{e:3T}) for all $x$ and $y$.

Let us show that the representation~(\ref{e:3T}) is surjective in the sense
that for every $z \in U(1)$ we can choose the dynamical gauge potentials on the
line segment $\overline{xy}$ such that $W_L$ is of the form~(\ref{e:3T}) for
this given $z$. To this end, we take the determinant of~(\ref{e:3T}, \ref{e:3O}),
\[ \overline{z}^{7-2p} \;=\; \det W_L \;=\; \det \left( \:\intL_x^y \intR_y^x
\right) \:. \]
Using that the determinant is multiplicative, we obtain from~(\ref{e:3h}) that
\[ \overline{z}^{7-p} \;=\; \exp \left( -i \int_0^1 \Tr \left( A(\tau y +
(1-\tau) x)\: d\tau \right) \right) \:, \]
where we again set $A = (A^L_j - A^R_j)\: \xi^j$. This shows that the phase of
$z$ is simply additive along the line segment $\overline{xy}$. It follows
immediately that this phase can take arbitrary values, provided that there is
an $A \in \pi(\g)$ with non-zero trace. Indeed, it follows from~(\ref{e:3R}) and
the fact that $A$ is an odd-dimensional matrix that $\Tr(A) \neq 0$ for all $A
\neq 0$.

We finally return to the expansion~(\ref{e:3U}). Writing the chiral potentials
as block matrices,
\[ A^c_j \:\xi^j \;=\; \left( \begin{array}{cc} a^c_{11} & a^c_{12} \\
a^c_{21} & a^c_{22} , \end{array} \right), \]
and using that both $W_L(x+\varepsilon \xi, y)$ and $W_L(x,y)$ are of the
form~(\ref{e:3T}) with phases denoted by $z=z_\varepsilon$ and $z=z_0$,
respectively, we obtain
\begin{eqnarray*}
\left( \begin{array}{cc} z_\varepsilon \: \1_p & 0 \\ 0 &
\overline{z_\varepsilon}\: \1_q \end{array} \right) &=&
\left( \begin{array}{cc} z_0 \: \1_p & 0 \\ 0 &
\overline{z_0}\: \1_q \end{array} \right) \\
&&+\: i \varepsilon
\left( \begin{array}{cc} z_0 \: (a^L_{11} - a^R_{11}) & \overline{z_0}\: a^L_{12}
- z_0 \: a^R_{12} \\ z_0\: a^L_{21} - \overline{z_0}\: a^R_{21} &
\overline{z_0}\: (a^L_{22} - a^R_{22}) \end{array} \right)
\:+\: O(\varepsilon^2) \:.
\end{eqnarray*}
Since $z_0 \in U(1)$ can take arbitrary values, it follows that
\[ a^L_{11} - a^R_{11} \;=\; \lambda\: \1_p \:,\;\;\;\;\;
a^L_{22} - a^R_{22} \;=\; -\lambda\: \1_q \:,\;\;\;\;\; a^c_{12}
\;=\; 0 \;=\; a^c_{21}\:. \]
Chiral potentials of this form correspond precisely to the dynamical
gauge group ${\mathcal{G}}_p$ in~(\ref{e:3i}). \QED

We finally make three remarks which give a better justification of the ansatz
for the vector/axial potentials in~(\ref{e:3a}), of the formalism used and
of the variational principles to which this formalism applies.

\begin{Remark} \label{rem32} (The chiral potentials on the generations) \em
Compared to the most general ansatz for the vector and axial potentials,
\begin{equation}
C \;=\; C^{(a \alpha)}_{(b \beta)} \:,\spc
E \;=\; E^{(a \alpha)}_{(b \beta)} \:,   \label{e:3r1}
\end{equation}
the potentials in~(\ref{e:3c}) are restricted in that they must be the same
for the three generations. We shall now justify the ansatz in~(\ref{e:3c})
from the form of the gauge terms.

Recall that in~{\S}\ref{esec21} we combined the
regularization functions of the three generations to new
``effective'' regularization functions in each
sector~(\ref{e:26z}). Here we write this procedure symbolically as
\[ T^{(n)}_\circ \;=\; \sum_{\alpha=1}^3 T^{(n)}_{\alpha\: \circ} \:, \]
where $T^{(n)}_{\alpha\: \circ}$ involves the regularization functions for a
single Dirac sea in generation $\alpha$. Let us consider how the analog of
the gauge term~(\ref{e:3d}) looks like. In the case when the potentials are
diagonal on the generations, i.e.
\begin{equation}
C \;=\; (C^\alpha)^a_b\: \delta^\alpha_\beta \:,\spc
E \;=\; (E^\alpha)^a_b\: \delta^\alpha_\beta \:,    \label{e:3rA}
\end{equation}
the generalization of~(\ref{e:3d}) is straightforward, namely
\begin{equation}
P_0(x,y) \;=\; \sum_{\alpha=1}^3
\left( \chi_L \:X_L\: \intL_x^y + \chi_R \:X_R \intR_x^y \right)_\alpha\:
\frac{i}{2}\:\xi\slsh\: T^{(-1)}_{\alpha\: [0]}(x,y)\:,    \label{e:3rB}
\end{equation}
where the index $\alpha$ of the brackets means that we take the ordered
exponentials of the chiral potentials in the corresponding generation.
This gauge term involves relative phase shifts of the individual
Dirac seas. If we substitute it into the EL equations, we get many
contributions involving these relative phases, and if we want these
contributions to drop out, we must introduce additional regularization
conditions for certain polynomials in $T^{(n)}_{\alpha\: \circ}$,
$\alpha=1,2,3$. Thus unless we impose very strong additional conditions on the
regularization, the only way to fulfill the EL equations is to set all the
relative phases to zero. This gives precisely our ansatz~(\ref{e:3c}).

If the potentials $C$ and $E$ are not diagonal on the generations, the form
of the gauge terms is not obvious because there is no longer a canonical
way to put in the
factors $T^{(-1)}_{\alpha\: \circ}$. This point could be clarified by
generalizing the regularized causal perturbation expansion
of Appendix~\ref{pappB} to the case of systems of Dirac seas involving
different regularizations, but we do not want to get into these technical details
here. Qualitatively speaking, it is clear that if already the
potentials~(\ref{e:3rA}) lead to strong additional conditions in the
EL equations, this will be even more the case for the general
ansatz~(\ref{e:3r1}). \em
\end{Remark}

\begin{Remark} (The vector component is null on the light
  cone)\index{vector component!is null on the light cone} \em
\label{vcnlc}
In~{\S}\ref{psec25} we introduced the regularization condition that the
vector component should be null on the light cone. We remarked that this
condition need not be imposed ad hoc, but that it actually follows from the
equations of discrete space-time. We are now in a position to justify this
regularization condition from the EL equations.

In our formula for the perturbation of the
eigenvalues~(\ref{e:3v1}, \ref{e:3v2}) we omitted all contributions
involving factors $T^{(n)}_{\{.\}}$, assuming that they are of lower degree on
the light cone. This is the only point where we used that the vector component is
null on the light cone. Without imposing these regularization conditions, we get
for $\Delta \lambda^{xy}_{nc-}$ the additional contributions
\begin{eqnarray*}
&& -\nu_{nc} \left( T^{(0)}_{[2]}\: \overline{T^{(1)}_{\{0\}}} +
T^{(-1)}_{[0]}\: \overline{T^{(2)}_{\{2\}}} \right) \\
&&+ \frac{T^{(0)}_{[1]}\: \overline{T^{(-1)}_{[0]}} - T^{(-1)}_{[0]}\:
\overline{T^{(0)}_{[1]}}}{\overline{\lambda^{xy}_{nc-}} - \lambda^{xy}_{nc-}}
\left( \nu_{nc}\; T^{(0)}_{[1]}\: \overline{T^{(1)}_{\{0\}}}\:-\:
\overline{\nu_{nc}}\; T^{(1)}_{\{0\}}\: \overline{T^{(0)}_{[1]}} \right) .
\end{eqnarray*}
This leads to an additional contribution to $H_{nc}$, (\ref{e:3w}), of the form
\[ \left( \nu_{nc}\:K + \overline{\nu_{nc}}\: \overline{K} + L +
\overline{L} \right) \: \overline{T^{(0)}_{[0]}}^{-1} \]
with polynomials $K$ and $L$, where $K$ is given explicitly by
\[ K \;=\; T^{(0)}_{[1]}\: T^{(0)}_{[1]}\: \overline{T^{(-1)}_{[0]}\:
T^{(1)}_{\{0\}}} - T^{(-1)}_{[0]}\: T^{(1)}_{\{0\}}\:
\overline{T^{(0)}_{[0]}\: T^{(0)}_{[1]}}\:. \]
The monomials appearing here have a different homogeneity in the ``large''
light-cone coordinate $l$ than those in~(\ref{e:3B}) (more precisely, they
involve an additional factor of $l$; note that the scaling in $l$ is given by
the upper index of $T^{(n)}_\circ$, see~(\ref{p:D1}--\ref{p:D3})). Using
the different scaling behavior in $l$, we can distinguish between the contributions
involving $K$ and $L$ in the EL equations in the sense that both of these
contributions must vanish separately. But this means that we can just as well
omit $K$ in the EL equations, exactly as it was done
in~(\ref{e:3v1}, \ref{e:3v2}) under the assumption that the vector component
is null on the light cone. This argument applies similarly to other contributions
to the EL equations, to every degree on the light cone. \em
\end{Remark}

\begin{Remark} \label{remark31} (n-point actions)\index{action!n-point} \em
We now discuss some difficulties which arise in the study of actions other
than two-point actions. These difficulties are the reason why we do not
consider such actions here. Let $S$ be a general $n$-point action
\[ S \;=\; \sum_{x_1,\ldots,x_n \in M} {\mathcal{L}}[P(x_1,x_2) \cdots
P(x_{n-1}, x_n)\: P(x_n, x_1)] ] \:,\spc n \geq 1. \]
If $n=1$, the corresponding EL equations are of the form~(\ref{e:2d}) with
\begin{equation}
Q(x,y) \;=\; \delta_{xy}\: f[P(x,x)] \:,    \label{e:30}
\end{equation}
where $f$ is a functional depending only on $P(x,x)$. Expressions
like~(\ref{e:30}) do not have a well-defined continuum limit
because the methods of Chapter~\ref{psec2} apply to composite
expressions only away from the origin (i.e.\ for $x \neq y$). Even
if one succeeded in giving~(\ref{e:30}) a mathematical meaning,
this expression is local and does not involve any ordered
exponentials of the chiral potentials. As a consequence, we would
have no gauge terms, and the only constraint for the chiral
potentials would be the causality compatibility condition. The
resulting dynamical gauge group ${\mathcal{G}}=U(8)_L \times
U(7)_R$ would be too large for physical applications. For these
reasons, one-point actions do not seem worth considering.

If on the other hand $n>2$, the operator $Q$ in the EL equations
takes the form
\begin{eqnarray*}
Q(x_1, x_2) &=& \sum_{x_2,\ldots,x_{n-1} \in M} f[P(x_1,x_2) \cdots
P(x_{n-1}, x_n)\: P(x_n, x_1)] \\[-.8em]
&& \hspace*{5cm} \times\; P(x_2, x_2) \cdots P(x_{n-1}, x_n)\:.
\end{eqnarray*}
where $f$ is a functional on the closed chain. Again, it is not clear how to
make mathematical sense of this expression in the continuum limit, but in
contrast to~(\ref{e:30}) it now seems possible in principle to adapt the
methods of Chapter~\ref{psec2}. We disregard these technical difficulties here
and merely discuss the form of the gauge terms in the simplest example of a
single Dirac sea and a $U(1)$ vector potential $A$. It might be that the only
relevant contributions to the EL equations comes about when the points
$x_1,\ldots,x_n$ all lie on a straight line. Generally speaking, the situation
in this case would be quite similar to that for a two-point action, and does not
seem to give anything essentially new (although the quantitative details would
clearly be different). In particular, the gauge terms of type~(\ref{e:3d}) drop
out in the closed chain, in agreement with the fact that the $U(1)$ corresponds
to an unbroken local gauge symmetry. However, the situation is much different if
we assume that the points $x_1,\ldots,x_n$ do not necessarily lie on a straight
line. Namely, in this case the phase shifts in the closed chain add up to an integral
along the polygon $C$ with vertices $x_1,\ldots,x_n$,
\[ e^{-i \int_{x_1}^{x_2} A_j\: (x_2-x_1)^j} \cdots
e^{-i \int_{x_n}^{x_1} A_k\: (x_1-x_n)^k} \;=\; \exp \left( -i
\oint_C A_i \: ds^i \right) \:. \]
Stokes' theorem allows us to write this line integral as a surface integral.
More precisely, choosing a two-dimensional surface $S$ with $\partial S = C$,
\[ \exp \left( -i \oint_C A_i \: ds^i \right) \;=\; \exp \left( -i \int_S
F_{ij}\: d\sigma^{ij} \right) \:, \]
where $F=dA$ is the field tensor and $d\sigma$ is the area form on $S$. This
simple consideration shows that the phase shift in the closed chain is in
general not zero; indeed, it is zero for any position of the points $x_k$ if
and only if the field tensor vanishes identically. Thus in the EL equations
we now expect additional contributions which involve surface integrals of the
gauge field tensor; we refer to such contributions as {\em{surface terms}}.
\index{surface term}
The appearance of surface terms seems a problem because they give constraints even
for those gauge potentials which correspond to a local symmetry of the system. \em
\end{Remark}

\chapter{Spontaneous Block Formation}
\setcounter{equation}{0} \label{esec4} The dynamical gauge group
introduced in the previous chapter cannot be identified with the
physical gauge group, because the results of Theorem~\ref{thm32}
are not compatible with the gauge groups in the standard model.
Namely, if we allow for two relations between the basic fractions
(case~{{(3)}}), the resulting dynamical gauge group $(U(7)
\times U(1))_L \times (U(7) \times U(1))_R$ is too large. The
cases~{{(1)}} and~{{(2)}}, on the other hand, seem too
restrictive because either no chiral gauge fields are
allowed~(\ref{e:3F}), or else the chiral gauge fields must be
Abelian and are diagonal on the sectors~(\ref{e:3I}), in contrast
to the weak $SU(2)$ gauge fields in the standard model.
Fortunately, these seeming inconsistencies disappear when scalar
potentials are taken into account, and it is indeed possible in
case~{{(2)}} to model an interaction involving non-Abelian
chiral gauge fields. The point is that if scalar potentials are
included, the dynamical mass matrices $Y_{L\!/\!R}$,
(\ref{e:3dyn}), are in general not diagonal on the sectors. Using
a local transformation of the fermionic projector, one can
reformulate the interaction such that the dynamical mass matrices
become diagonal, but then the resulting chiral fields have
off-diagonal contributions and can be identified with so-called
``effective'' non-Abelian gauge fields. In this chapter we study
the EL equations for an interaction involving both chiral and
scalar potentials. After the preparations of~{\S}\ref{esec41},
we show in~{\S}\ref{esec42} that the EL equations imply that
the fermionic projector splits globally into four so-called
blocks, which interact with each other only via free gauge fields.
We can distinguish between three quark blocks and one lepton
block; these will be analyzed in more detail in~{\S}\ref{esec43}.
In Chapter~\ref{esec5} we finally give the
transformation to the corresponding effective interaction.

Since including the scalar potentials may give further constraints for the
dynamical gauge potentials, we cannot expect that the dynamical gauge potentials of
the previous chapter will all be admissible here. Therefore, we merely assume
that the dynamical gauge potentials present in the system correspond to a
subgroup of the dynamical gauge group of Theorem~\ref{thm32}.
In order not to get lost in analytical details which are of no physical
relevance, we make the following additional assumptions.
\begin{itemize}
\item[{{(I)}}] The system should contain chiral dynamical gauge fields.
\item[{{(II)}}] The chiral Dirac particles should enter the EL equations.
\end{itemize}
From the physical point of view, the last assumption is trivial because
otherwise the chiral Dirac particles (=neutrinos) would be unobservable.
Furthermore, we need to assume that our system involves several gauge fields which
are sufficiently ``independent'' from each other. This assumption could be made
precise in many different ways; our formulation seems particularly convenient.
\begin{itemize}
\item[{{(III)}}] The free gauge fields should distinguish the chiral and massive Dirac
particles in the sense that for every pair of a chiral and a massive Dirac
particle there is a free dynamical gauge field which couples to the two
particles differently.
\end{itemize}
For the interactions in the standard model, the last assumption is clearly
satisfied because the electromagnetic field couples to all massive Dirac particles,
but not to the neutrinos. Assumption~{{(III)}} could be weakened, but this
would make it necessary to rule out a number of exceptional cases in the analysis,
and we do not want to consider this here.
We finally give our guideline for dealing with the regularization.
\begin{itemize}
\item[{{(IV)}}] Impose as few relations between the basic fractions as
possible such that {{(I)}}--{{(III)}} can be fulfilled.
\end{itemize}
This method will uniquely determine all relations between the basic
fractions.

\section{The Partial Trace and the Dynamical Mass Matrices}
\setcounter{equation}{0} \label{esec41}
We want to analyze the EL equations in the presence
of chiral and scalar potentials~(\ref{e:3dir}) to the degree 4 on
the light cone. Thus the only difference to the setting of
Theorem~\ref{thm32} is that, instead of a constant matrix $Y$, we
now allow more generally for dynamical mass matrices $Y_L(x)$ and
$Y_R(x)$. One difficulty is that the scalar potentials may depend
in a complicated way on the generation index (in contrast to the
chiral potentials which we assumed to be constant on the
generations; see~(\ref{e:3c})). In particular, the partial
trace~(\ref{part}) becomes a non-trivial operation when dynamical
mass matrices are present. In this section, we give a few general
considerations on the partial trace of the dynamical mass
matrices.

We first introduce a convenient notation. In our calculations so far, we
omitted the mass matrix $Y$ in all contributions to the fermionic projector.
Now that we are working with the dynamical mass matrices $Y_{L\!/\!R}$, these
clearly have to be written out. In composite expressions, we need to make clear
how the partial traces are to be taken. To this end, we denote the sums over
the upper and lower generation index by the tildes $\acute{\;\:}$ and
$\grave{\;\:}$, respectively. Thus we introduce the matrices
\label{acuteY_LR} \label{graveY_LR} \label{hatY_LR}
\begin{eqnarray*}
\acute{Y}_{L\!/\!R} \;:\; \C^{8 \times 3} \to \C^8 \:,\spc
(\acute{Y}_{L\!/\!R})^a_{(b \beta)} &=& \sum_{\alpha=1}^3 (Y_{L\!/\!R})^{(a
\alpha)}_{(b \beta)} \\
\grave{Y}_{L\!/\!R} \;:\; \C^8 \to \C^{8 \times 3} \:,\spc
(\grave{Y}_{L\!/\!R})^{(a \alpha)}_b &=& \sum_{\beta=1}^3 (Y_{L\!/\!R})^{(a
\alpha)}_{(b \beta)} \:.
\end{eqnarray*}
Similarly, we denote the sum over both generation indices by the accent
$\hat{\;\:}$,
\[ \hat{Y}_{L\!/\!R} \;:\; \C^8 \to \C^8 \:,\spc
(\hat{Y}_{L\!/\!R})^a_b \;=\; \sum_{\alpha, \beta=1}^3 (Y_{L\!/\!R})^{(a
\alpha)}_{(b \beta)} \:. \]
Clearly, $(\acute{Y}_{L\!/\!R})^* = \grave{Y}_{R\!/\!L}$ and
$(\hat{Y}_{L\!/\!R})^* = \hat{Y}_{R\!/\!L}$. In a contribution to the fermionic
projector which involves a product of dynamical mass matrices, the partial
trace leads us to label the first and last factor $Y_{L\!/\!R}$ by
$\acute{\;\:}$ and $\grave{\;}$, respectively.
For example, in the presence of a homogeneous scalar potential, we write
the light-cone expansion of the left-handed component of the fermionic projector
in analogy to~(\ref{e:2o}) as
\begin{equation}
\chi_L\: P(x,y) \;=\; \chi_L \left( X\: \frac{i \xi\slsh}{2}\:
T^{(-1)}_{[0]}\:+\: \hat{Y}_L\: T^{(0)}_{[1]} \:+\: \frac{i \xi\slsh}{2}\:
\acute{Y}_L\: \grave{Y}_R\: T^{(0)}_{[2]}\:+\: \cdots \right) . \label{e:30h2}
\end{equation}
Furthermore, we denote the contraction in the sector index by $\Tr_S$,
\[ \Tr_S B \;\equiv\; \sum_{n=1}^8 B^n_n\:. \]
One should keep in mind that the partial trace is {\em{not}} cyclic, because we
sum over the upper and lower index independently. For example,
\begin{equation}
\Tr_S \acute{Y}_L \grave{Y}_R \;\stackrel{\mbox{\tiny{in general}}}{\neq}\;
\Tr_S \acute{Y}_R \grave{Y}_L \:.    \label{e:4nc}
\end{equation}
But both terms are clearly real and non-negative.

The EL equations are formulated in terms of the fermionic projector, which is
obtained from the auxiliary fermionic projector by taking the partial
trace~(\ref{part}). Therefore we regard the fermionic projector as a physical
object only after the partial trace has been taken. Thus it is a reasonable
point of view that we do not need to worry about noncausal line integrals in
the light-cone expansion as long as the corresponding contributions to the
auxiliary fermionic projector drop out when the partial trace is taken. This
leads us to weaken the causality compatibility condition~(\ref{89}) by imposing
a condition only on the partial trace of the spectral projectors.

\begin{Def} \label{defwccc}
The Dirac operator is {\bf{weakly causality
    compatible}}\index{causality compatible!weakly} if\[ \sum_{\alpha, \beta=1}^3  (X\: (\tilde{p} - \tilde{k}))^{(a \alpha)}_{(b \beta)}
\;=\; \sum_{\alpha, \beta=1}^3  ((\tilde{p} - \tilde{k})\: X^*)^{(a \alpha)}_{(b \beta)}
\:. \]
Under this assumption, the fermionic projector is defined
canonically by
\begin{equation}
P^a_b(x,y) \;=\; \sum_{\alpha, \beta=1}^3
X\: \frac{1}{2}\:(\tilde{p} - \tilde{k}))^{(a \alpha)}_{(b \beta)}(x,y) \:. \label{e:42x}
\end{equation}
\end{Def}
In what follows, we shall assume that the weak causality compatibility condition
is satisfied for all contributions to the fermionic projector which are of
relevance to the degree on the light cone under consideration.

Our point of view that the fermionic projector has a physical meaning only
after taking the partial trace also implies that we
should consider different choices of dynamical mass matrices as being
equivalent if taking the partial trace~(\ref{part}) gives the same fermionic
projector. Furthermore, for this equivalence it is not necessary
that the fermionic projectors be identical, but it suffices that all
contributions to the fermionic projector which enter the EL equations are the
same. More specifically, to the degree 4 on the light cone the EL equations
will involve at most quadratic terms in $m$, and so every factor $Y_{L\!/\!R}$
carries an accent. Thus two dynamical mass matrices can be considered as being
equivalent if their partial traces coincide.  In other words, the dynamical mass
matrices are determined only modulo the equivalence relation
\[ B_1 \simeq B_2 \spc {\mbox{if}} \spc {\mbox{$\acute{B}_1=\acute{B}_2$
and $\grave{B}_1 = \grave{B}_2$}}. \]
This arbitrariness in choosing the dynamical mass matrices can be used to
simplify these matrices. For example, we will set the matrix entries to
zero whenever possible by applying for every $a, b \in \{1,\ldots,8\}$ and $c
\in \{L, R\}$ the rule
\[ (\acute{Y}_c)^a_{(b.)} = 0 = (\grave{Y}_c)^{(a.)}_b
\;\;\;\;\Longrightarrow\;\;\;\; (Y_c)^{(a.)}_{(b.)} = 0 \:. \]
Here the dot means that we are using a matrix notation in the generation index,
i.e.\ $(\acute{Y}_c)^a_{(b.)}$ is (for fixed $a$, $b$) a $3$-vector and $(Y_c)^{(a.)}_{(b.)}$ a $3 \times 3$ matrix.
We refer to this method of simplifying the
dynamical mass matrices that we {\em{choose a convenient representation}} of
$Y_c$.

In order to rule out pathological cases, we need to impose a condition on
the dynamical mass matrices. Note that in the vacuum the mass matrices
are block diagonal in the sense that $(Y_{L\!/\!R})^{(a.)}_{(b.)} = \delta^a_b
Y^a_{L\!/\!R}$ for suitable $3 \times 3$ matrices $Y^a_{L\!/\!R}$. Thus the
off-diagonal elements $(Y_{L\!/\!R})^{(a.)}_{(b.)}$, $a \neq b$, contain
scalar potentials. It would be too restrictive to assume that there are
no cancellations when the partial trace is taken; i.e.\ we do want to allow for
the possibility that $(\acute{Y}_c)^a_{(b.)}=0$ or $(\grave{Y}_c)^{(a.)}_b=0$
although $(Y_c)^{(a.)}_{(b.)} \neq 0$ (for some $a \neq b$). But such cancellations
should occur only with a special purpose, for example in order to ensure that the Dirac
operator be weakly causality compatible or in order to arrange that certain
terms drop out of the EL equations. For such a purpose, it is not sufficient
that one off-diagonal element of $\acute{Y}_c$ vanishes, but
all the off-diagonal elements in the same row should be zero. This is the
motivation for the following definition.
\begin{Def} \label{def40}
The dynamical mass matrices\index{mass matrix!non-degenerate dynamical}
are {\bf{non-degenerate}} if for all $a, b \in
\{1,\ldots,8\}$, $a \neq b$ and $c \in \{L, R\}$,
\[ (\grave{Y}_c)^{(a.)}_b \neq 0 {\mbox{ and }} (\acute{Y}_c)^a_{(b.)} =0
\;\;\;\;\Longrightarrow\;\;\;\; (\acute{Y}_c)^a_{(d.)} = 0 {\mbox{ for all $d
\neq a$}}. \]
\end{Def}

The freedom to choose a convenient representation of the dynamical mass
matrices reduces our problem to revealing the structure of the matrices
$\acute{Y}_c$ and $\grave{Y}_c$. One difficulty is that the EL equations
involve these matrices only in products of the form $\acute{Y}_{L\!/\!R}(y)
\grave{Y}_{L\!/\!R}(x)$. The following elementary lemma will allow us to use
information on the matrix product to derive properties of the individual
factors.
\begin{Lemma} {\bf{(Uniform Splitting Lemma)}} \label{lemma40}
Let ${\mathcal{B}} \subset {\mbox{\rm{Mat}}}(\C^{p_1}, \C^{p_2})$ be a
set of $(p_2 \times p_1)$ matrices with the property that for all
$B_1, B_2 \in {\mathcal{B}}$ there is $\lambda \in \C$ such that
\begin{equation}
B_1^*\: B_2 \;=\; \lambda\: \1_{\sC^{p_1}}\:.    \label{e:4z}
\end{equation}
Then there is a unitary $(p_2 \times p_2)$ matrix $U$ and an
integer $r \geq 0$ with $r p_1 \leq p_2$ such that every $B \in
{\mathcal{B}}$ can be written in the form
\begin{equation}
B \;=\; U \left( \begin{array}{cc}
\overbrace{b \oplus \cdots \oplus b}^{\mbox{\tiny{$p_1$ summands}}} \\
0 \end{array} \right)
\begin{array}{l}  \} {\mbox{\footnotesize{$r p_1$ rows}}} \\[.5em]
\} {\mbox{\footnotesize{$p_2 - r p_1$ rows}}} \end{array}
\end{equation}
for a suitable $(r \times 1)$ matrix $b$.
\end{Lemma}
We mention for clarity that $b \oplus \cdots \oplus b$ is a $(r
p_1 \times p_1)$ matrix; it could also be written as a block
matrix with diagonal entries $b$. The word ``uniform'' in the name
of the lemma refers to the fact that the unitary transformation
$U$ is independent of $B \in {\mathcal{B}}$. In our applications,
this will mean that $U$ is constant in space-time. Such constant
unitary transformations are irrelevant (e.g.\ they could be
absorbed into a more general definition of the partial trace), and
we can often simply ignore them. \\[.5em]
{\em{Proof of Lemma~\ref{lemma40}. }}
Let $(e_1,\ldots,e_{p_1})$ be an orthonormal basis of $\C^{p_1}$. We introduce
the subspaces
\[ E_i \;=\; \bra \{ B e_i {\mbox{ with }} B \in {\mathcal{B}} \} \ket \;\subset\;
\C^{p_2} \]
and the mappings
\[ \pi_i \;:\; {\mathcal{B}} \rightarrow E_i \;:\; B \mapsto B e_i \:. \]
The property~(\ref{e:4z}) implies that for all $B_1, B_2 \in
{\mathcal{B}}$,
\begin{equation}
\langle B_2 \:e_i,\: B_1\: e_j \rangle \;=\; \langle B_1^*\:B_2 \:e_i,\: e_j \rangle
\;=\; \lambda(B_1, B_2) \: \delta_{ij}\:.    \label{e:4y}
\end{equation}
If $i \neq j$, this relation shows that the subspaces $(E_i)_{i=1,\ldots,p_1}$
are orthogonal. In the case $i=j$, (\ref{e:4y}) yields that the inner product
$\langle \pi_i(B_1),\: \pi_i(B_2) \rangle$ is independent of $i$. Thus the
mappings $\pi_i$ are unitarily equivalent, and so we can arrange by a unitary
transformation that the $\pi_i$ all have the same matrix representation
$\pi(B)=b$.
\QED

\section{Analysis of Degeneracies}
\setcounter{equation}{0} \label{esec42}
The operator $Q$ corresponding to the Dirac
operator~(\ref{e:3dir}) is again given by~(\ref{e:2c},
\ref{e:39z}),
\begin{equation}
Q(x,y) \;=\; \frac{1}{2} \;\sum_{k=1}^{K_{xy}} \frac{\partial
{\mathcal{L}}(\lambda^{xy})}{\partial \lambda^{xy}_k}\: F^{xy}_k\;
P(x,y) \:.    \label{e:30c}
\end{equation}
Following~{{(I)}} and~{{(IV)}}, we can restrict attention to
case~{{(2)}} of Theorem~\ref{thm32}. In this case, the
eigenvalues of $A$ are highly degenerate. We must take into
account that these degeneracies will in general be removed by the
scalar perturbation. This subtle problem is treated in a more
general context in Appendix~\ref{appC}. We now specialize the
obtained results using a notation which is adapted to the
dynamical gauge group ${\mathcal{G}}_p$, (\ref{e:3I}). We let
$\uparrow$ and $\downarrow$ be the sets
\[ \uparrow \:\;=\; \{1,\ldots,p\} \:,\spc \downarrow \:\;=\; \{p+1,\ldots,7\} \]
and introduce the corresponding projectors $I_{\uparrow \!/\!
  \downarrow}$ \label{I_arrows} by
\[ I_\uparrow \;=\; \sum_{n \in \uparrow} I_n \:,\spc
I_\downarrow \;=\; \sum_{n \in \downarrow} I_n \:, \] where
$(I_n)^a_b = \delta^a_b\: \delta^a_n$ are the projectors on the
sectors (in the case $p=7$, we set $\downarrow = \emptyset$ and
$P_\downarrow = 0$). To the highest degree on the light cone
(i.e.\ if the eigenvalues are treated as in Theorem~\ref{thm31}),
the chiral gauge fields corresponding to~${\mathcal{G}}_p$ lead to
five distinct eigenvalues of $A_{xy}$, one of which is zero. The
spectral projectors corresponding to the kernel and the non-zero
eigenvalues are given by $I_8$ and
\begin{equation}
(\chi_c\: I_\uparrow + \chi_{\bar{c}}\: I_\downarrow)\: F_s \spc {\mbox{with
$c=L\!/\!R$, $s=\pm$,}}
    \label{e:30a}
\end{equation}
respectively. To the next lower degree on the light cone, we need to take into
account the perturbation of $A$ by gauge terms and the scalar
potentials. Theorem~\ref{thmB2} shows that the dimension of the kernel of $A$
is not affected by the perturbation, and thus it suffices to consider the
non-zero eigenvalues. According to Theorem~\ref{thmB1}, the degeneracy of
the non-zero eigenvalues is in general removed.
In order to describe the splitting of the eigenvalues in the massive
sectors, we first associate to each spectral projector~(\ref{e:30a}) a
projector on an invariant subspace of $A$ (which is no longer necessarily an
eigenspace), and the perturbed eigenvalues are then obtained by diagonalizing
$A$ on these invariant subspaces (see~{\S}\ref{appC1} and~{\S}\ref{appC4}
for details). It is the main result of Theorem~\ref{thmB1} that the perturbation
is block diagonal on the left- and right-handed components of the invariant
subspaces. This means more precisely that the left- and right-handed
components of $(F_k)_{k=2,\ldots,K}$, i.e. the image of the eight projectors
\begin{equation}
\chi_c\: I_\uparrow\: F_s {\mbox{ and }} \chi_c\: I_\downarrow\: F_s
\spc {\mbox{with $c=L\!/\!R$, $s=\pm$}}, \label{e:30b}
\end{equation}
can be perturbed to obtain invariant subspaces of $A$, and thus it
suffices to analyze $A$ on these smaller subspaces. Since each of
these subspaces carries fixed indices $(c,s)$, a basis on each
subspace may be labeled by the sector index $n$. We choose a basis
such that $A$ is diagonal on the invariant spaces. We denote the
corresponding eigenvalues (counting multiplicities) by
$(\lambda_{ncs}+ \Delta \lambda_{ncs})$ and the spectral projectors
by $(F_{ncs} + \Delta F_{ncs})$. For clarity, we point out that the
unperturbed spectral projectors $F_{ncs}$ appearing here may differ
from those in~(\ref{e:3k}) in that we are using a different basis on
the sectors, which need not be orthogonal and may depend on $c,s$
and $x,y$. This slight abuse of notation cannot lead to confusion
because in~(\ref{e:3k}) we are free to choose any basis on the
degenerate subspaces.

For our choice of the Lagrangian~(\ref{e:2N}) and the dynamical
gauge group according to~(\ref{e:3I}), the factors $\partial
{\mathcal{L}}/\partial \lambda_k$ in~(\ref{e:30c}) vanish
identically to the highest degree, see~(\ref{e:3van}). Thus it
suffices to take into account the perturbation of these factors.
Using the above notation, we obtain
\begin{equation}
Q(x,y) \;=\; \frac{1}{2} \sum_{n,c,s} \left( \Delta
\frac{\partial {\mathcal{L}}(\lambda^{xy})}{\partial
\lambda^{xy}_{ncs}} \right) F^{xy}_{ncs}\: P(x,y)
\:+\:(\deg < 4)\:. \label{e:41a}
\end{equation}
Note that the perturbation of the spectral projectors $\Delta F_{ncs}$ does not
appear here; this is a major simplification. Computing the
perturbation of the Lagrangian, one sees that our task
is to compute terms of the form
\begin{eqnarray}
&& \sum_{n,c,s} {\mathcal{P}}(\lambda^{xy}_{ncs},
\overline{\lambda^{xy}_{ncs}}) \; \Delta \lambda^{xy}_{ncs}\:
F^{xy}_{ncs}\: P(x,y) \label{e:30d} \\
&& \sum_{n,c,s} {\mathcal{P}}(\lambda^{xy}_{ncs},
\overline{\lambda^{xy}_{ncs}}) \; \overline{\Delta
\lambda^{xy}_{ncs}}\: F^{xy}_{ncs}\: P(x,y)\:, \label{e:30e}
\end{eqnarray}
where ${\mathcal{P}}$ stands for a function in both arguments.
The subtle point in computing expressions of the
form~(\ref{e:30d}, \ref{e:30e}) is to carry out the sums over $n
\in \:\uparrow$ and $n \in \:\downarrow$ (for fixed $c, s$),
because the corresponding indices $\{ (ncs), n \in \:\uparrow
\!/\! \downarrow \}$ label our basis vectors on the invariant
subspaces associated to the projectors $\chi_c I_{\uparrow \!/\!
\downarrow} F_s$. We shall now give a procedure for explicitly
computing these sums. First of all, it is helpful that the
unperturbed eigenvalues do not depend on $n \in \:\uparrow$ or
$\downarrow$. Thus the polynomials ${\mathcal{P}}$
in~(\ref{e:30d}, \ref{e:30e}) may be taken out of the sums. It is a
complication that $\Delta \lambda_{ncs}$ and $P(x,y)$ involve the
gauge potentials corresponding to the free gauge group
${\mathcal{F}}_p$. To bypass this difficulty, we choose $x$ and
$y$ on a fixed null line ${\mathcal{L}}$ in Minkowski space,
\begin{equation}
    x,y \in {\mathcal{L}} = u + \R\: v \spc {\mbox{with $v^2=0$}}
    \label{e:4line}
\end{equation}
and arrange by a gauge transformation that the free gauge
potentials vanish identically on ${\mathcal{L}}$ (this is possible
because free gauge transformations are local unitary
transformations; see page~\pageref{e:3l}).  After this
transformation, the chiral potentials are Abelian on $\overline{xy}$
and diagonal in the sector index.

We first state the formulas for the perturbation of the eigenvalues in
full generality; we shall discuss and analyze these formulas afterwards
beginning with simple special cases. In order to keep the notation as
simple as possible, we restrict attention to the case $c=L$ and
$n \in \:\uparrow$, and we shall give symbolic
replacement rules with which the analogous formulas are obtained in all
other cases.
\begin{Def} \label{def36}
Let $\nu$, $\mu$ and $\nu_8$, $\mu_8$ be the phase factors
\begin{eqnarray}
\nu &=& \Tr_S \left( I_7\: \intR_x^y \intL_y^x \right) \:,\spc\:\:
\mu \;=\; \Tr_S \left( I_1\: \intL_x^y \right)\: \Tr_S
\left( I_7\: \intR_y^x \right) \label{e:4f} \\
\nu_8 &=& \Tr_S \left( I_8\: \intR_x^y \intL_y^x \right) \:,\spc
\mu_8 \;=\; \Tr_S \left( I_1\: \intL_x^y \right)\: \Tr_S
\left( I_8\: \intR_y^x \right) . \label{e:4g}
\end{eqnarray}
We introduce the $p \times p$ matrix $\Lambda$ by
\begin{eqnarray}
\hspace*{-1cm}\Lambda &=& \nu \:\int_x^y dz \;I_\uparrow\:\acute{Y}_L\:\grave{Y}_R \;
 I_\uparrow\;T^{(0)}_{[2]} \: \overline{T^{(0)}_{[0]}} \label{e:3b1} \\
&&+\nu \: \int_y^x dz \; I_\uparrow\: \acute{Y}_R\:\grave{Y}_L
\: I_\uparrow\; T^{(-1)}_{[0]}\:\overline{T^{(1)}_{[2]}} \label{e:3b2} \\
&& \!\!\! \left. \begin{array}{l}
+I_\uparrow\: \hat{Y}_L(y)\:I_\uparrow\:\hat{Y}_L(x)\: I_\uparrow \;
T^{(0)}_{[1]}\:\overline{T^{(0)}_{[1]}} \\[.6em]
- I_\uparrow\: \acute{Y}_R(y)\:I_\uparrow\:\grave{Y}_L(x) \:I_\uparrow\;
T^{(-1)}_{[0]}\:\overline{T^{(1)}_{[2]}} \\[.6em]
\displaystyle -\frac{1}{\nu \lambda_- - \overline{\nu} \lambda_+}
\;I_\uparrow \left( \hat{Y}_L(y)\:T^{(0)}_{[1]}
\:\overline{T^{(-1)}_{[0]}} \:-\: \hat{Y}_R(y)\: T^{(-1)}_{[0]}\:
\overline{T^{(0)}_{[1]}} \right) \\[.9em]
\hspace*{1.96cm} \times \:I_\uparrow
\left(\nu\:\hat{Y}_R(x) \; T^{(0)}_{[1]}
\:\overline{T^{(0)}_{[0]}} \:-\: \overline{\nu}\:
\hat{Y}_L(x)\; T^{(0)}_{[0]}\: \overline{T^{(0)}_{[1]}} \right)
I_\uparrow \end{array} \right\} \spc\label{e:3b3} \\
&& \!\!\! \left. \begin{array}{l}
+\mu \nu\: I_\uparrow\: \hat{Y}_L(y)\:I_\downarrow\:\hat{Y}_L(x)\: I_\uparrow
\;T^{(0)}_{[1]}\:\overline{T^{(0)}_{[1]}} \\[.6em]
- \mu \nu\: I_\uparrow\: \acute{Y}_R(y)\:I_\downarrow\:\grave{Y}_L(x)
\:I_\uparrow\; T^{(-1)}_{[0]}\:\overline{T^{(1)}_{[2]}} \\[.6em]
\displaystyle -\frac{\mu \nu}{\lambda_- - \lambda_+} \;I_\uparrow
\left( \hat{Y}_L(y)\: T^{(0)}_{[1]} \:\overline{T^{(-1)}_{[0]}} \:-\:
\hat{Y}_R(y)\: T^{(-1)}_{[0]}\: \overline{T^{(0)}_{[1]}} \right) \\[.9em]
\hspace*{1.53cm} \times \:I_\downarrow
\left(\hat{Y}_R(x) \: T^{(0)}_{[1]} \:\overline{T^{(0)}_{[0]}} \:-\:
\hat{Y}_L(x)\; T^{(0)}_{[0]}\: \overline{T^{(0)}_{[1]}} \right)
I_\uparrow \end{array} \right\} \label{e:3b6} \\
&&-\mu_8 \nu_8\: I_\uparrow\: \acute{Y}_R(y)\:I_8\:\grave{Y}_L(x)
\:I_\uparrow\; T^{(-1)}_{[0]}\:\overline{T^{(1)}_{[2]}}\:. \label{e:3b9}
\end{eqnarray}
We denote the spectral adjoint of $\Lambda$ (as defined
in~(\ref{e:2Abar})) by $\overline{\Lambda}$.
\end{Def}

\begin{Lemma} \label{lemma37}
Up to contributions of degree $<4$,
\begin{eqnarray}
\sum_{n \in \uparrow} \Delta \lambda^{xy}_{nL+} \: F^{xy}_{nL+}\: P(x,y) &=& 0
\;=\; \sum_{n \in \uparrow} \overline{\Delta \lambda^{xy}_{nL+}}
\: F^{xy}_{nL+}\: P(x,y) \label{e:3a1} \\
\sum_{n \in \uparrow} \Delta \lambda^{xy}_{nL-} \: F^{xy}_{nL-}\: P(x,y) &=&
\Lambda\: P(x,y) \label{e:3a2} \\
\sum_{n \in \uparrow} \overline{\Delta \lambda^{xy}_{nL-}} \: F^{xy}_{nL-}\: P(x,y)
&=& \overline{\Lambda}\: P(x,y) \:. \label{e:3a3}
\end{eqnarray}
The corresponding formulas for the opposite chirality are obtained by the
symbolic replacements
\begin{equation}
L \:\longleftrightarrow\: R \:,\;\;\;\;\;
\nu \:\longleftrightarrow\: \overline{\nu} \:,\;\;\;\;\;
\nu_8 \:\longleftrightarrow\: \overline{\nu_8} \:,\;\;\;\;\;
\mu_8 \:\longleftrightarrow\: \overline{\nu} \nu_8\: \mu_8 \:.
\label{e:30f}
\end{equation}
In the case $p<7$, we may furthermore perform the replacements
\begin{equation}
\uparrow \;\longleftrightarrow\; \downarrow \:,\;\;\;\;\;
\nu \:\longleftrightarrow\: \overline{\nu} \:,\;\;\;\;\;
\mu \:\longleftrightarrow\: \overline{\mu} \spc {\mbox{and}} \spc
\mu_8 \:\longleftrightarrow\: \overline{\mu \nu}\: \mu_8\:. \label{e:30g}
\end{equation}
\end{Lemma}
{\Proof}
According to~(\ref{e:2s2}), to the highest degree on the light cone we have
the identity $F_{nc+} P(x,y) = 0$.
This gives~(\ref{e:3a1}), and~(\ref{e:3a2}) follows directly from
Theorem~\ref{thmB1} and the results of~{\S}\ref{appC3}. In a basis where
$\Lambda$ is diagonal, (\ref{e:3a3}) is an immediate consequence
of~(\ref{e:3a2}).

To derive the replacement rule~(\ref{e:30f}), we first note that in the case
$p=7$, the projector $I_\downarrow$ vanishes, and thus all contributions to
$\Lambda$ involving $\mu$ are equal to zero. In the case $p<7$,
\begin{eqnarray*}
\mu &\!\!\!\longrightarrow\!\!\!&
\Tr_S \left( I_1\: \intR_x^y \right)\: \Tr_S \left( I_7\: \intL_y^x \right)
\;=\; \overline{\nu} \: \Tr_S \left( I_1\: \intL_x^y \right) \; \nu\:
\Tr_S \left( I_7\: \intR_y^x \right) \;=\; \mu \\
\mu_8 &\!\!\!\longrightarrow\!\!\!&
\Tr_S \left( I_1\: \intR_x^y \right)\: \Tr_S \left( I_8\: \intL_y^x \right)
\;=\; \overline{\nu} \: \Tr_S \left( I_1\: \intL_x^y \right) \; \nu_8\:
\Tr_S \left( I_8\: \intR_y^x \right) \;=\; \overline{\nu} \nu_8\: \mu \:.
\end{eqnarray*}
Using the relations
\[ \Tr_S \left( I_7\: \intL_x^y \right)\: \Tr_S \left( I_1\: \intR_y^x \right)
\;=\; \overline{\mu} \quad {\mbox{and}}\quad
\Tr_S \left( I_7\: \intL_x^y \right)\: \Tr_S \left( I_8\: \intR_y^x \right)
\;=\; \overline{\mu \nu}\: \mu_8\:, \]
the replacement rule~(\ref{e:30g}) is straightforward.
\QED

A straightforward calculation using~(\ref{e:41a}) and
Lemma~\ref{lemma37} shows that for our Lagrangian~(\ref{e:2N}), the EL
equations yield the conditions
\begin{equation}
\left[ \overline{\lambda^{xy}_{\uparrow L -}}\: \Lambda
\:+\: \lambda^{xy}_{\uparrow L -}\: \overline{\Lambda}
\right] \:
\frac{P(x,y)}{\lambda^{xy}_{\uparrow L -}} \;=\; f(x,y)\:
I_\uparrow\: P(x,y)\:+\: (\deg < 4) \:,
    \label{e:30l}
\end{equation}
where we set $\lambda_{\uparrow cs} = \lambda_{ncs}$, $n \in \:\uparrow$.
Here $f(x,y)$ can be any scalar function; it takes into account that the
average of all eigenvalues drops out in the EL equations when we take the
difference of the contributions resulting from the two terms
in~(\ref{e:2N}).
Similar conditions for the opposite chirality and for $\uparrow$ replaced by
$\downarrow$ are obtained from~(\ref{e:30l}) by applying the
rules~(\ref{e:30g}, \ref{e:30f}). We point out that the resulting four
equations must clearly be satisfied for the same function $f(x,y)$. These
four equations together are even equivalent to the EL equations to
degree $4$.

The remaining problem is to analyze the obtained equations of
types~(\ref{e:30l}). At first sight, this seems a difficult problem
because the matrix $\Lambda$ has a
complicated explicit form (see Def.~\ref{def36}) and because taking the
spectral adjoints makes it necessary to diagonalize these
matrices. Fortunately, the requirement that the EL equations be mathematically
consistent will give us strong restrictions on the form of $\Lambda$,
and this will indeed make it possible to reveal a relatively simple global
structure of the admissible interactions.
In order to explain how the mathematical consistency conditions come about, we
first recall that for polynomial Lagrangians~(\ref{e:2B}) we saw
after~(\ref{e:2C}) that the resulting operator $Q$ is a polynomial in the
fermionic projector and is thus well-defined within the formalism of the
continuum limit. However, the situation is different for our
Lagrangian~(\ref{e:2O}) because the spectral weight is an operation which
does not necessarily make sense in the continuum limit.
More specifically, the mathematical problem in~(\ref{e:30l}) is to make sense
of the spectral adjoint. For clarity, we explain the difficulty and our basic
argument in the simple example
\begin{equation}
\overline{B_1\: M_1 \:+\: B_2\: M_2}\:, \label{e:31a}
\end{equation}
where $B_1$ and $B_2$ are matrices depending on the macroscopic potentials, and
$M_{1\!/\!2}$ are two monomials. The monomials can be considered as scalar
functions which are highly singular on the light cone, and which we can control
in the continuum limit only in the weak sense. To form the spectral adjoint
in~(\ref{e:31a}), we need to know the eigenvalues and spectral projectors of
the matrix $B_1 M_1 + B_2 M_2$. In general, the spectral decomposition of this
matrix will depend nonlinearly on $M_1$ and $M_2$, because the zeros of the
characteristic polynomials involve roots of the monomials. In this generic
situation, the spectral adjoint is ill-defined in the formalism of the
continuum limit. The only case in which the eigenvalues are linear in $M_1$
and $M_2$ is when the eigenvectors can be chosen independent of the monomials.
This is possible iff the matrices $B_1$ and $B_2$ have a common eigenvector
basis, or equivalently, if they commute,
\[ [B_1, B_2] \;=\; 0 \:. \]
This simple argument shows that the requirement that the spectral adjoint be
well-defined leads to commutator relations for the macroscopic potentials.
In the next lemma we apply this argument to the matrix $\Lambda$.
By the {\em{contributions to $\Lambda$}} we mean the
individual summands obtained by multiplying out all the terms
in~(\ref{e:3b1}--\ref{e:3b9}).
\begin{Lemma} \label{lemma35}
For any $x, y \in {\mathcal{L}}$ there is a basis on the sectors
such that the contributions to $\Lambda$ are all diagonal
matrices.
\end{Lemma}
{\Proof} Clearly, our argument after~(\ref{e:31a}) applies in the same way
to the spectral adjoint of a finite sum. Thus in order to make mathematical
sense of the spectral adjoint $\overline{\Lambda}$, we need to assume that
the contributions to $\Lambda$ all commute with each other. Hence we can
choose a basis such that these contributions are all diagonal.
In particular, one sees that in this basis the matrix products
$\hat{Y}_{c_1}(x)$ and $\hat{Y}_{c_2}(y)$ are diagonal for all $c_1,
c_2 \in \{L, R\}$.
\QED

We proceed by analyzing the EL equations~(\ref{e:30l}) for special choices
of $x$ and $y$, for which the matrix $\Lambda$ becomes particularly simple.
We begin with the situation where we choose $x$ such that the scalar
potentials vanish at $x$, i.e.
\begin{equation}
Y_L(x) \;=\; Y \;=\; Y_R(x) \label{e:32d}
\end{equation}
with $Y$ the mass matrix of the vacuum (for example, we can choose $x$
close to infinity). Then the matrices $Y_{L\!/\!R}(x)$ are diagonal in the
sector index and on the massive sectors are a multiple of the identity.
Thus the ``off-diagonal'' contributions~(\ref{e:3b6}, \ref{e:3b9})
to $\Lambda$ vanish. In the ``diagonal''
contributions~(\ref{e:3b1}--\ref{e:3b3}), on the other hand, we can
simplify our notation by omitting the factors $\hat{Y}_{L\!/\!R}(x)$.
Then the matrix $\Lambda$ takes the form
\begin{eqnarray*}
\Lambda &=& \nu \:\int_x^y dz \;I_\uparrow\:\acute{Y}_L\:\grave{Y}_R \;
 I_\uparrow\;T^{(0)}_{[2]} \: \overline{T^{(0)}_{[0]}} \\
&&+\nu \: \int_y^x dz \; I_\uparrow\: \acute{Y}_R\:\grave{Y}_L
\: I_\uparrow\; T^{(-1)}_{[0]}\:\overline{T^{(1)}_{[2]}} \\
&&+I_\uparrow\: \hat{Y}_L(y)\:I_\uparrow \;
T^{(0)}_{[1]}\:\overline{T^{(0)}_{[1]}} \\
&&- I_\uparrow\: \acute{Y}_R(y)\:\grave{Y}\:I_\uparrow \;
T^{(-1)}_{[0]}\:\overline{T^{(1)}_{[2]}} \\
&& -\frac{\nu\: T^{(0)}_{[1]} \:\overline{T^{(0)}_{[0]}} \:-\: \overline{\nu}\:
T^{(0)}_{[0]}\: \overline{T^{(0)}_{[1]}}}
{\nu \lambda_- - \overline{\nu} \lambda_+}
\;I_\uparrow \left( \hat{Y}_L(y)\:T^{(0)}_{[1]}
\:\overline{T^{(-1)}_{[0]}} \:-\: \hat{Y}_R(y)\: T^{(-1)}_{[0]}\:
\overline{T^{(0)}_{[1]}} \right) I_\uparrow \:.
\end{eqnarray*}
Evaluating the EL equations~(\ref{e:30l}) for this choice of $\Lambda$ yields
the following result.
\begin{Lemma} \label{lemma34}
Suppose that {{(I)}} holds. Without introducing any relations
between the basic fractions (besides those of
Theorem~\ref{thm32}), we can choose for any $y \in {\mathcal{L}}$
suitable parameters $a, b \in \R$ and $c \in \C$ such that at $y$,
\begin{eqnarray}
I_\uparrow\: \acute{Y}_L \grave{Y}_R\: I_\uparrow &=&
a\: I_\uparrow \:,\spc\quad
I_\uparrow\: \acute{Y}_R \grave{Y}_L\: I_\uparrow \;=\;
b\: I_\uparrow \label{e:32e} \\
I_\uparrow\: \hat{Y}_L(y)\: I_\uparrow &=& c\: I_\uparrow \;\:,\spc\;\;\,
I_\uparrow\:\hat{Y}_R(y)\: I_\uparrow \;=\; \overline{c}\: I_\uparrow \:.
\label{e:32f}
\end{eqnarray}
The analogous formulas for $I_\uparrow$ interchanged by $I_\downarrow$ are obtained
by the replacements
\begin{equation}
\uparrow \;\longleftrightarrow\; \downarrow \spc {\mbox{and}} \spc
L \:\longleftrightarrow\: R    \label{e:4rep}
\end{equation}
with the parameters $a$, $b$ and $c$ unchanged.
\end{Lemma}
{\Proof}
The above $\Lambda$ contains contributions which are scalar multiples of
the matrices $I_\uparrow \hat{Y}_L(y) I_\uparrow$ and
$I_\uparrow \hat{Y}_R(y) I_\uparrow$.
Thus in the basis of Lemma~\ref{lemma35}, these matrices are both diagonal.
Since one is the adjoint of the other, we conclude that these matrices are
normal, and thus their spectral adjoints coincide with the usual adjoints,
\begin{equation}
\overline{ I_\uparrow\: \hat{Y}_L\: I_\uparrow} \;=\;
I_\uparrow\: \hat{Y}_R\: I_\uparrow \:,\spc
\overline{ I_\uparrow\: \hat{Y}_R\: I_\uparrow} \;=\;
I_\uparrow\: \hat{Y}_L\: I_\uparrow \:. \label{e:31b}
\end{equation}
The matrices $\acute{Y}_L \grave{Y}_R$ and $\acute{Y}_R \grave{Y}_L$, on the
other hand, are Hermitian and thus spectrally selfadjoint,
\begin{equation}
\overline{ \acute{Y}_L \:\grave{Y}_R} \;=\; \acute{Y}_L \:\grave{Y}_R \:,\spc
\overline{ \acute{Y}_R \:\grave{Y}_L} \;=\; \acute{Y}_R \:\grave{Y}_L \:.
\label{e:31c}
\end{equation}
Applying the relations~(\ref{e:31b}) and~(\ref{e:31c}), a straightforward
calculation gives
\begin{eqnarray*}
\lefteqn{ \overline{\lambda_{\uparrow L -}} \:\Lambda +
\lambda_{\uparrow L -}\: \overline{\Lambda} } \\
&=& \int_x^y dz \;I_\uparrow\:\acute{Y}_L\:\grave{Y}_R \;
 I_\uparrow\: \left( T^{(0)}_{[0]}\:T^{(0)}_{[2]} \: \overline{
T^{(-1)}_{[0]}\:T^{(0)}_{[0]}} \:+\:
T^{(-1)}_{[0]}\:T^{(0)}_{[0]}\: \overline{T^{(0)}_{[0]}\:T^{(0)}_{[2]}}
\right) \\
&&+\int_y^x dz \; I_\uparrow\: \acute{Y}_R\:\grave{Y}_L
\: I_\uparrow\: \left( T^{(-1)}_{[0]}\:T^{(0)}_{[0]}\:
\overline{T^{(-1)}_{[0]}\: T^{(1)}_{[2]}} \:+\:
T^{(-1)}_{[0]}\: T^{(1)}_{[2]}\: \overline{T^{(-1)}_{[0]}\:T^{(0)}_{[0]}}
\right) \\
&&+\overline{\nu} \;I_\uparrow\: \hat{Y}_L\:I_\uparrow \;
T^{(0)}_{[0]} \:T^{(0)}_{[1]}\:\overline{T^{(-1)}_{[0]} \:T^{(0)}_{[1]}}
\:+\: \nu \;I_\uparrow\: \hat{Y}_R\:I_\uparrow \;
T^{(-1)}_{[0]} \:T^{(0)}_{[1]} \:\overline{T^{(0)}_{[0]} \:T^{(0)}_{[1]}} \\
&&-\overline{\nu} \: I_\uparrow\: \acute{Y}_R \:\grave{Y}\:I_\uparrow \;
T^{(-1)}_{[0]}\:T^{(0)}_{[0]} \:\overline{T^{(-1)}_{[0]} \:T^{(1)}_{[2]}}
\:-\: \nu \: \overline{I_\uparrow\: \acute{Y}_R \:\grave{Y}\:I_\uparrow} \;
T^{(-1)}_{[0]} \:T^{(1)}_{[2]} \:\overline{T^{(-1)}_{[0]}\:T^{(0)}_{[0]}} \\
&&+ \left(\nu\: T^{(0)}_{[1]} \:\overline{T^{(0)}_{[0]}} \:-\: \overline{\nu}\:
T^{(0)}_{[0]}\: \overline{T^{(0)}_{[1]}} \right)
\;I_\uparrow \left( \hat{Y}_L\:T^{(0)}_{[1]}
\:\overline{T^{(-1)}_{[0]}} \:-\: \hat{Y}_R\: T^{(-1)}_{[0]}\:
\overline{T^{(0)}_{[1]}} \right) I_\uparrow \\
&=& \int_x^y dz \;I_\uparrow\:\acute{Y}_L\:\grave{Y}_R \;
 I_\uparrow\: \left( T^{(0)}_{[0]}\:T^{(0)}_{[2]} \: \overline{
T^{(-1)}_{[0]}\:T^{(0)}_{[0]}} \:+\:
T^{(-1)}_{[0]}\:T^{(0)}_{[0]}\: \overline{T^{(0)}_{[0]}\:T^{(0)}_{[2]}}
\right) \\
&&+\int_y^x dz \; I_\uparrow\: \acute{Y}_R\:\grave{Y}_L
\: I_\uparrow\: \left( T^{(-1)}_{[0]}\:T^{(0)}_{[0]}\:
\overline{T^{(-1)}_{[0]}\: T^{(1)}_{[2]}} \:+\:
T^{(-1)}_{[0]}\: T^{(1)}_{[2]}\: \overline{T^{(-1)}_{[0]}\:T^{(0)}_{[0]}}
\right) \\
&&+\nu\: \left( I_\uparrow \:\hat{Y}_L\: I_\uparrow
\: T^{(0)}_{[1]}\: T^{(0)}_{[1]}\: \overline{T^{(-1)}_{[0]}\: T^{(0)}_{[0]}}
\:-\: \overline{I_\uparrow\: \acute{Y}_R\:\grave{Y}\:I_\uparrow} \;
T^{(-1)}_{[0]} \:T^{(1)}_{[2]} \:\overline{T^{(-1)}_{[0]}\:T^{(0)}_{[0]}}
\right) \\
&&+\overline{\nu} \left( I_\uparrow \:\hat{Y}_R\: I_\uparrow
\: T^{(-1)}_{[0]}\: T^{(0)}_{[0]}\: \overline{T^{(0)}_{[1]}\: T^{(0)}_{[1]}}
\:-\: I_\uparrow\: \acute{Y}_R\:\grave{Y}\:I_\uparrow \;
T^{(-1)}_{[0]}\:T^{(0)}_{[0]} \:\overline{T^{(-1)}_{[0]} \:T^{(1)}_{[2]}}
\right) \:,
\end{eqnarray*}
where for simplicity the arguments $y$ were omitted.
We substitute this formula into (\ref{e:30l}). Since we do not allow for
additional relations between the basic fractions, we can simplify the
resulting simple fractions only by applying~(\ref{e:3C}). This implies
that~(\ref{e:30l}) is satisfied for suitable $f(x,y)$ if and only if the
following five matrices are multiples of $I_\uparrow$,
\begin{eqnarray}
&&\hspace*{1.85cm}
\int_x^y I_\uparrow\:\acute{Y}_L\:\grave{Y}_R \: I_\uparrow \:,\spc
\int_y^x I_\uparrow\:\acute{Y}_R\:\grave{Y}_L \: I_\uparrow \label{e:32g} \\
&&\left. \begin{array}{clc}
\overline{\nu} \left( I_\uparrow\: \hat{Y}_R\: I_\uparrow -
I_\uparrow\: \acute{Y}_R\:\grave{Y}\: I_\uparrow \right)
&\!\!\!\!\!,\quad& \nu\: I_\uparrow\: \hat{Y}_L\: I_\uparrow
- \overline{\nu}\: I_\uparrow\:\acute{Y}_R\:\grave{Y}\: I_\uparrow\:, \\[.5em]
\hspace*{2.5cm} \nu\: \overline{I_\uparrow\: \acute{Y}_L\: I_\uparrow}
- \overline{\nu}\: I_\uparrow\:\acute{Y}_R\: I_\uparrow \:. \hspace*{-3cm}
\end{array} \right\} \qquad \label{e:32h}
\end{eqnarray}

We can assume that $y \neq x$, because otherwise~(\ref{e:32e})
and~(\ref{e:32f}) follow immediately from~(\ref{e:32d}).
Differentiating~(\ref{e:32g}) with respect to $y$ along the line
${\mathcal{L}}$ gives~(\ref{e:32e}) ($a$ and $b$ are real because
the matrices on the left of~(\ref{e:32e}) are Hermitian).
According to~{{(I)}}, the phase factor $\nu$ can take any value
on the unit circle. Thus in~(\ref{e:32h}) the contributions
involving $\nu$ and $\overline{\nu}$ must separately be multiples
of $I_\uparrow$. This gives the left relation in~(\ref{e:32f}),
and the relation on the right is obtained by taking the adjoint.

The analogous relations for $I_\uparrow$ replaced by $I_\downarrow$ are
derived  in the same way. The replacements~(\ref{e:4rep}) leave the phase
factor $\nu$ unchanged (see~(\ref{e:4f}) and~(\ref{e:4g})).
Thus the EL equation~(\ref{e:30l}) remains valid under~(\ref{e:4rep})
for the same function $f$ only if the parameters $a$, $b$,
and $c$ are unchanged.
\QED

Next we consider the the degeneracies in the limit $y \to x$. In this
case, the formulas of Definition~\ref{def36} simplify in that all phase factors
drop out. We obtain the following result.
\begin{Lemma} \label{lemma46}
Without introducing any relations between the basic fractions
(besides those of Theorem~\ref{thm32}), the dynamical mass matrices must
satisfy the relations
\begin{equation}
I_\uparrow \:\hat{Y}_L\: I_\downarrow \;=\; 0 \;=\;
I_\uparrow \:\hat{Y}_R\: I_\downarrow \:. \label{e:4b}
\end{equation}
\end{Lemma}
{\Proof}
According to the replacement rule~(\ref{e:30f}), it suffices to derive the
second part of~(\ref{e:4b}). We compute the matrix $\Lambda$ modulo
scalar multiples of $I_\uparrow$. Using~(\ref{e:32e}) and~(\ref{e:2qa}),
we obtain
\begin{eqnarray*}
\lefteqn{ \Lambda \;=\;
I_\uparrow\: \hat{Y}_L\:I_\downarrow\:\hat{Y}_L\: I_\uparrow
\;T^{(0)}_{[1]}\:\overline{T^{(0)}_{[1]}} } \\
&& -\frac{1}{\lambda_- - \lambda_+} \:I_\uparrow
\left( \hat{Y}_L\: T^{(0)}_{[1]} \:\overline{T^{(-1)}_{[0]}} -
\hat{Y}_R\: T^{(-1)}_{[0]}\: \overline{T^{(0)}_{[1]}} \right) \\
&&\qquad \times\;
I_\downarrow \left(\hat{Y}_R \: T^{(0)}_{[1]} \:\overline{T^{(0)}_{[0]}} -
\hat{Y}_L\; T^{(0)}_{[0]}\: \overline{T^{(0)}_{[1]}} \right) I_\uparrow \\
&=& \left( I_\uparrow\: \hat{Y}_L\:I_\downarrow\:\hat{Y}_L\: I_\uparrow
+ I_\uparrow\: \hat{Y}_R\:I_\downarrow \:\hat{Y}_R\: I_\uparrow \right)\;
\frac{\lambda_-}{\lambda_- - \lambda_+}\:
T^{(0)}_{[1]} \:\overline{T^{(0)}_{[1]}} \\
&&-I_\uparrow\: \hat{Y}_L\:I_\downarrow\:\hat{Y}_R\: I_\uparrow \:
\:\frac{1}{\lambda_- - \lambda_+} \:T^{(0)}_{[1]}\:T^{(0)}_{[1]}\:
\overline{ T^{(-1)}_{[0]}\: T^{(0)}_{[0]} } \\
&&-I_\uparrow\: \hat{Y}_R\:I_\downarrow\:\hat{Y}_L\: I_\uparrow \:
\frac{1}{\lambda_- - \lambda_+} \:T^{(-1)}_{[0]}\: T^{(0)}_{[0]}\:
\overline{ T^{(0)}_{[1]}\:T^{(0)}_{[1]} } \:.
\end{eqnarray*}

Next we compute the square bracket in~(\ref{e:30l}),
\begin{eqnarray*}
\lefteqn{ \overline{\lambda_{\uparrow L -}} \: \Lambda +
\lambda_{\uparrow L -}\: \overline{\Lambda} } \\
&=&-\frac{I_\uparrow\: \hat{Y}_L\:I_\downarrow\:\hat{Y}_R\: I_\uparrow}{\lambda_- - \lambda_+} \:\left( \lambda_+\:
T^{(0)}_{[1]}\:T^{(0)}_{[1]}\: \overline{ T^{(-1)}_{[0]}\: T^{(0)}_{[0]} }
\:-\: \lambda_-\: T^{(-1)}_{[0]}\: T^{(0)}_{[0]}\:
\overline{T^{(0)}_{[1]}\:T^{(0)}_{[1]}} \right) \\
&&+\frac{I_\uparrow\: \hat{Y}_R\:I_\downarrow\:\hat{Y}_L\: I_\uparrow}{\lambda_- - \lambda_+} \:\left( \lambda_-\:
T^{(0)}_{[1]}\:T^{(0)}_{[1]}\: \overline{ T^{(-1)}_{[0]}\: T^{(0)}_{[0]} }
\:-\: \lambda_+\: T^{(-1)}_{[0]}\: T^{(0)}_{[0]}\:
\overline{T^{(0)}_{[1]}\:T^{(0)}_{[1]}} \right) .
\end{eqnarray*}
Since $I_\downarrow$ projects onto a subspace of dimension $7-p<p$, the
rank of the matrices $I_\uparrow \hat{Y}_R I_\downarrow \hat{Y}_L I_\uparrow$
and $I_\uparrow \hat{Y}_L I_\downarrow \hat{Y}_R I_\uparrow$ is smaller than
$p$, and therefore these matrices cannot be scalar multiples of $I_\uparrow$.
Thus the EL equations have a well-defined continuum limit only if the
factors $(\lambda_- - \lambda_+)^{-1}$ in the above expression drop out.
This is the case only if
\[ I_\uparrow\: \hat{Y}_R\:I_\downarrow\:\hat{Y}_L\: I_\uparrow
\;=\; I_\uparrow\: \hat{Y}_L\:I_\downarrow\:\hat{Y}_R\: I_\uparrow \:. \]
If these necessary conditions are satisfied, the above formula simplifies to
\[ \overline{\lambda_{\uparrow L -}} \: \Lambda +
\lambda_{\uparrow L -}\: \overline{\Lambda}
\;=\; I_\uparrow\: \hat{Y}_L\:I_\downarrow\:\hat{Y}_R\: I_\uparrow \:
\left( T^{(0)}_{[1]}\:T^{(0)}_{[1]}\:
\overline{ T^{(-1)}_{[0]}\: T^{(0)}_{[0]} }
\:+\:T^{(-1)}_{[0]}\: T^{(0)}_{[0]}\:
\overline{T^{(0)}_{[1]}\:T^{(0)}_{[1]}} \right) \:. \]
Now the EL equations have a well-defined continuum limit, and assuming for
the regularization parameters only the relation~(\ref{e:3C}), we conclude
that
\begin{equation}
I_\uparrow\: \hat{Y}_R\:I_\downarrow\:\hat{Y}_L\: I_\uparrow \;=\; 0\:.
\label{e:4a}
\end{equation}
The matrix product in this equation can be written in the form $B B^*$ with
$B \equiv I_\uparrow \hat{Y}_R I_\downarrow$. Hence~(\ref{e:4a})
implies that $B=0$.
\QED

The previous two lemmas simplify considerably the structure of the perturbation on the
degenerate subspaces. Namely, we can write $\Lambda$ in the form
\[ \Lambda \;=\; \rho(\nu, \overline{\nu})\: I_\uparrow
\:-\: I_\uparrow\: \acute{Y}_R(y) \left( I_\uparrow + \mu \nu\:
I_\downarrow + \mu_8 \nu_8\: I_8 \right)\: \grave{Y}_L(x)\: I_\uparrow\;
T^{(-1)}_{[0]}\: \overline{T^{(1)}_{[2]}}\:, \]
where $\rho$ is a complex function which is invariant under the
replacements~(\ref{e:4rep}). A short calculation yields
\begin{eqnarray}
\lefteqn{ \overline{\lambda_{\uparrow L -}}\: \Lambda
+ \lambda_{\uparrow L -}\: \overline{\Lambda}
\;=\; (a + \nu \:b + \overline{\nu} \:\overline{b})\: I_\uparrow } \label{e:4AA} \\
&&- \overline{\nu}\;
I_\uparrow\: \acute{Y}_R(y) \:I_\uparrow\: \grave{Y}_L(x) \:I_\uparrow\;
\overline{N} \:-\: \nu\;
\overline{I_\uparrow\: \acute{Y}_R(y) \:I_\uparrow\: \grave{Y}_L(x) \:I_\uparrow}\:
N \label{e:4A} \\
&&-\mu\; I_\uparrow\: \acute{Y}_R(y) \:I_\downarrow\: \grave{Y}_L(x) \:I_\uparrow
\; \overline{N} \:-\: \overline{\mu}\;
\overline{I_\uparrow\: \acute{Y}_R(y) \:I_\downarrow\: \grave{Y}_L(x) \:I_\uparrow}
\; N \label{e:4B} \\
&&- \mu_8\: \overline{\nu} \nu_8 \;
I_\uparrow\: \acute{Y}_R(y) \:I_8\: \grave{Y}_L(x) \:I_\uparrow
\; \overline{N} \:-\:
\overline{\mu_8}\: \nu \overline{\nu_8}\;
\overline{I_\uparrow\: \acute{Y}_R(y) \:I_8\: \grave{Y}_L(x) \:I_\uparrow}
\; N \:, \spc \label{e:4C}
\end{eqnarray}
where the complex functions $a$ and $b$ are invariant
under~(\ref{e:4rep}), and $N$ is the monomial
\begin{equation}
N \;=\; T^{(-1)}_{[0]}\: T^{(1)}_{[2]}\: \overline{T^{(-1)}_{[0]}\:
T^{(0)}_{[0]}}\:.    \label{e:4H}
\end{equation}
We split up the analysis of the EL equations corresponding
to~(\ref{e:4AA}--\ref{e:4C}) into several lemmas.  We say that the
summands~(\ref{e:4B}) or~(\ref{e:4C}) are {\em{non-trivial}} if there are
admissible dynamical mass matrices such that this summand or one of the
expressions obtained by applying the replacements~(\ref{e:30f})
and/or~(\ref{e:30g}) are non-zero.  Furthermore, we refer to two phase
functions $\alpha, \beta \in S^1$ as being {\em{independent}} if for $\alpha$
fixed, $\beta$ can take any value in $S^1$ and vice versa\footnote{{\textsf{Online version}:}
The arguments in the following lemmas need to be modified if we allow for
a local chiral transformation as considered in the book~\cite{cfs}
(listed in the references in the preface to the second online edition).}

\begin{Lemma} \label{lemma4A}
Under the assumptions~{{(I)}}--{{(III)}}, $\nu$ is independent of the
phase functions $\mu$, $\mu_8$, $\nu_8$, and $\overline{\mu} \mu_8$.
The term~(\ref{e:4C}) is non-trivial.
\end{Lemma}
{\Proof} Suppose that the dynamical mass matrices were zero in the neutrino
sector, i.e.
\begin{equation}
Y_L\: I_8 \;\equiv\; 0 \;\equiv\; Y_R\: I_8    \label{e:41}
\end{equation}
Then the Dirac operator, and thus also the fermionic projector, would be
invariant on the neutrino sector. As a consequence, the chiral Dirac particles
would drop out of all composite expressions due to chiral cancellations, in
contradiction to~{{(II)}}. We conclude that~(\ref{e:41}) is false. Since we
are free to choose a convenient representation of the dynamical mass matrices,
we can assume that the matrices
\begin{equation}
\acute{Y}_L\: I_8 \:,\;\;\;\;\; \acute{Y}_R\: I_8 \:,\;\;\;\;\;
\grave{Y}_L\: I_8 \:,\;\;\;\;\; \grave{Y}_R\: I_8 \label{e:4obs}
\end{equation}
do not all vanish identically. The contributions to the fermionic projector
which involve the matrix products $I_8 \acute{Y}_{L\!/\!R}$ or
$\grave{Y}_{L\!/\!R} I_8$ enter only the perturbation calculation for the
kernel of $P(x,y)\: P(y,x)$, and according to Theorem~\ref{thmB2} they drop out
of the EL equations. Thus~{{(II)}} is satisfied only if
\[ (I_\uparrow + I_\downarrow)\: \acute{Y}_L\: I_8 \;\not \equiv\; 0
\spc {\mbox{or}} \spc
(I_\uparrow + I_\downarrow)\: \acute{Y}_R\: I_8 \;\not \equiv\; 0 \:. \]
This shows that~(\ref{e:4C}) is non-trivial.

According to~{{(III)}}, there is a free dynamical gauge field which couples
differently to the Dirac particles in the sectors $n=1$ and $n=8$. The
corresponding free gauge potentials describe relative phase shifts of the
fermionic projector on~${\mbox{Im}}\: I_1$ and~${\mbox{Im}}\: I_8$.
These relative phases are captured by $\mu_8$ and $\mu_8 \nu_8$ (see~(\ref{e:4g})).
Since the free gauge potentials on the line segment $\overline{xy}$ can be
chosen arbitrarily, it follows that $\nu$ is independent of $\mu_8$ and $\mu_8
\nu_8$. A similar argument for $I_7$ instead of $I_1$ shows that $\nu$ and
$\overline{\mu} \mu_8$ are independent.
$\;{\mbox{ }}$ \QED

\begin{Lemma} \label{lemma4B}
Imposing at most one additional relation between the basic fractions (besides
those of Theorem~\ref{thm32}), $\nu$ and $\mu$ are independent. The
term~(\ref{e:4B}) is non-trivial.
\end{Lemma}
{\Proof}
Assume to the contrary that $\nu$ and $\mu$ are dependent or that~(\ref{e:4B})
is trivial. Then the phases in~(\ref{e:4AA}--\ref{e:4B}) are all dependent
on $\nu$. The independence of the phases established in
Lemma~\ref{lemma4A} yields that the EL equations must be satisfied separately
for~(\ref{e:4C}). Imposing at most one additional relation between
the basic fractions, we cannot arrange that~(\ref{e:4C}) drops out of the EL
equations. We thus obtain that for a suitable complex $\kappa$,
\begin{equation}
I_\uparrow \:\acute{Y}_R(y) \:I_8\: \grave{Y}_L(x) \:I_\uparrow \;=\;
\kappa(x,y)\: I_\uparrow \:,    \label{e:42}
\end{equation}
and this condition must also be satisfied after the replacements~(\ref{e:30f})
and/or (\ref{e:30g}) for the same $\kappa$. Since the rank of $I_8$ is smaller
than that of $I_\uparrow$, the lhs of~(\ref{e:42}) is a singular matrix, and
thus $\kappa$ vanishes identically. This implies that the lhs of~(\ref{e:42})
is trivial (i.e. vanishes also after the replacements~(\ref{e:30f}, \ref{e:30g})),
in contradiction to Lemma~\ref{lemma4A}.
\QED

Having established that the phases in~(\ref{e:4AA}) and~(\ref{e:4A}) are
independent of those in~(\ref{e:4B}) and~(\ref{e:4C}), we can now apply the
uniform splitting lemma to~(\ref{e:4A}).
\begin{Lemma} \label{lemma4C}
Imposing at most one additional relation between the basic fractions, we can
arrange by a constant unitary transformation that for all $a,b=1,\ldots,p$ and
$c,d=p+1,\ldots,7$,
\begin{equation}
(I_\uparrow\: \grave{Y}_{L\!/\!R}\: I_\uparrow)^{(a \alpha)}_b \;=\; \delta^a_b\:
u^\alpha_{L\!/\!R} \:,\spc
(I_\downarrow\: \grave{Y}_{R\!/\!L}\: I_\downarrow)^{(c \alpha)}_d \;=\; \delta^c_d\:
u^\alpha_{L\!/\!R} \label{e:4D}
\end{equation}
with $u_{L\!/\!R}(x) \in \C^3$.
\end{Lemma}
{\Proof} It clearly suffices to consider one chirality. Since $\nu$ is
independent of $\mu$ and $\mu_8 \nu_8$, the EL equations imply that
\begin{equation}
I_\uparrow \:\acute{Y}_R(y)\: I_\uparrow \:\grave{Y}_L(x)\: I_\uparrow \;=\;
\lambda(x,y)\: I_\uparrow \:.    \label{e:4F}
\end{equation}
The dynamical mass matrices can be chosen independently at $x$ and
$y$. Denoting the class of admissible matrices $I_\uparrow
\grave{Y}_L I_\uparrow$ by ${\mathcal{B}}$, we are in the setting
of Lemma~\ref{lemma40} with $p_1=p$ and $p_2=3p$. Since $p_2$ is
divisible by $p_1$, we can, possibly after increasing $r$, assume
that $p_2 - r p_1 =0$, and thus
\[ I_\uparrow \:\grave{Y}_L\: I_\uparrow \;=\; U\: (\underbrace{u_L \oplus
\cdots \oplus u_L}_{\mbox{\scriptsize{$p$ summands}}}) \]
with $u_L(x) \in \C^3$. Omitting the constant unitary transformation and writing
out the components, this is just the lhs of~(\ref{e:4D}).
Under the replacement~(\ref{e:4rep}), $\nu$ as well as $\alpha$ and $\beta$ are
unchanged. As a consequence, also the function $\lambda$ in~(\ref{e:4F}) is
invariant under~(\ref{e:4rep}), and this implies that the mappings $\pi_i$ of
Lemma~\ref{lemma40} obtained for $B=I_\uparrow \grave{Y}_L I_\uparrow$ and $B =
I_\downarrow \grave{Y}_R I_\downarrow$ are all unitarily equivalent. This
proves the rightmost equation of~(\ref{e:4D}).
\QED

It remains to analyze~(\ref{e:4B}) and~(\ref{e:4C}).
\begin{Lemma} \label{lemma4D}
The EL equations to degree 4 can be satisfied only if we impose at least one
additional relation between the basic fractions.
\end{Lemma}
{\Proof}
In the limit $y \to x$, the matrices
$I_\uparrow \acute{Y}_R(y) \:I_.\: \grave{Y}_L(x)\: I_\uparrow$ can be written
in the form $B^* B$ with $B=I_.\: \grave{Y}_L \: I_\uparrow$ and are therefore
Hermitian and positive semidefinite.
This shows that~(\ref{e:4B}) and~(\ref{e:4C}) cannot
cancel each other identically. According to Lemma~\ref{lemma4B}, (\ref{e:4B})
is non-trivial. It suffices to consider the case that~(\ref{e:4B}) does not
vanish identically (in the other cases when~(\ref{e:4B}) is non-zero after
applying~(\ref{e:30f}, \ref{e:30g}) the argument is analogous).
Then we can arrange a contribution to~(\ref{e:4A}--\ref{e:4C})
of the form $(\mu A N + \overline{\mu A N})$ with a matrix $A \neq 0$. The same
contribution must be present after performing the replacements~(\ref{e:4rep}).
Since these replacements transform $\mu$ into $\overline{\mu}$
(see~(\ref{e:30f}) and~(\ref{e:30g})), we obtain a condition of the form
\begin{equation}
\mu A N + \overline{\mu A N} \;=\; \overline{\mu} B N + \mu \overline{B N}
\spc {\mbox{for all $\mu \in S^1$}}    \label{e:4I}
\end{equation}
with $B$ a matrix.
Without introducing an additional relation between the basic fractions, we must
treat $N$ and $\overline{N}$ as being independent, and thus~(\ref{e:4I}) has no
solution.
{\mbox{$\;\;\;\hfill$}} \QED
Using that $A$ and $B$ go over to positive matrices as $y \to x$, one
sees that in order to arrange that~(\ref{e:4I}) has a solution, we need to
impose that $N$ and $\overline{N}$ coincide in the EL equations,
i.e.\footnote{{\textsf{Online version}:}
For a difficulty to realize this relation between the basic fractions
by a suitable regularization see the proof of Lemma~3.10.3
in the book~\cite{cfs} (listed in the references in the preface to the second online edition).
One should keep in mind that, following the consideration after~\cite[eq.~(3.7.13)]{cfs},
the relation must still hold if we replace the factors~$T^{(1)}_{[2]}$ and~$\overline{T^{(1)}_{[2]}}$
by a non-zero real constant.}
\begin{equation}
(N - \overline{N})\: \left(\overline{T^{(0)}_{[0]}}\right)^{-1}
\;=\; 0 \:.   \label{e:4rel}
\end{equation}
The next lemma is again an application of the uniform splitting lemma and uses
the non-degeneracy assumption of Def.~\ref{def40}.
\begin{Lemma} \label{lemma402}
Suppose that the basic fractions satisfy (in addition to the conditions of
Theorem~\ref{thm32}) the relation~(\ref{e:4rel}) with $N$ according
to~(\ref{e:4H}).  Then the parameter $p$ in~(\ref{e:3I}) is equal to $4$.
The phase factors in the neutrino sector are determined by
\begin{equation}
\nu_8 \;=\; \nu \spc {\mbox{and}}\spc \mu_8 \;=\; \mu {\mbox{ or }} \overline{\mu}\:.
    \label{e:4N}
\end{equation}
We can arrange by constant unitary transformations that for $a, b=1,2,3$,
\begin{equation}
(I_\downarrow\: \grave{Y}_{L\!/\!R} \: I_\uparrow)^{(a+4\: \alpha)}_b
\;=\; \delta^a_b\: v^\alpha_{L\!/\!R} \:,\spc
(I_\uparrow\: \grave{Y}_{R\!/\!L}\: I_\downarrow)^{(a \alpha)}_{b+4}
\;=\; \delta^a_b\: \overline{v^\alpha_{L\!/\!R}} \label{e:4E}
\end{equation}
with $v_{L\!/\!R}(x) \in \C^3$. In the two cases for $\mu_8$ in~(\ref{e:4N}),
\begin{equation}
(I_8\: \grave{Y}_{L\!/\!R}\: I_\uparrow)^{(8 \alpha)}_4 \;=\;
v^\alpha_{L\!/\!R} {\mbox{ or }} \overline{v^\alpha_{L\!/\!R}} \:,
\label{e:48a}
\end{equation}
respectively. Furthermore,
\begin{equation}
I_8\: \grave{Y}_{R\!/\!L}\: I_\downarrow \;=\; 0\:.     \label{e:4R}
\end{equation}
\end{Lemma}
{\Proof}
Imposing~(\ref{e:4rel}) and using~(\ref{e:4D}), the EL
equations~(\ref{e:30l}) reduce to the conditions
\begin{eqnarray}
\lefteqn{ \lambda(x,y)\: I_\uparrow \;=\;
\mu \: I_\uparrow\: \acute{Y}_R(y)\: I_\downarrow\: \grave{Y}_L(x)\: I_\uparrow
\:+\: \overline{\mu \: I_\uparrow\: \acute{Y}_R(y)\: I_\downarrow\: \grave{Y}_L(x) \: I_\uparrow} } \nonumber \\
&&+ \mu_8\: \overline{\nu} \nu_8\: I_\uparrow\: \acute{Y}_R(y)\: I_8\:
\grave{Y}_L(x)\: I_\uparrow \:+\:
\overline{\mu_8}\: \nu \overline{\nu_8}\:
\overline{I_\uparrow\: \acute{Y}_R(y)\: I_8\:\grave{Y}_L(x)\:
I_\uparrow} \:.\quad \label{e:4M}
\end{eqnarray}
We first prove that the phase factors must be dependent in the sense that
\begin{equation}
\mu_8\: \overline{\nu} \nu_8 \;=\; \mu {\mbox{ or }} \overline{\mu}\:.
\label{e:4chir}
\end{equation}
Assuming the contrary, we must treat the four summands in~(\ref{e:4M}) as being
independent, and thus
\begin{equation}
I_\uparrow\: \acute{Y}_R(y)\: I_\downarrow\: \grave{Y}_L(x)\: I_\uparrow \;=\;
\kappa(x,y)\: I_\uparrow\:.    \label{e:4J}
\end{equation}
Performing the replacement~(\ref{e:4rep}) and using that $\mu$ transforms
to $\overline{\mu}$, we obtain furthermore that
\begin{equation}
I_\downarrow\: \acute{Y}_L(y)\: I_\uparrow\: \grave{Y}_R(x)\: I_\downarrow
\;=\; \kappa(x,y)\: I_\downarrow    \label{e:4K}
\end{equation}
with the same $\kappa$ as in~(\ref{e:4J}). We apply Lemma~\ref{lemma40}
to~(\ref{e:4J}) (with $p_1=p$ and $p_2=3(7-p)$) and to~(\ref{e:4K}) (with
$p_1=7-p$ and $p_2=p$). Leaving out the constant unitary transformations, we
obtain the representations
\begin{equation}
I_\downarrow\: \grave{Y}_L\: I_\uparrow \;=\; \left( \begin{array}{cc}
\overbrace{b \oplus \cdots \oplus b}^{\mbox{\tiny{$p$ summands}}} \\
0 \end{array} \right) \:,\spc
I_\uparrow\: \grave{Y}_R\: I_\downarrow \;=\; \left( \begin{array}{cc}
\overbrace{\overline{b} \oplus \cdots \oplus \overline{b}
}^{\mbox{\tiny{$7-p$ summands}}} \\
0 \end{array} \right)\:,    \label{e:4L}
\end{equation}
where $\overline{b}$ is the complex conjugate of the vector $b \in \C^3$.
According to Lemma~\ref{lemma4B}, (\ref{e:4B}) is non-trivial. Since the
contributions to the EL equations involving $\mu$ are unchanged when
applying the replacements~(\ref{e:30f}) and~(\ref{e:30g}), we can arrange
that~(\ref{e:4J}) does not vanish, and thus $b \neq 0$. On the lhs
of~(\ref{e:4L}), the inequality $r p_1 \leq p_2$ implies that $r<3$. Thus on
the rhs of~(\ref{e:4L}), the number of zero rows is $3p-r(7-p)>3$. Therefore,
$I_p \grave{Y}_R I_\downarrow=0$, or, equivalently, by taking the adjoint and
in components,
\[ (\acute{Y}_L)^a_{(d.)} \;=\; 0 \spc {\mbox{for $d=p$ and
$a=p+1,\ldots,7$}}. \]
On the other hand, the lhs of~(\ref{e:4L}) implies that
\[ (\grave{Y}_L)^a_{(d.)} \;\neq\; 0 \spc {\mbox{for $d=p$ and
$a=p+1,\ldots,7$}}. \]
The non-degeneracy assumption of Def.~\ref{def40} allows us to conclude that
\[ (\acute{Y}_L)^a_{(d.)} \;=\; 0 \spc {\mbox{for all $a=p+1,\ldots,7$ and
$d \neq a$.}} \]
This implies that $I_\downarrow \acute{Y}_L I_\uparrow = 0$, in contradiction
to the rhs of~(\ref{e:4L}) and the fact that $b \neq 0$.

Repeating the above argument for the opposite chirality gives in analogy
to (\ref{e:4chir}) that
\begin{equation}
\mu_8 \;=\; \mu {\mbox{ or }} \overline{\mu}\:.    \label{e:4chir2}
\end{equation}
Using that $\mu$ and $\nu$ are independent according to Lemma~\ref{lemma4B},
(\ref{e:4chir}) and~(\ref{e:4chir2}) are equivalent to~(\ref{e:4N}).

In the case $\mu_8=\mu$, the EL equations~(\ref{e:30l}) reduce to the
conditions
\begin{equation}
I_\uparrow \:\acute{Y}_R(y)\: (I_\downarrow + I_8)\: \grave{Y}_L(x)\:
I_\uparrow \;=\; \kappa(x,y)\: I_\uparrow\:.   \label{e:4P}
\end{equation}
After the replacement~(\ref{e:4rep}), the phase factors in~(\ref{e:4M})
are no longer dependent~(cf.~(\ref{e:30f}) and~(\ref{e:30g})), and thus
we get the conditions
\begin{eqnarray}
I_\downarrow\: \acute{Y}_L(y)\: I_\uparrow\: \grave{Y}_R(x)\: I_\downarrow &=&
\overline{\kappa(x,y)}\: I_\downarrow \label{e:4Q} \\
I_\downarrow\: \acute{Y}_L(y)\: I_8\: \grave{Y}_R(x)\: I_\downarrow &=& 0\:.
\label{e:4QQ}
\end{eqnarray}
The last relation implies~(\ref{e:4R}). Applying the above argument
for~(\ref{e:4J}) and (\ref{e:4K}) to~(\ref{e:4P}) and~(\ref{e:4Q}), we
again get a contradiction unless ${\mbox{Rg}}\: I_\uparrow = {\mbox{Rg}}\:
(I_\downarrow + I_8)$. This shows that $p=4$. Possibly after increasing $r$, we
obtain in analogy to~(\ref{e:4L}) the representations
\begin{equation}
(I_\downarrow + I_8)\: \grave{Y}_L\: I_\uparrow \;=\; b \oplus b \oplus b \oplus b
\:,\spc I_\uparrow \:\grave{Y}_R\: I_\downarrow \;=\; \overline{b}  \oplus
\overline{b} \oplus \overline{b} \:. \label{e:40z}
\end{equation}
Writing these relations in components gives~(\ref{e:4E}, \ref{e:48a}).

In the case $\mu_8 = \overline{\mu}$, we obtain in analogy to~(\ref{e:4P}) the
condition
\begin{equation}
I_\uparrow \:\acute{Y}_R(y)\: I_\downarrow\: \grave{Y}_L(x)\:
I_\uparrow \:+\: \overline{I_\uparrow \:\acute{Y}_R(y)\: I_8\:
\grave{Y}_L(x)\: I_\uparrow}
\;=\; \kappa(x,y)\: I_\uparrow\:,   \label{e:49z}
\end{equation}
and after the replacement~(\ref{e:4rep}) again the conditions~(\ref{e:4Q},
\ref{e:4QQ}). The lhs of~(\ref{e:4Q}) can be split into a product of
matrices of the form $A(y)\: B(x)$. Since the equations~(\ref{e:4Q}) and~(\ref{e:49z}) involve the same function $\kappa(x,y)$, the matrices
on the lhs of~(\ref{e:49z})
must split in the same way. To this end, the matrix $I_\uparrow\:
\acute{Y}_R(y) \:I_8\: \grave{Y}_L(x) \:I_\uparrow$ must (possibly after a
constant unitary transformation) be diagonal for all $x$ and $y$, so that
the spectral adjoint reduces to the complex conjugate (i.e.\ to taking the
complex conjugate of all matrix entries). After taking this complex conjugate,
we can proceed exactly as in the case $\mu_8 = \mu$ above. The only
difference is that we obtain a representation not for the matrix
$I_8 \grave{Y}_L I_\uparrow$ but for its complex conjugate, and this leads
to the complex conjugate in~(\ref{e:48a}).
\QED
We remark that the fact that the partial trace is
non-cyclic, (\ref{e:4nc}), is essential for the above construction to work.
Namely, according to the lhs of~(\ref{e:4E}),
\begin{equation}
I_\uparrow\: \acute{Y}_L\: I_8\: \grave{Y}_R\: I_\uparrow \;
\stackrel{\mbox{\tiny{in general}}}{\neq}\; 0\:.    \label{e:4R1}
\end{equation}
On the other hand, the weak causality compatibility condition,
Def.~\ref{defwccc}, implies that
\begin{equation}
I_8\: \acute{Y}_R\: I_\uparrow\: \grave{Y}_L\: I_8 \;=\;
X_R\: I_8\: \acute{Y}_R\: I_\uparrow\: \grave{Y}_L\: I_8 \;=\; 0\:.
    \label{e:4S}
\end{equation}
If the partial trace were cyclic, (\ref{e:4R1}) and~(\ref{e:4S}) would be
inconsistent.

Combining the previous lemmas and choosing a convenient representation for the
dynamical mass matrices gives the main result of this section.
\begin{Thm} {\bf{(spontaneous block formation)}}
  \label{thm41}\index{spontaneous block formation}
We consider the EL equations corresponding to the Lagrangian~(\ref{e:2P}) in
the presence of chiral and scalar potentials~(\ref{e:3a}--\ref{e:3dir}) to
the degree 4 on the light cone. We assume that the Dirac operator is weakly
causality compatible and that the dynamical mass matrices are non-degenerate
(see Defs.~\ref{defwccc} and~\ref{def40}). Then, following~{{(IV)}},
we need to introduce two relations between the basic fractions. Imposing that
\begin{equation}
(M - \overline{M})\: \overline{T^{(0)}_{[0]}}^{-1}
\;=\; 0 \;=\; (N - \overline{N})\: \overline{T^{(0)}_{[0]}}^{-1} \label{e:4rmon}
\end{equation}
with $M$ and $N$ according to~(\ref{e:3B}, \ref{e:4H}), we can arrange by
constant unitary transformations that the Dirac operator is of the following
form,
\begin{eqnarray}
i \Pdd \!\!\!\!\!\!\!\!\!&& -\:m\: \chi_L \left(
Y^q_R \oplus Y^q_R \oplus Y^q_R \oplus Y^l_R \right)
\:-\: m\: \chi_R \left(Y^q_L \oplus Y^q_L \oplus Y^q_L \oplus Y^l_L \right)
\label{e:471} \\
&&+\: (\chi_R\: \Aslsh_L + \Aslsh_V) \left(
\sigma^3 \oplus \sigma^3 \oplus \sigma^3 \oplus \sigma^3 \right)
\label{e:47a} \\
&&+\: (\Aslsh^q\: \1) \oplus 0_{\sC^2} \:+\: 0_{\sC^6} \oplus (\Aslsh^l\:\1
+ \Aslsh^s\: \sigma^3) \:.
\end{eqnarray}
Here $Y^{q \!/\!l}_{L\!/\!R}$ are $2 \times 2$ matrices on the sectors which
depend also on the generations, i.e.\ in components
\[ Y^{q / l}_c \;=\; (Y^{q \!/\!l}_c)^{(a \alpha)}_{(b \beta)} \spc
{\mbox{with $a,b=1,2$, $\alpha, \beta = 1,2,3, c=L\!/\!R$.}} \]
The chiral and vector potentials are trivial on the generations and depend only
on the sector index.  $A_L$, $A_V$ and $A^l$ are vector fields, and $A^q$ is a
$3 \times 3$ matrix potential ($\1$ and $\sigma^3$ are Pauli matrices).
The vector field $A^s$ is a function of $A_L$ and $A_R$; the two possible choices
are
\begin{equation}
A^s \;\equiv\; 0 \spc {\mbox{or}} \spc A^s \;\equiv\; -A_L - 2 A_V\:.
    \label{e:4cam}
\end{equation}
The dynamical gauge groups (see Def.~\ref{def1}) are given by
\begin{equation}
{\mathcal{G}} \;=\; U(1)_L \times {\mathcal{F}} \:,\spc
{\mathcal{F}} \;=\; U(1)_V \times U(3)^q \times U(1)^l \:,
\label{e:B3}
\end{equation}
where the indices clarify to which potentials in the Dirac operator the groups
correspond.
\end{Thm}
{\Proof} Lemmas~\ref{lemma34} and~\ref{lemma46} do not immediately apply
here because they are based on the assumption that we have only one relation
between the basic fractions. But it is straightforward to check that if in
these lemmas we allowed for an additional relation between the basic fractions,
the argument of Lemma~\ref{lemma4D} would still go through, thus making it
necessary to introduce a third relation between the basic fractions.

Collecting the results of Lemmas~\ref{lemma34}--\ref{lemma402} and
choosing a convenient representation for the dynamical mass
matrices, we obtain that the dynamical mass matrices are block
diagonal as in~(\ref{e:471}). Thus it remains to derive the
dynamical gauge group and the form of the corresponding gauge
potentials. Possibly after reordering the sectors, the $U(1)_L$ is
precisely the group ${\mathcal{B}}_4$ in Def.~\ref{def32}. The
free gauge group is obtained by taking the maximal subgroup of
${\mathcal{F}}_4$ for which the gauge potentials respect the phase
conditions~(\ref{e:4N}). In the two cases in~(\ref{e:4cam}), the
$U(1)_V$ shifts the phases of $\mu$ and $\mu_8$ by the same or the
opposite amount, respectively. This corresponds to the two cases
in~(\ref{e:4N}). The other potentials must leave the phase
functions unchanged, and thus they must coincide on the
sectors which are mapped into each other by the dynamical mass
matrices~(\ref{e:4E}). This gives the group $U(3)^q \times
U(1)^l$. \QED We point out that, except for the potentials $A^q$,
the Dirac operator splits into four direct summands. The first
three summands are identical and involve massive Dirac particles,
whereas the chiral Dirac particles are contained in the last
summand. The gauge potentials $A^q$ describe an interaction
between the Dirac particles in the three identical summands. In
analogy to the standard model, it is natural to identify the
fermions in the first three and the last summands with the quarks
and leptons, respectively. In order to make these notions precise,
we first observe that for the fermionic projector, the above
splitting means that for all contributions considered so
far\footnote{We remark for clarity that that the contributions to
the fermionic projector which involve the $U(3)^q$ gauge fields or
currents, which have not been considered so far, do not split in
the form~(\ref{e:44}).}, \label{P^q} \label{P^l}
\begin{equation}
P(x,y) \;=\; U(x,y) \left( P^q \oplus P^q \oplus P^q \oplus P^l \right) ,
\label{e:44}
\end{equation}
where $U$ is the generalized phase transformation by the potentials $A^q$,
\begin{equation}
U(x,y) \;=\; \Pexp \left( -i \int_0^1 d\tau\; A^q_j(\tau y + (1-\tau)x)\:
(y-x)^j \right) .    \label{e:45}
\end{equation}
The unitary transformation~(\ref{e:45}) clearly commutes with the direct sum
and thus drops out of the closed chain,
\begin{equation}
A_{xy} \;=\; A^q_{xy} \oplus A^q_{xy} \oplus A^q_{xy} \oplus A^l_{xy}
\spc{\mbox{with}}\spc A^{q/l}_{xy} \;\equiv\; P^{q/l}(x,y)\:
P^{q/l}(y,x)\:.    \label{e:46}
\end{equation}
\begin{Def}
The first three direct summands in~(\ref{e:44}) and~(\ref{e:46}) are referred
to as the {\bf{quark blocks}}\index{quark block}. The last direct
summand is the {\bf{lepton block}}\index{lepton block}.
\end{Def}

\section{The Dynamical Mass Matrices in the Quark and Neutrino Blocks}
\setcounter{equation}{0} \label{esec43}
We now specify the dynamical mass matrices in the
quark and neutrino blocks.
\begin{Thm} \label{thm42}
Under the assumptions of Theorem~\ref{thm41}, the EL equations are satisfied
to degree 4 on the light cone if and only if the matrices $Y^q_{L\!/\!R}$
and $Y^l_{L\!/\!R}$ in~(\ref{e:471}) have (after suitable constant unitary
transformations) the following properties at all space-time points,
\begin{eqnarray}
\hat{Y}^q_L &=& (\hat{Y}^q_R)^* \;=\; \left( \begin{array}{cc} c & 0 \\
0 & \overline{c} \end{array} \right) \label{e:4f0} \\
\hat{Y}^l_L &=& (\hat{Y}^l_R)^* \;=\; \left( \begin{array}{cc} c & 0 \\
0 & 0 \end{array} \right) \label{e:4f00} \\
\grave{Y}^q_L &=& \left( \begin{array}{cc} a & {\mathcal{V}}_L \:\overline{b} \\
{\mathcal{U}}_L \:b & \overline{a} \end{array} \right) \:,\spc
\grave{Y}^q_R \;=\; \left( \begin{array}{cc} \overline{a} &
{\mathcal{V}}_R \:\overline{b} \\
{\mathcal{U}}_R \:b & a \end{array} \right) \label{e:4f1}
\end{eqnarray}
and in the two cases in~(\ref{e:4cam}),
\begin{equation}
\grave{Y}^l_L \;=\; \left( \begin{array}{cc} a & 0 \\
{\mathcal{W}}_L\:b & 0
\end{array} \right) \:,\spc
\grave{Y}^l_R \;=\; \left( \begin{array}{cc} \overline{a} & * \\
{\mathcal{W}}_R\:b & *
\end{array} \right) \label{e:4f2}
\end{equation}
and
\begin{equation}
\grave{Y}^l_L \;=\; \left( \begin{array}{cc} a & 0 \\
{\mathcal{W}}_L\:\overline{b} & 0
\end{array} \right) \:,\spc
\grave{Y}^l_R \;=\; \left( \begin{array}{cc} \overline{a} & * \\
{\mathcal{W}}_R\: \overline{b} & * \end{array} \right) \:,
\label{e:4f3}
\end{equation}
respectively. Here we use a matrix notation in the sector index.
In~(\ref{e:4f1}--\ref{e:4f3}), the matrix entries are vectors in
$\C^3$ (and this takes into account the dependence on the
generations). The parameter $c$ is complex, $a,b \in \C^3$, and
the stars stand for any vectors in $\C^3$. The off-diagonal
elements are non-trivial in the sense that there is a space-time
point where $b \neq 0$. The matrices ${\mathcal{U}}_{L\!/\!R}$,
${\mathcal{V}}_{L\!/\!R}$, ${\mathcal{W}}_{L\!/\!R} \in U(3)$ are
constant unitary transformations.
\end{Thm}
{\Proof}
We only consider the first case in~(\ref{e:4cam}); the second is obtained
in the same way keeping track of the complex conjugates.
The weak causality condition of Def.~\ref{defwccc} implies that $\grave{Y}_L
I_8=0$. On the other hand, we already observed after~(\ref{e:4obs}) that
the matrix product $\grave{Y}_R I_8$ enters only the perturbation
calculation for the kernel, which is trivial according to
Theorem~\ref{thmB2}. This explains the zeros and stars in~(\ref{e:4f00},
\ref{e:4f2}). Then~(\ref{e:4f0}, \ref{e:4f00}) follow
immediately from Lemma~\ref{lemma34} and Lemma~\ref{lemma46}. A short
calculation using~(\ref{e:4f0}, \ref{e:4rmon}, \ref{e:4N}) yields that the EL equations to degree 4 reduce to
the conditions
\begin{eqnarray}
\int_x^z dz\; I_\uparrow \:\acute{Y}_L \grave{Y}_R\: I_\uparrow
&=& \alpha(x,y)\: I_\uparrow \label{e:4C1} \\
\overline{\nu}\: I_\uparrow \:\acute{Y}_R(y)\: I_\uparrow\:
\grave{Y}_L(x)\: I_\uparrow \:+\: \nu\:
\overline{I_\uparrow \:\acute{Y}_R(y)\: I_\uparrow\:
\grave{Y}_L(x)\: I_\uparrow} &=& \beta(x,y)\: I_\uparrow \label{e:4C2} \\
\mu\: I_\uparrow\: \acute{Y}_R(y)\: (I_\downarrow+I_8)\:
\grave{Y}_L(x)\: I_\uparrow \nonumber \\
\:+\:\overline{\mu}\: \overline{I_\uparrow\: \acute{Y}_R(y)\:
(I_\downarrow+I_8)\: \grave{Y}_L(x)\: I_\uparrow}
&=& \gamma(x,y)\: I_\uparrow \label{e:4C3}
\end{eqnarray}
as well as to the conditions obtained by the replacements
\begin{equation}
L \:\longleftrightarrow\: R \:,\spc \nu \:\longleftrightarrow\:
\overline{\nu} \label{e:4C4}
\end{equation}
and/or
\begin{equation}
I_\uparrow + I_8 \:\longrightarrow\: I_\downarrow \:,\spc
I_\downarrow \:\longrightarrow\: I_\uparrow \:,\spc
\mu \:\longleftrightarrow\: \overline{\mu} \:,\spc
\nu \:\longleftrightarrow\: \overline{\nu} \label{e:4C5}
\end{equation}
with the complex functions $\alpha, \beta$, and $\gamma$ unchanged. We first
substitute~(\ref{e:4D}) into (\ref{e:4C2}). Comparing with the relation
obtained by applying~(\ref{e:4C4}), one sees that $\langle u_L, u_L
\rangle = \overline{ \langle u_R, u_R \rangle}$, and thus we can arrange
with a constant unitary transformation that $u_L=\overline{u_R}$. This
explains the diagonal entries in~(\ref{e:4f1}, \ref{e:4f2}).
Substituting~(\ref{e:4E}) and~(\ref{e:48a}) into~(\ref{e:4C3}) and
comparing with the relation obtained by applying~(\ref{e:4C5}), we obtain
similarly that $\langle u_R, u_R \rangle = \overline{ \langle u_L, u_L
\rangle}$ and thus, up to a constant unitary transformation,
$v_L=\overline{v_L}$. Since we have already used the freedom in choosing
orthogonal bases in order to arrange that $u_L = \overline{u_R}$, we
now need to take into account these unitary transformations. This gives
the off-diagonal elements in~(\ref{e:4f1}, \ref{e:4f2}).
We conclude that~(\ref{e:4f0}--\ref{e:4f2}) are necessary conditions.
Substituting~(\ref{e:4f0}--\ref{e:4f2})
into~(\ref{e:4C1}--\ref{e:4C3}) and applying~(\ref{e:4C4}, \ref{e:4C5}),
one verifies immediately that these conditions are also sufficient.
The last statement in Lemma~\ref{lemma4B} implies that $b$ is non-trivial.
\QED

In Chapters~\ref{esec3} and~\ref{esec4} we always restricted
attention to our model variational principle~(\ref{e:2O}). We now
make a few general comments on how our methods could be extended
to other two-point actions, and which features of the Lagrangian
are important for getting a physically interesting continuum
limit.

The methods of Chapter~\ref{esec3} immediately apply to other
two-point actions; the only obstruction is that the gauge terms in
the operator $Q(x,y)$ must be simple fractions in $T^{(n)}_\circ$ and
$\overline{T^{(n)}_\circ}$. The general mechanism is that the
eigenvalues of $A_{xy}$ are influenced by the gauge terms (cf.\
(\ref{e:3k}, \ref{e:3v1}, \ref{e:3v2})). When analyzed in
the EL equations, this leads to conditions for the eigenvalues of
the ``phase matrices'' $W_c$ (see~(\ref{e:3n}) or~(\ref{e:3L})),
and these conditions can finally be translated into constraints
for the dynamical gauge fields. In this last step one uses
crucially that the EL equations are nonlocal in the sense that
they yield relations between the chiral potentials even at distant
points (see e.g.\ (\ref{e:3U})). This gives rise to {\em{global
constraints}}\index{global constraints}, i.e.\ conditions which must hold in all of
space-time. For example, Theorem~\ref{thm32} states that the
dynamical gauge group in case~{{(2)}} must be contained in one
of the groups $({\mathcal{G}}_p)_{p=0,\ldots,3}$ in the whole
space-time, but it cannot be the group ${\mathcal{G}}_{p_1}$ in
one region of space-time and a different group
${\mathcal{G}}_{p_2}$ in another region (as one sees by
considering line integrals which join the two regions).

For the spontaneous block formation, it is essential that the EL equations
are satisfied only if the eigenvalues of $A_{xy}$ are highly degenerate.
The requirement that these degeneracies should be respected by the
scalar potentials can then be used to show that the potentials must split
globally into a direct sum.

While this general mechanism should occur similarly for most other
La\-gran\-gians, the details
depend sensitively on the particular form of the action. Our model
Lagrangian has the special feature that it involves only the absolute squares
of the eigenvalues of $A_{xy}$. This is the reason why Theorem~\ref{thm31}
involves only the absolute squares of $\nu_{nc}$, (\ref{e:3n}), leading to the
relatively weak constraint for the dynamical gauge group~(\ref{e:3gg}) (if we
had, for example, considered instead the polynomial Lagrangian~(\ref{e:2C}),
the gauge terms to highest degree would have led to conditions also for the
phases of $\nu_{nc}$, giving rise to much stronger conditions).
To the next lower degree on the light cone, the phases of $\nu_{nc}$ do
enter the analysis. But since perturbing the absolute square gives rise
to a real part, $\Delta |\lambda_{ncs}|^2 = 2 {\mbox{Re}}(\overline{
\lambda_{ncs}} \: \Delta \lambda_{ncs})$, we can easily arrange that
only the real part of $\nu_{ncs}$ comes into play, and so the phases are
fixed only up to signs. This is a major advantage of our action over e.g.\
polynomial actions, where the same flexibility for the phases can be
arranged only for a large degree of the polynomial. Another action which has
the nice property that it depends only on the absolute squares of the
eigenvalues is the determinant action~(\ref{e:2M}).
Working with the spectral trace leads to the specific problem that one
must handle spectral adjoints. This is clearly a technical complication,
but we do not consider it to be essential for the spontaneous block
formation.

\begin{Remark} \label{remark417} (Massive neutrinos) \em
\index{massive neutrino}
In the analysis of Chapters~\ref{esec3} and~\ref{esec4}, the
structure of the neutrino sector was used several times: In the
vacuum, the chiral cancellations were useful because, as a
consequence, the EL equations were trivially satisfied in the
neutrino sector (see~(\ref{e:26a}) and~(\ref{e:2triv})). In
Theorem~\ref{thm31}, the chiral cancellations in the neutrino
sector are the reason why the dynamical gauge fields are not
allowed to describe a mixing between the neutrinos and the massive
fermions (see the argument leading to~(\ref{e:3r})). In the proof
of Theorem~\ref{thm32}, it was essential that the number of
massive sectors is odd (see~(\ref{e:3S})). Finally, in the
analysis of the degeneracies we always treated the neutrino sector
separately.

Generally speaking, the chiral fermions lead to complications in the
case with interaction, because the dynamical gauge fields were not
allowed to describe a mixing between the massive and the chiral
fermions, and this made it necessary to take scalar potentials into
account. Also in view of recent experimental observations, it thus
seems tempting to consider a neutrino sector which is built up of
massive chiral Dirac seas. This is indeed possible, although we see
the following difficulties. First, it is not clear how chiral
fermions should be described in Minkowski space (for details see
{\S}\ref{sec21}). Furthermore, building in massive chiral fermions
is certainly not easy. Namely, if the resulting neutrino sector does
not give rise to chiral cancellations, we must extend the Lagrangian
in order to arrange that the EL equations are satisfied in the
vacuum. The analysis of the interaction would also be considerably
different. Finally, one should keep in mind that the recent
experiments do not measure the mass of the neutrinos directly, but
merely observe neutrino oscillations, i.e.\ a mixing of the
neutrinos in different generations. This mixing could also be
explained for massless neutrinos if the interaction of the neutrinos
were suitably modified. For these reasons, we feel that before
moving on to massive neutrinos, one should first get a better
understanding of variational principles for a massless neutrino
sector. \em
\end{Remark}

\chapter{The Effective Gauge Group} \label{secegg}
\setcounter{equation}{0} \label{esec5} In this chapter we will
reformulate the interaction of the Dirac particles with chiral and
scalar fields as specified in Theorems~\ref{thm41} and~\ref{thm42}
as an interaction via ``effective'' non-Abelian gauge fields.
Before working out the details in~\S\ref{esec51}
and~\S\ref{esec52}, we begin by explaining the general construction. Consider
the Dirac equation in the presence of chiral and scalar
potentials~(\ref{e:3dir}). Since the dynamical mass matrix
$Y_L=(Y_L)^{(a \alpha)}_{(b \beta)}$ (with $a,b=1,\ldots, 8$ and
$\alpha, \beta = 1,2,3$) need not be Hermitian, we cannot
diagonalize it by a unitary transformation. But using the polar
decomposition, we can at least represent $Y_L$ in the form
\begin{equation}
Y_L \;=\; U_L\: Y^\eff\: U_R^{-1} \label{e:5A}
\end{equation}
with two unitary matrices $U_{L\!/\!R} \in U(3 \times 8)$ and $Y^\eff$
a diagonal matrix with real non-negative entries\footnote{For the reader
not familiar with the polar decomposition we outline the construction.
For a matrix $A \in {\mbox{Mat}}(\sC^n)$ we introduce the Hermitian and
positive semidefinite matrix $R = \sqrt{A^* A}$. A short calculation shows
that the matrix $V$ defined by
\[ V\: u \;=\; \left\{ \begin{array}{cl} u & {\mbox{for }}
u \in {\mbox{Ker}}\: R \\
A\: R^{-1}\: u & {\mbox{for }} u \in ({\mbox{Ker}}\: R)^\perp
\end{array} \right. \]
is unitary and satisfies the relation $A=VR$. Diagonalizing $R$ by a
unitary transformation, i.e.\ $R=WDW^{-1}$ with $D$ diagonal and $W$ unitary,
we obtain the desired representation $A=U_1 D U_2^{-1}$ with $U_1 \equiv VW$
and $U_2 \equiv W$.}.
We introduce the so-called {\em{chiral transformation}}\index{chiral transformation} $V$ by
\begin{equation}
V \;=\; \chi_L\: U_R \:+\: \chi_R\: U_R\:. \label{e:5E}
\end{equation}
Note that the adjoint of $V$,
\[ V^* \;=\; \chi_R\: U_L^{-1} \:+\: \chi_L\: U_R^{-1} \:, \]
is in general different from its inverse, which we denote by a bar,
\label{overlineV}
\[ \overline{V} \;\equiv\; V^{-1} \;=\;
\chi_L\: U_L^{-1} \:+\: \chi_R\: U_R^{-1} \:. \]
Thus the chiral transformation need not be unitary. The chiral transformation
of the Dirac operator is defined by and computed to be
\[ \overline{V}^*\: \left( i \Pdd + \chi_L (\Aslsh_R - mY_R) + \chi_R
(\Aslsh_L - mY_L) \right) \overline{V}
\;=\; i \Pdd + \chi_L \:\Aslsh_R^\eff + \chi_R\: \Aslsh_L^\eff - m Y^\eff \]
with $Y^\eff$ as in~(\ref{e:5A}) and
\begin{equation}
A_c^\eff \;=\; U_c^{-1} A_c U_c \:+\: i U_c^{-1}\: (\partial U_c) \:,
\spc c \in \{L, R\}. \label{e:5D}
\end{equation}
Finally, the effective fermionic projector is obtained from the
auxiliary fermionic projector by the chiral transformation
\begin{equation}
P^\eff \;=\; V\:P\: V^* \:. \label{e:5C}
\end{equation}
It satisfies the effective Dirac equation
\begin{equation}
\left( i \Pdd + \chi_L \:\Aslsh_R^\eff + \chi_R\: \Aslsh_L^\eff - m Y^\eff
\right) P \;=\; 0\:. \label{e:5B}
\end{equation}

Since the chiral transformation is one-to-one, the effective fermionic
projector gives an equivalent formulation of the physical system. The
advantage of the effective description is that the effective mass matrix
$Y^\eff$ is diagonal. This means that if we interpret the
sector index after the chiral transformation as labeling the
different types of Dirac particles (like $u$, $d$, $e$, $\nu_e$, etc.),
the effective scalar potentials describe a dynamical shift of the mass
of each type of fermion, whereas the interaction between different
types of fermions is described only by the effective chiral potentials.
Thus, apart from the fact that we allow for dynamical mass shifts, the
Dirac particles interact as in the standard model via chiral fields.

In general, the effective potentials have locally the form of non-Abelian
gauge potentials. But they cannot be chosen at every point independently,
because it must be possible to represent them in the form~(\ref{e:5D})
with $A_c$ the Abelian gauge potentials of Theorem~\ref{thm41}.
We refer to~(\ref{e:5D}) as the {\em{gauge condition}}\index{gauge condition}.

For clarity, we point out that the unitary transformations in the polar
decomposition~(\ref{e:5A}) are not uniquely determined. Thus, similar
to the freedom of choosing different gauges, there is a certain
arbitrariness in the choice of $U_L$ and $U_R$. Since at infinity the
dynamical mass matrices go over to the mass matrix $Y$ of the vacuum,
we can and will always choose $U_{L\!/\!R}$ such that
\begin{equation}
\lim_{x \to \infty} U_{L\!/\!R}(x) \;=\; \1 \:. \label{e:5N}
\end{equation}

\section{The Chiral Transformation in the Quark Blocks}
\setcounter{equation}{0} \label{esec51}
Using the splitting~(\ref{e:46}), we may disregard
the $U(3)^q$ potentials and can analyze the chiral transformation
in the quark and lepton blocks separately. In this section, we
consider a quark block and for ease in notation omit the
superscript $q$. According to Theorem~\ref{thm42}, the EL
equations to degree $4$ give information only on the partial
traces of the dynamical mass matrices. Therefore, the dynamical
mass matrices, and as a consequence also the chiral transformation
and the effective potentials, are not completely determined. This
means that we have a certain freedom to arbitrarily change these
objects, and we shall use this freedom to make the following
assumption on the form of the effective chiral gauge potentials.
\begin{Def} \label{def51}
The effective chiral potential $A_c$, $c \in \{L, R\}$, has
{\bf{unitary mixing}}\index{unitary mixing} if for every space-time point $x$ there is a unitary
matrix $W_c \in U(3)$ and a $U(2)$ potential $a_c$ such that at $x$,
\begin{equation}
A_c^\eff \;=\; \left( \begin{array}{cc} \1 & 0 \\ 0 & W_c \end{array}
\right) a_c \left( \begin{array}{cc} \1 & 0 \\ 0 & W_c^{-1} \end{array}
\right) \;=\; \left( \begin{array}{cc} a_c^{11} & a_c^{12}\: W_c^{-1} \\
a_c^{21}\: W_c & a_c^{22} \end{array} \right) \label{e:5F}
\end{equation}
(here as in Theorem~\ref{thm42} we use a matrix notation in the sectors).
The matrix $W_c$ is referred to as the {\bf{mixing
    matrix}}\index{mixing matrix}.
\end{Def}
Thus we impose that the effective chiral potentials be trivial on the
generations except for a unitary mixing of the generations in the
off-diagonal matrix elements. This assumption is clearly satisfied for the
gauge potentials in the standard model if we choose $W_R \equiv \1$ and
$W_L$ equal to the CKM mixing matrix. Our ansatz is more general in that
we allow for both left- and right-handed mixing matrices and that
$W_c = W_c(x)$ need not be a constant matrix. Ultimately, the assumption
of unitary mixing should be justified from the EL equations. But this makes
it necessary to consider the EL equations to the degree $3$ on the
light cone, and we do not want to enter this analysis here. Therefore,
we simply take Def.~\ref{def51}
as a physically reasonable technical simplification.

Our first lemma characterizes those chiral transformations which respect
the condition of Def.~\ref{def51}.
\begin{Lemma} \label{lemma51}
The effective chiral potential has unitary mixing if and only if the
unitary transformation $U_c$ in~(\ref{e:5E}) is for all $x$ of the form
\begin{equation}
U_c \;=\; \left( \begin{array}{cc} u_c^{11} & u_c^{12}\: W_c^{-1} \\
u_c^{21}\: W_c & u_c^{22} \end{array} \right) \spc {\mbox{with
$u_c \in U(2)$.}} \label{e:5F2}
\end{equation}
Furthermore, the mixing matrix is constant.
\end{Lemma}
If $U_c$ is of the form~(\ref{e:5F2}) with $W_c$ a constant matrix, it is
obvious that the corresponding effective chiral potential~(\ref{e:5D})
has unitary mixing. In order to show that the converse is also true, we must
analyze the differential equation for $U_c$ and use the boundary conditions
at infinity~(\ref{e:5N}). \\[.5em]
{\em{Proof of Lemma~\ref{lemma51}. }} It suffices to prove the ``only if''
part. Thus we assume that $A_c^\eff$ has unitary mixing and shall derive
that $U_c$ is of the form~(\ref{e:5F2}). For ease in notation we omit
the subscript $c$. According to Theorem~\ref{thm41}, $A$ is diagonal and
can thus be written as $A=\alpha \1 + \beta \sigma^3$ with real functions
$\alpha$ and $\beta$. When substituting into~(\ref{e:5D}), $\alpha$ yields
a contribution to $A^\eff$ with unitary mixing, independent of the form
of $U$. Thus $\alpha$ is irrelevant for the following argument, and we
can assume that $A$ is a multiple of $\sigma^3$.

Let $\Omega$ be the set where the field tensor $F = dA - i A \land A$ is
non-zero,
\[ \Omega \;=\; \left\{ x \:|\: F(x) \neq 0 \right\} . \]
We shall first prove that on each connected component $\Omega_C$ of
$\Omega$, $U$ is for all $x \in \overline{\Omega_C}$ of the form
\begin{equation}
U(x) \;=\; \left( \begin{array}{cc} V_1 & 0 \\ 0 & V_2 \end{array} \right)
u(x) \left( \begin{array}{cc} \1 & 0 \\ 0 & W^{-1} \end{array} \right)
\label{e:5I}
\end{equation}
with $u \in U(2)$ and constant unitary matrices $V_1, V_2, W \in U(3)$.
To this end, we differentiate~(\ref{e:5D}) and~(\ref{e:5F}) to obtain
\begin{equation}
U^{-1} \:F\: U \;=\; F^\eff \;=\; \left( \begin{array}{cc}
f^{11} & f^{12}\: W^{-1} \\ f^{21}\: W & f^{22} \end{array} \right)
\label{e:5G}
\end{equation}
with $f = da + a \land a$ (these relations can be understood immediately
from the behavior of the field tensor under gauge transformations).
At $x \in \Omega_c$, $0 \neq F \sim \sigma^3$. Using this fact
in~(\ref{e:5G}) shows that $U(x)$ must be of the form
\begin{equation}
U \;=\; \left( \begin{array}{cc} B_1 & 0 \\ 0 & B_2 \end{array} \right)
\left( \begin{array}{cc} \cos \varphi & \sin \varphi \\
-\sin \varphi & \cos \varphi \end{array} \right)
\left( \begin{array}{cc} \1 & 0 \\ 0 & W^{-1} \end{array} \right) \label{e:5H}
\end{equation}
with $B_1, B_2 \in U(3)$ and $\varphi \in \R$. Hence at $x$, the first
summand in~(\ref{e:5D}) is of the required form~(\ref{e:5F}), and thus
the second summand must also be of this form. Computing $i U^{-1} (\partial U)$
for $U$ according to~(\ref{e:5H}), one sees that this term is of the
form~(\ref{e:5F}) only if at $x$,
\begin{equation}
\partial W \;=\; 0 \spc {\mbox{and}} \spc \partial B_{1\!/\!2}
\;\sim\; B_{1\!/\!2} \label{e:5K}
\end{equation}
(in the special case $\sin \varphi =0$, we merely obtain that
$\partial (B_2 W^{-1}) \sim B_2 W^{-1}$, but since in this case $U$
only involves the product $B_2 W^{-1}$, we can arrange that $\partial W=0$).
Integrating~(\ref{e:5K}) gives~(\ref{e:5I}).

Let $\Lambda = \R^4 \setminus \Omega$ be the set where $F$ vanishes. We
next prove that on each connected component $\Lambda_C$ of $\Lambda$,
$U$ is of the form
\begin{equation}
U(x) \;=\; \left( \begin{array}{cc} e^{i \phi(x)} & 0 \\ 0 & e^{-i \phi(x)}
\end{array} \right) V\: u_\eff(x)
\left( \begin{array}{cc} \1 & 0 \\ 0 & W^{-1} \end{array} \right)
\label{e:5J}
\end{equation}
with real $\phi$, $u_\eff \in U(2)$ and constant matrices
$V \in U(6)$ and $W \in U(3)$.
In order to derive this formula, we first use that $F=0$ on $\Lambda_C$ to
represent $A$ as a pure gauge potential, i.e.
\begin{equation}
A \;=\; i B^{-1}\: (\partial B) \spc {\mbox{with}} \spc
B \;=\; \left( \begin{array}{cc} e^{-i \phi} & 0 \\ 0 & e^{i\phi}
\end{array} \right) \label{e:5K2}
\end{equation}
and a real function $\phi$. According to the first part of~(\ref{e:5G}),
$F^\eff$ also vanishes. Let us consider what this tells us about the
terms in~(\ref{e:5F}). Using that the phase factors can be absorbed
into $a^{21}$, we can arrange that $W^{-1} (\partial W)$ is trace-free.
Then the contributions to $F^\eff$ involving $\partial W$ and $\partial a$
are linearly independent. From this we conclude that $W$ is constant on
$\Lambda_C$ and that $A^\eff$ can be represented as
\begin{equation}
A^\eff \;=\; i \left( \begin{array}{cc} \1 & 0 \\ 0 & W \end{array} \right)
u_\eff^{-1}\: (\partial u_\eff) \left( \begin{array}{cc}
\1 & 0 \\ 0 & W^{-1} \end{array} \right) \label{e:5L}
\end{equation}
with $u_\eff \in U(2)$. On the other hand, substituting~(\ref{e:5K2})
into~(\ref{e:5D}) gives
\begin{equation}
A^\eff \;=\; i (BU)^{-1}\: \partial (BU)\:. \label{e:5M}
\end{equation}
Differentiating the unitary matrix
\[ B\:U \left( \begin{array}{cc} \1 & 0 \\ 0 & W \end{array} \right)
u^{-1}_\eff \]
and using~(\ref{e:5L}) and~(\ref{e:5M}), one sees that this matrix is
constant on $\Lambda_C$, proving~(\ref{e:5J}).

Note that the representation~(\ref{e:5J}) poses a weaker constraint
on $U$ than equation~(\ref{e:5I}). We shall now prove that on $\Lambda_C$
even~(\ref{e:5I}) holds. If $\Lambda_C$ extends to infinity, we can
according to~(\ref{e:5N}) assume that
\[ \lim_{\Lambda_C \ni x \to \infty} \phi \;=\; 0 \:,\spc
\lim_{\Lambda_C \ni x \to \infty} u_\eff \;=\; \1
\spc {\mbox{and}} \spc V \;=\;
\left( \begin{array}{cc} \1 & 0 \\ 0 & W \end{array} \right) . \]
Then~(\ref{e:5J}) indeed goes over to~(\ref{e:5I}). If conversely
$\Lambda_C$ is compact, we choose $y \in \partial \Lambda_C$. Then at $y$
both~(\ref{e:5I}) and~(\ref{e:5J}) hold, and comparing these formulas
one sees that $V$ must be a diagonal matrix. This implies that on
$\Lambda_C$, (\ref{e:5J}) reduces to~(\ref{e:5I}).

We just showed that for all $X$, $U$ can be represented in the
form~(\ref{e:5I}), where $V$ is constant on each connected component
of $\Omega$ and $\Lambda$. Possibly after multiplying $U$
by piecewise constant unitary transformations and/or absorbing a constant
unitary transformation from $u$ into $V_1$, $V_2$, or $W$, we can assume
that all matrices in~(\ref{e:5I}) are continuous. The asymptotics at
infinity~(\ref{e:5N}) finally yields that $V_1=\1$ and $V_2=W$.
$\;{\mbox{ }}$ \QED

Using the result of the previous lemma in~(\ref{e:5A}), we can now
compute the dynamical mass matrices and analyze the conditions of
Theorem~\ref{thm42}. We restrict attention to the special case, which will
be of relevance later, that the right-handed chiral transformation is trivial.
\begin{Lemma} \label{lemma52}
Assume that $U_R \equiv \1$. If~(\ref{e:4f0}) and~(\ref{e:4f1}) are
satisfied, the mixing matrix and the potential $u_L$ in~(\ref{e:5F})
has the following properties,
\begin{eqnarray}
\acute{Y} \: \grave{W}_L &=& 0 \;=\; \acute{W}_L\: \grave{Y}
\label{e:5O} \\
|Y\: \grave{W}_L| &=& |\grave{Y}| \;=\; |\acute{W}_L\: Y| \label{e:5P} \\
u_L &\in& SU(2)\:. \label{e:5Q}
\end{eqnarray}
Furthermore, (\ref{e:4f0}) and~(\ref{e:4f1}) are also satisfied if
we leave $U_L$ unchanged and set the effective scalar potentials to zero,
\begin{equation}
Y^\eff \;\equiv\; Y\:. \label{e:5R}
\end{equation}
Conversely, if $U_R \equiv \1$ and~(\ref{e:5O}--\ref{e:5R}) are
satisfied, then~(\ref{e:4f0}) and~(\ref{e:4f1}) hold.
\end{Lemma}
{\Proof} Evaluating~(\ref{e:5A}) for $U_L$ according to~(\ref{e:5F}) and
$U_R \equiv \1$ gives
\[ Y_L \;=\; \left( \begin{array}{cc} u_L^{11}\: Y^\eff_1 &
u_L^{12}\: W_L^{-1}\: Y^\eff_2 \\
u_L^{21}\: W_L\: Y^\eff_1 & u_L^{22}\: Y^\eff_2 \end{array} \right) \]
with $Y^\eff = {\mbox{diag}}(Y^\eff_1, Y^\eff_2)$. Assume that~(\ref{e:4f0})
and~(\ref{e:4f1}) are satisfied. Let us evaluate these relations for the
off-diagonal elements of $Y_L$. Since $b$ in~(\ref{e:4f1}) is non-trivial,
the function $u_L^{21}(x)$ does not vanish identically, but clearly it is
zero at infinity. As usual, we implicitly assume that $u^{21}_L$ decays
asymptotically at infinity, without necessarily being zero outside a
compact set. Then we can apply a perturbation argument to the lower left
matrix element of $Y_L$. Namely, (\ref{e:4f0}) yields that
$\acute{W}_L \grave{Y}^\eff_1 = 0$, and taking the asymptotic limit gives
the rhs of~(\ref{e:5O}). The rhs of~(\ref{e:5O}) is obtained similarly
from the upper right matrix element of $Y_L$. Applying the above perturbation
argument to the off-diagonal terms in~(\ref{e:4f1}) yields~(\ref{e:5P}).

We next evaluate~(\ref{e:4f0}) for the diagonal elements of $Y_L$. Since
$Y^\eff$ is a positive matrix, $\hat{Y}^\eff_1$ and $\hat{Y}^\eff_2$ are
real and $\geq 0$. Furthermore, $u_L$ satisfies as a $U(2)$ matrix the
relation $|u^{11}_L| = |u^{22}_L|$. From~(\ref{e:4f0}) we conclude that
$u_L^{11} = \overline{u_L^{22}}$, and thus $u \in SU(2)$.

Finally, it is straightforward to check that~(\ref{e:5O}--\ref{e:5R})
imply the~(\ref{e:4f0}) as well as~(\ref{e:4f1}).
$\;{\mbox{ }}$ \QED

In the remainder of this section we shall analyze and discuss the gauge
condition~(\ref{e:5D}). First, we substitute~(\ref{e:5F2}) and pull
the constant mixing matrix outside,
\[ A^\eff_c \;=\; \left( \begin{array}{cc} \1 & 0 \\ 0 & W_c \end{array}
\right) \left( u_c^{-1} A_c u_c \:+\: i u_c^{-1}\: (\partial u_c)
\right) \left( \begin{array}{cc} \1 & 0 \\ 0 & W_c^{-1} \end{array}
\right) . \]
Next, we decompose the potential and the unitary transformation into
the $U(1)$ and $SU(2)$ parts, i.e.
\[ A_c \;=\; \alpha \:\1 + a\: \sigma^3 \spc {\mbox{and}} \spc
u_c \;=\; e^{-i \phi}\: v \]
with real functions $\alpha, a, \phi$ and $v \in SU(2)$. This gives
\begin{equation}
A^\eff_c \;=\; (\alpha + \partial \phi)\:\1 \:+\:
\left( \begin{array}{cc} \1 & 0 \\ 0 & W_c \end{array} \right)
\left[ a \:v^{-1} \sigma^3 v \:+\: i \:v^{-1} (\partial v) \right]
\left( \begin{array}{cc} \1 & 0 \\ 0 & W_c^{-1} \end{array}
\right) . \label{e:5V}
\end{equation}
Thus $\phi$ describe a usual $U(1)$ gauge transformation. The square bracket
can be regarded as an $SU(2)$ potential, and the matrix $W_c$ introduces
a unitary mixing in the off-diagonal elements. The remaining question is
in which way the expression in the square brackets gives a constraint for
the $SU(2)$ potential.
\begin{Def} \label{def52}
The field tensor $F=dA -i A \land A$ of an $SU(2)$ potential\index{potential!$SU(2)$} $A$ is
{\bf{simple}}\index{potential!simple} if for every $x$ there is a real-valued $2$-form $\Lambda$
and $s \in su(2)$ such that
\begin{equation}
F(x) \;=\; \Lambda\: s \:. \label{e:5T}
\end{equation}
\end{Def}

\begin{Lemma} \label{lemma53}
An $SU(2)$ potential $A$ can be represented in the form
\begin{equation}
A \;=\; a\: v^{-1} \sigma^3 v \:+\: i\: v^{-1} (\partial v) \label{e:5S}
\end{equation}
with $a(x) \in \R$ and $v(x) \in SU(2)$ if and only if its field tensor
is simple.
\end{Lemma}
{\Proof} If $A$ is of the form~(\ref{e:5S}), its field tensor is given by
\[ F \;=\; f\: v^{-1} \sigma^3 v \]
with $f=da$, and this is obviously simple.

Assume conversely that $F$ is simple. We choose $v_1 \in SU(2)$ such that
\[ v_1 \:s\: v_1^{-1} \;=\; \lambda\: \sigma^3 \]
with $\lambda \in \R$ and introduce the gauge potential $\tilde{A}$ by
\begin{equation}
\tilde{A} \;=\; v_1 \:A\: v_1^{-1} \:-\: i\: v_1 (\partial v_1^{-1}) \:.
\label{e:5U}
\end{equation}
From~(\ref{e:5T}) one sees that the corresponding field tensor is
\[ \tilde{F} \;=\; f\: \sigma^3 \]
with the real-valued $2$-form $f=\lambda \Lambda$. From the fact that
$\tilde{F}$ is closed we conclude that $df=0$, and thus there is a
$1$-form $a$ with $f=da$ (note that we are working in Minkowski
space, which is clearly simply connected). By construction, the $SU(2)$ potentials
$\tilde{A}$ and $a \sigma^3$ have the same field tensor $\tilde{F}$.
As a consequence, they are related to each other by an
$SU(2)$ transformation, i.e.
\begin{equation}
\tilde{A} \;=\; a\: v_2^{-1} \sigma^3 v_2 \:+\: i\: v_2^{-1}
(\partial v_2) \label{e:5V2}
\end{equation}
with $v_2 \in SU(2)$. Substituting~(\ref{e:5V2}) into~(\ref{e:5U}) and
solving for $A$ gives~(\ref{e:5S}) with $v=v_2 v_1$.
\QED
With this lemma we have reformulated the gauge condition~(\ref{e:5V})
as a structure condition for the effective field tensor. This makes it
possible to regard the effective chiral potentials as locally defined
objects. More precisely, we shall treat the effective chiral potentials
as local gauge potentials, which are constrained only by local conditions
like e.g.\ that the effective field tensor be simple, but we shall not consider
the corresponding chiral transformation (which involves integrating the
effective potentials and is therefore defined in a nonlocal way).
In particular, when we have conditions between the effective potentials
in the quark and neutrino sectors, we shall always satisfy them by
\label{5locr}
{\em{local relations}}, i.e.\ by algebraic or differential equations
involving the effective potentials. This procedure corresponds to the
usual requirement of locality in physics. It could be further justified
later by the fact that the EL equations yield differential
equations for the effective potentials (the ``field equations''), and
it seems impossible to satisfy such differential equations if the
effective potentials obey nonlocal constraints.

\section{The Chiral Transformation in the Lepton Block}
\setcounter{equation}{0} \label{esec52}
We come to the analysis in the lepton block; for
ease in notation the superscript $l$ will be omitted. As a
consequence of the chiral massless fermions, the dynamical
matrices are different in the lepton and quark blocks. More
precisely, $Y_L$ and $Y_R$ must now be of the form~(\ref{e:4f00})
and~(\ref{e:4f2}, \ref{e:4f3}). We shall first show that these
conditions are incompatible with a unitary mixing and then resolve
this problem by modifying the mixing in the right-handed
component. Let us assume that~(\ref{e:4f2}) or~(\ref{e:4f3}) are
satisfied for dynamical mass matrices of the form~(\ref{e:5A})
with $U_{L\!/\!R}$ according to Lemma~\ref{lemma51}. Then,
choosing a space-time point where $b \neq 0$, we have
\begin{equation}
(U_L \:Y^\eff\: \grave{U}^{-1}_R)\: I_2 \;=\; 0 \spc{\mbox{but}}\spc
I_2\: (U_R\: Y^\eff\: \grave{U}_L^{-1}) \;\neq\; 0\:,
\label{e:5aa}
\end{equation}
where $I_{1\!/\!2}$ are again the projectors on the two sectors.
Introducing the unit vectors $n=3^{-\frac{1}{2}}\: (1,1,1) \in \C^3$
and $u=(0,n) \in \C^6$, we can write the first condition in~(\ref{e:5aa})
without a partial trace as
\begin{equation}
U_L\: Y^\eff\: U_R^{-1}\:u \;=\; 0\:. \label{e:5b}
\end{equation}
In the vacuum, the mass matrix $Y^\eff=Y$ is strictly positive in the
first sector. A perturbation argument yields that, at least for weak fields,
the effective mass matrix is of the form $Y^\eff = {\mbox{diag}}(Y^\eff_1,
Y^\eff_2)$ with $Y^\eff_1>0$. Therefore, the condition~(\ref{e:5aa}) can
only be satisfied if $U^{-1}_R u$ vanishes in the first sector. Thus,
using~(\ref{e:5F2}),
\[ \left( \begin{array}{cc} v^{11} & v^{12}\: W_c^{-1} \\
v^{21}\: W_c & v^{22} \end{array} \right) \left( \begin{array}{c}
0 \\ n \end{array} \right) \;=\; \left( \begin{array}{c}
0 \\ {\mbox{$*$}} \end{array} \right) \]
with $v=u_R^{-1}$. This implies that $v^{12}=0$ and thus $U_R \equiv 1$.
Using this result in~(\ref{e:5b}), we obtain that $Y^\eff u=0$,
and since $Y^\eff$ is a diagonal matrix with non-negative entries, we
conclude that $Y^\eff_2=0$. Finally, the relations $U_R=\1$ and
$Y^\eff_2=0$ imply that $I_2 U_R Y^\eff=0$, in contradiction to
the second equation in~(\ref{e:5aa}).

In order to avoid the above contradiction, the vector $U_R^{-1} u$ must
vanish identically in the first sector without $U_R$ being trivial. The
natural way to arrange this is to replace the unitary matrix $W_R$
in~(\ref{e:5F}) by a matrix which is zero on $\bra n \ket$ and is
unitary on $\bra n \ket^\perp$. In analogy to the procedure in the
previous section, we first introduce the corresponding effective potentials
and determine $U_R$ afterwards. We let $\Pi$ be the projector
\begin{equation}
\Pi \;=\; |n \ket \bra n | \spc {\mbox{with}} \spc
n \;=\; \frac{1}{\sqrt{3}}\: (1,1,1)\:. \label{e:52a}
\end{equation}

\begin{Def} \label{def53}
The effective potential $A_R^\eff$ has {\bf{projected
    mixing}}\index{projected mixing} if for every
space-time point there is a unitary matrix $W_R \in U(3)$ with
\[ W_R\: n \;=\; n \:, \]
as well as real functions $b^1_R$ and $b^2_R$
and a $U(2)$ potential $a_R$ such that at $x$,
\begin{equation}
A^\eff_R \;=\; \left( \begin{array}{cc} b^1_R & 0 \\ 0 & b^2_R \end{array}
\right) \:+\: (1-\Pi) \left( \begin{array}{cc}
a_R^{11} & a_R^{12}\: W_R^{-1} \\
A_R^{21}\: W_R & a_R^{22} \end{array} \right) \:. \label{e:52b}
\end{equation}
$W_R$ is called the {\bf{mixing matrix}}\index{mixing matrix}.
\end{Def}

\begin{Lemma} \label{lemma54}
The effective potential $A_R^\eff$ has projected mixing if and only if
the unitary transformation $U_R$ in~(\ref{e:5E}) is for all $x$ of the form
\begin{equation}
U_R \;=\; \Pi \left( \begin{array}{cc} v_1 & 0 \\ 0 & v_2 \end{array}
\right) \:+\: (\1-\Pi) \left( \begin{array}{cc} u_{11} & u^{12}\: W_R^{-1} \\
u^{21}\: W_R & u^{22} \end{array} \right) \label{e:5cc}
\end{equation}
with $v_{1\!/\!2} \in U(1)$ and $u \in U(2)$. Furthermore, the matrix $W_R$
is constant.
\end{Lemma}
{\Proof} We consider the effective potential on $\bra n \ket$ and
$\bra n \ket^\perp$ separately. On $\bra n \ket^\perp$, $A^\eff_R$ is of
the form as in Def.~\ref{def51}, and so Lemma~\ref{lemma52} applies.
On $\bra n \ket$, on the other hand, $A^\eff_R$ is a diagonal potential,
and integrating the differential equation for $U_R$ as in the proof of
Lemma~\ref{lemma51} shows that $U_R$ is also diagonal.
\QED

The effective potentials in the second summand in~(\ref{e:52b}) have
different properties than usual gauge fields. First, one should keep in
mind that the right-handed potential $A^R_\eff$ does not couple to the
left-handed massless fermions, and therefore the off-diagonal elements
in~(\ref{e:52b}) cannot be regarded as describing an interaction between
the massive leptons and the neutrinos. Indeed, one must be careful about
associating any physical interaction to the second summand in~(\ref{e:52b}),
because the factor $(\1-\Pi)$ gives zero when the partial trace is taken,
and also because some degrees of freedom of the corresponding potentials
drop out of the fermionic projector when $\tilde{t}$ is multiplied by
the chiral asymmetry matrix~(\ref{e:42x}). For these reasons, we regard
the off-diagonal elements in~(\ref{e:52b}) as describing a new type
of interaction whose physical significance is not clear at the moment.
We refer to an effective potential which involves a factor $(\1-\Pi)$
as a \label{5nil} {\em{nil potential}}\index{potential!nil}.

The next lemma is very useful because it allows us to compute the vectors
$a$ and $b$ in Theorem~\ref{thm42} without specifying $U_R$. In this way,
we can get around the detailed analysis of the nil potential.
\begin{Lemma} \label{lemma55}
Suppose that $A^\eff_L$ and $A^\eff_R$ have unitary and projected mixing,
respectively. Then $U_L$ and $U_R$ can be chosen such that
\[ \grave{Y}_L \;=\; U_L\: \grave{Y}^\eff \:. \]
\end{Lemma}
{\Proof} Since $(\1-\Pi) n=0$, the partial trace of the second summand
in~(\ref{e:5C}) is zero, whereas in the first summand the factor $\Pi$
drops out. Thus $\grave{Y}_L = U_L Y^\eff \grave{V}$ with $V$ a diagonal
$U(2)$ matrix. This matrix commutes with $Y^\eff$ and can thus be absorbed
into $U_L$.
\QED

\section{Derivation of the Effective Gauge Group}
\setcounter{equation}{0} \label{esec53}
We are now ready to prove the main result of this chapter.
\begin{Thm} \label{thm51}
We consider the EL equations corresponding to the Lagrangian (\ref{e:2N})
under the assumptions of Theorem~\ref{thm41}. Assume furthermore that the
right-handed effective potentials in the lepton block have projected
mixing and that all other effective potentials have unitary mixing
(see Defs.~\ref{def51} and~\ref{def53}). Imposing local relations between
the effective potentials (as explained on page~\pageref{5locr}), the right-handed
chiral transformation is trivial in the quark blocks, $U^q_R \equiv \1$.
The mixing matrices are constant and satisfy the relations
\begin{eqnarray}
\acute{Z}\: \grave{W}^q_L &=& \acute{W}^q_L\: \grave{Z} \;=\;
\acute{Z}\: \grave{W}^l_L \;=\; 0 \label{e:5a} \\
|Z\: \grave{W}^q_L| &=& |\acute{W}^q_L\: Z| \;=\; |Z\: \grave{W}^l_L|
\;=\; |\grave{Z}| \:, \label{e:5bv}
\end{eqnarray}
where $Z=\frac{1}{m}\: {\mbox{diag}}(m_1, m_2, m_3)$ is the mass matrix
of the massive fermions. The effective Dirac operator is of the following
form,
\begin{eqnarray}
i \Pdd \!\!\!\!\!\!\!\!\!&& -\:m \left(
Y^\eff_q \oplus Y^\eff_q \oplus Y^\eff_q \oplus Y^\eff_l \right)
\label{e:5c} \\
&&+\: \chi_R  \left( \Aslsh^\eff_L \oplus \Aslsh^\eff_L \oplus
\Aslsh^\eff_L \oplus \Aslsh^\eff_L \right) \:+\:
\chi_L\: \Aslsh_R \left( \sigma^3 \oplus \sigma^3 \oplus \sigma^3 \oplus
\sigma^3 \right) \label{e:5d} \\
&&+\: (\Aslsh^q\: \1) \oplus \left( \Aslsh^l\: \1 \:+\: (\1 - \Pi)\:
\Aslsh^{\mbox{\scriptsize{nil}}}_R \right) \:. \label{e:5e}
\end{eqnarray}
Here $A^\eff_L$ is a $2 \times 2$ matrix potential, $A^q$ is a
$3 \times 3$ matrix potential, $A_R$ and $A^l$ are vector fields,
and $A^{\mbox{\scriptsize{nil}}}$ is a nil potential. The effective gauge group is
\begin{equation}
{\mathcal{G}}^\eff \;=\; SU(2)^\eff_L \times U(1)_R \times
U(3)^q \times U(1)^l\:. \label{e:5g}
\end{equation}
The only constraint for the chiral potentials is that the field tensor
corresponding to $A^\eff_L$ must be simple (see Def.~\ref{def52}).
The EL equations to degree $4$ are satisfied for the same effective potentials
if the effective scalar potentials are set to zero,
\begin{equation}
Y^\eff \;=\; Y\:. \label{e:5h}
\end{equation}
\end{Thm}
{\Proof}
We rewrite~(\ref{e:4C2}) and~(\ref{e:4C3}) in terms of the effective
potentials. According to Lemma~\ref{lemma55}, $\grave{Y}^l_L$ is
independent of $U_R$, and thus $\beta$ and $\gamma$ can be expressed in
terms of~$Y^\eff_l$, the Abelian potentials $A_V$, $A_s$ in~(\ref{e:47a}),
and the non-Abelian effective potential $A^\eff_L$ in the lepton block.
Furthermore, (\ref{e:4C2}) and~(\ref{e:4C3}) can be satisfied by
local relations between the effective potentials only if the non-Abelian
effective gauge fields coincide in the quark and lepton blocks. This
yields the effective chiral gauge group~(\ref{e:5g}) and the form of the
corresponding potentials in~(\ref{e:5d}) and~(\ref{e:5e}).
Lemma~\ref{lemma53} shows that the gauge conditions~(\ref{e:5D})
are satisfied if and only if the field tensor corresponding to
$A^\eff_L$ is simple.

According to Lemmas~\ref{lemma51} and~\ref{lemma54}, the mixing matrices
are constant. Using Lemma~\ref{lemma52}, we obtain~(\ref{e:5a}) and~(\ref{e:5bv}) in the quark blocks. The corresponding relations in the lepton block are
obtained similarly from~(\ref{e:4f00}) and~(\ref{e:4f2}). Finally,
(\ref{e:5h}) follows immediately from Lemma~\ref{lemma52} and an
analogous perturbation argument in the lepton block.
\QED

Note that~(\ref{e:5a}) and~(\ref{e:5bv}) are not satisfied if $W_L$
is equal to the identity matrix. Thus the EL equations imply that the
off-diagonal components of the effective gauge fields involve a non-trivial
mixing of the generations. The fact that we may set $Y^\eff$ equal to $Y$,
(\ref{e:5h}), means that the effective scalar potentials are irrelevant
for the derivation of the effective gauge group. But this does not answer
the question whether effective scalar potentials may occur in the system
or not; to this end one must analyze the EL equations to lower degree on
the light cone.

Finally we point out that Theorem~\ref{thm51} only gives necessary
conditions for the effective potentials. But it is to be expected the
analysis of the EL equations to degree $3$ will give
further constraints for the effective potentials. Taking this into
account, the results of Theorem~\ref{thm51} are in perfect
agreement with physics: The $SU(3)^q$ and $SU(2)^\eff_L$ can be
identified with the strong and weak gauge groups, respectively.
The coupling of the corresponding gauge potentials to the fermions
is exactly as in the standard model. The $SU(3)^q$ is a free gauge
group (see the discussion on page~\pageref{3fgg}), and this implies that the
corresponding gauge fields are necessarily massless. However, the
$SU(2)^\eff_L$ is spontaneously broken. The electromagnetic
potential corresponds to a linear combination of the potentials of
the subgroup $SU(2)^\eff_L \times U(1)_R \times U(1)^q \times
U(1)^l \subset {\mathcal{G}}^\eff$, characterized by the property
that it is a traceless vector potential. In order to make the
connection to the standard model more precise, it remains to
explain why only this particular linear combination occurs, and
furthermore one must analyze the masses of the spontaneously
broken gauge fields in the resulting field equations.
To answer these questions, one needs to analyze the EL equations to
degree $3$ on the light cone; this is an interesting project for the
future.

\begin{appendix}
\include{pfp3}
\end{appendix}

\backmatter
\bibliographystyle{amsalpha}

\vspace*{.3cm}%
\noindent
{\footnotesize{%
{{NWF I - Mathematik, Universit\"at Regensburg, D-93040 Regensburg, Germany}} \\
Email: {\tt{Felix.Finster@mathematik.uni-regensburg.de}}}

\printindex
\cleardoublepage
\markboth{NOTATION INDEX}{}
\addcontentsline{toc}{chapter}{Notation Index}
\twocolumn[\vspace*{2.8cm} \centerline{\bf{\Large{Notation Index}}} \vspace*{1.9cm}]

\begin{itemize}
\item[] $(M,\langle .,.\rangle)$ \,\pageref{Mrangle}
\item[] $\mathrm{E}$ \,\pageref{E}
\item[] $J_x$ \,\pageref{Jx}
\item[] $J$, $J^\lor_x$, $J^\land_x$ \,\pageref{Jlorx}
\item[] $I$, $I^\lor_x$, $I^\land_x$ \,\pageref{Ilorx}
\item[] $L$, $L^\lor_x$, $L^\land_x$ \,\pageref{Elorx}
\item[] $T_pM$ \,\pageref{Tpm}
\item[] $g_{jk}$ \,\pageref{gjk}
\item[] $\nabla$ \,\pageref{nabla}
\item[] $R^i_{jkl}$ \,\pageref{Rijkl}
\item[] $T_{jk}$ \,\pageref{Tjk}
\item[] $S$ \,\pageref{S}, \pageref{S2}, \pageref{e:2a}
\item[] $\Box$ \,\pageref{Box}
\item[] $\{.,.\}$ \,\pageref{{.,.}}
\item[] $\gamma^j$ \,\pageref{gammaj}
\item[] $\vec{\sigma}$ \,\pageref{vecsigma}
\item[] $\Pdd$ \,\pageref{Pdd}
\item[] $\Sl .|. \Sr$ \,\pageref{Sl.|.Sr}
\item[] $A^\ast$ \,\pageref{Aast}
\item[] $J^k$ \,\pageref{Jk}
\item[] $\rho$ \,\pageref{rho}
\item[] $\epsilon_{jklm}$ \,\pageref{epsilonjklm}
\item[] $\chi_L$, $\chi_R$ \,\pageref{chiL}
\item[] $(.|.)$ \,\pageref{(.|.)}, \pageref{(.|.)b}
\item[] $h$ \,\pageref{h}
\item[] $\Psi_{\vec{p} s \epsilon}$ \,\pageref{Psivecpsepsilon}
\item[] $d\mu_{\vec{p}}$ \,\pageref{dmuvecp}
\item[] $\land$ \,\pageref{land}
\item[] $\mathcal{F}$, $\mathcal{F}^n$ \,\pageref{mathcalFn}
\item[] $\hat{\Psi}_{\vec{p} s\epsilon}$, $\hat{\Psi}^\dagger_{\vec{p} s \epsilon}$ \,\pageref{hatPsidagger}
\item[] $\mathcal{G}$ \,\pageref{mathcalG}
\item[] $\mathcal{D}$ \,\pageref{mathcalD}, \pageref{mathcalD2}
\item[] $\sigma^{jk}$ \,\pageref{sigmajk}
\item[] $\Tr$ \,\pageref{Tr}, \pageref{Tr2}
\item[] $D$ \,\pageref{gcd}, \pageref{D}
\item[] $\bra .|. \ket$ \,\pageref{langle.|.rangle}
\item[] $\mathcal{B}$ \,\pageref{mathcalB}
\item[] $s_m$ \,\pageref{sm}
\item[] $s^\lor_m$, $s^\land_m$ \,\pageref{slorm}
\item[] $s^F_m$ \,\pageref{sFm}
\item[] $P^{\mbox{\scriptsize{sea}}}$ \,\pageref{Psea}, \pageref{Psea2}, \pageref{84}
\item[] $p_m$ \,\pageref{pm}
\item[] $k_m$ \,\pageref{km}
\item[] $\tilde{k}_m$ \,\pageref{tildekm}
\item[] $\tilde{p}_m$ \,\pageref{tildepm}
\item[] $\tilde{s}_m$ \,\pageref{sm2}
\item[] $p$ \,\pageref{p}
\item[] $k$ \,\pageref{k}
\item[] $X$ \,\pageref{X}
\item[] $Y$ \,\pageref{Y}
\item[] $P(x,y)$ \,\pageref{1g}, \pageref{Pfut}, \pageref{P(x,y)}
\item[] $c_{\mbox{\scriptsize{norm}}}$ \,\pageref{1g}, \pageref{Pfut}, \pageref{cnorm}
\item[] $\Psi_{\mbox{\scriptsize{in}}}$, $\Psi_{\mbox{\scriptsize{out}}}$ \,\pageref{Psiin}
\item[] $\Aslsh_L$, $\Aslsh_R$ \,\pageref{AslshR}
\item[] $\Phi$ \,\pageref{Phi}
\item[] $\Xi$ \,\pageref{Xi}
\item[] $S^\lor_{m^2}$ \,\pageref{Slorm2}
\item[] $Y_L$, $Y_R$ \,\pageref{YL}
\item[] $S^{(l)}$ \,\pageref{Sl}
\item[] $\Pexp$ \,\pageref{Pexp}
\item[] $T_{m^2}$ \,\pageref{Tm2}
\item[] $T_{m^2}^{\mbox{\scriptsize{reg}}}$ \,\pageref{Tregm2}
\item[] $T^{(l)}$ \,\pageref{T(l)}, \pageref{T^(n)}
\item[] $P^{\mbox{\scriptsize{he}}}$ \,\pageref{Phe}
\item[] $P^{\mbox{\scriptsize{le}}}$ \,\pageref{Ple}
\item[] $X^i$ \,\pageref{X^i}, \pageref{X^i2}
\item[] $\vert x\alpha \ket$ \,\pageref{vertxalpha}
\item[] $\Sl \Psi \:|\: \Phi \Sr$ \,\pageref{SlPsi|PhiSr}
\item[] $M$ \,\pageref{M}
\item[] $E_x$ \,\pageref{E_x}
\item[] $\mathcal{L}$ \,\pageref{mathcalL}
\item[] $A$ \,\pageref{A}
\item[] $|.|$ \,\pageref{|.|}
\item[] $\overline{A}$ \,\pageref{overlineA}
\item[] ${\tr}$ \,\pageref{tr}
\item[] $Q(x,y)$ \,\pageref{Q}, \pageref{e:2c}
\item[] $\mathcal{B}$ \,\pageref{mathcalB2}
\item[] $\hat{P}$ \,\pageref{hatP}
\item[] $E_P$ \,\pageref{E_P}
\item[] $s$ \,\pageref{s}
\item[] $l$ \,\pageref{l}
\item[] $u$ \,\pageref{u}
\item[] $v$ \,\pageref{v}
\item[] $l_{\mbox{\scriptsize{max}}}$ \,\pageref{lmax}
\item[] $\alpha_{\mbox{\scriptsize{max}}}$ \,\pageref{alphamax}
\item[] $\varepsilon_{\mbox{\scriptsize{shear}}}$
  \,\pageref{varepsilon_shear}
\item[] $T^{(n)}_{[p]}$ \,\pageref{T^(n)_[p]}
\item[] $T^{(n)}_{\{p\}}$ \,\pageref{T^(n)_{p}}
\item[] $\xi$ \,\pageref{xi}
\item[] $L$ \,\pageref{L}
\item[] $c_{\mbox{\scriptsize{reg}}}$ \,\pageref{c_reg}
\item[] $\mathcal{M}$ \,\pageref{mathcalM}, \pageref{mathcalM2}
\item[] $\lambda_+$, $\lambda_-$ \,\pageref{lambda_pm}
\item[] $F_+$, $F_-$ \,\pageref{F_pm}
\item[] $z^{(n)}_{[r]}$ \,\pageref{z^(n)_[r]}
\item[] $\deg$ \,\pageref{deg}
\item[] $\hat{Q}(p)$ \,\pageref{hatQ(p)}
\item[] $\mathcal{C}$ \,\pageref{mathcalC}
\item[] $\mathcal{C}^\lor$, $\mathcal{C}^\land$ \,\pageref{mathcalClor}
\item[] $A_L$, $A_R$ \,\pageref{A_L/R}
\item[] $Y_L$, $Y_R$ \,\pageref{Y_L/R}
\item[] $W_L$, $W_R$ \,\pageref{W_c}
\item[] $\nu_{nc}$ \,\pageref{nu_nc}
\item[] $I_{nc}$ \,\pageref{I_nc}
\item[] $B_p$ \,\pageref{e:3f1}
\item[] $F_p$ \,\pageref{e:3f1}
\item[] $\mathcal{B}_p$ \,\pageref{mathcalB_p}
\item[] $\mathcal{F}_p$ \,\pageref{mathcalF_p}
\item[] $\mathcal{G}_p$ \,\pageref{e:3A}
\item[] $\acute{Y}_L$, $\acute{Y}_R$ \,\pageref{acuteY_LR}
\item[] $\grave{Y}_L$, $\grave{Y}_R$ \,\pageref{graveY_LR}
\item[] $\hat{Y}_L$, $\hat{Y}_R$ \,\pageref{hatY_LR}
\item[] $I_\uparrow$, $I_\downarrow$ \,\pageref{I_arrows}
\item[] $P^q$ \,\pageref{P^q}
\item[] $P^l$ \,\pageref{P^l}
\item[] $\overline{V}$ \,\pageref{overlineV}
\item[] $Y^\eff$ \,\pageref{e:5A}
\item[] $A_L^\eff$, $A_R^\eff$ \,\pageref{e:5D}
\item[] $K_a$ \,\pageref{K_a}
\item[] $H_a$ \,\pageref{H_a}
\item[] $S^\Join_a$ \,\pageref{SJoin_a}
\item[] $K^{(n)}$ \,\pageref{K^(n)}
\item[] $S^{(n)}_\Join$ \,\pageref{S^(n)Join}
\item[] $H^{(n)}$ \,\pageref{H^(n)}

\end{itemize}


\end{document}